\documentclass{article}

\usepackage{epsf,hyperref}
\usepackage{epsfig}
\usepackage{graphics}
\usepackage{graphicx}
\usepackage[centertags]{amsmath}
\usepackage{amsfonts}
\usepackage{amsthm}
\usepackage{amssymb}

\usepackage{url}
\usepackage{subfigure}
\usepackage{hyperref}
\begin{document}

\title{On patterns and dynamics of Rule 22 cellular automaton}

\author{Genaro J. Mart\'{\i}nez$^{1,2}$, Andrew Adamatzky$^{2}$, Rolf Hoffmann$^{3}$ \\ Dominique D\'{e}s\'{e}rable$^{4}$, Ivan Zelinka$^{5}$}

\date{29 March 2020\footnote{Published in: {\em Complex Systems 28(2), 125-174, 2019}. \url{https://www.complex-systems.com/abstracts/v28_i02_a01/}}}

\maketitle

\begin{centering}
$^1$ Laboratorio de Ciencias de la Computaci\'on, Escuela Superior de C\'omputo, Instituto Polit\'ecnico Nacional, M\'exico. (\url{genaro.martinez@uwe.ac.uk}) \\
$^2$ Unconventional Computing Lab, University of the West of England, Bristol, United Kingdom. (\url{andrew.adamatzky@uwe.ac.uk}) \\
$^3$ Technische Universit\"{a}t Darmstadt, Darmstadt, Hessen, Deutschland. (\url{hoffmann@informatik.tu-darmstadt.de}) \\
$^4$ Institut National des Sciences Appliqu\'{e}es, Rennes, France. (\url{domidese@gmail.com}) \\
$^5$ Fakulta Elektrotechniky a Informatiky, Technick\'a Univerzita Ostrava, Czechia. (\url{ivan.zelinka@vsb.cz}) \\
\end{centering}

\begin{abstract}
Rule 22 elementary cellular automaton (ECA) has a 3--cell neighborhood, binary cell states, where a cell takes state `1' if there is exactly one neighbor, including the cell itself, in state `1'.
In Boolean terms the cell-state transition is a {\sc xor} function of three cell states. 
In physico--chemical terms the rule might be seen as describing propagation of self-inhibiting quantities/species. 
Space-time dynamics of Rule 22 demonstrates non-trivial patterns and quasi-chaotic behavior. 
We characterize the phenomena observed in this rule using mean field theory, attractors, de Bruijn diagrams, subset diagrams, filters, fractals and memory.
\end{abstract}
%
%
%
\section{Introduction}
\label{section:Introduction}
%
%
%
\subsection{Rule 22: history}
%
%
Elementary cellular automata (ECA) 
\cite{Wolfram-1983a,Wolfram-1994}
are one--dimensional arrays of finite state machines, or cells, which take states `0' or `1' and update their state depending on their own current state and on the state of their two immediate neighbors. 
Rule 22 ECA has a simple cell--state transition function: a cell takes state `1' if exactly one of its neighbors, including the cell itself, is in state `1'; otherwise, the cell takes state `0'. When perturbed at a single site the automaton exhibits something similar to recurrent wave--fronts in excitable media, which develop into fractal structures of Sierpi\'{n}ski  gasket
\cite{Wolfram-1995,Redeker:Adamatzky:Martinez-2013}. 
Due to countless generation and annihilation of wave--fronts, a dynamics of Rule 22 is sometimes characterized as chaotic
\cite{Wolfram-1994}
which rather reflects its unpredictability than any relation to noise or bifurcations. 
Most results of studies about Rule 22 were in algebraic properties or statistical approximations of the automaton dynamics. 
Thus, Zabolitzky
\cite{Zabolitzky-1988} 
reported results of an extended probabilistic analysis estimating non-trivial behavior on very large arrays perturbed by configurations with low densities of state `1'. 
He discovered critical properties that cannot be reproduced when the automaton is perturbed by a random configuration. 
McIntosh provided a systematic analysis of small configurations emerging in Rule 22
\cite{McIntosh-2009}; 
he proposed similarities with configurations observable in Conway's Game of Life. A topological analysis linked to chaotic behavior of the rule can be found in 
\cite{Jin:Chen:Yang-2009}. 
%
%
\subsection{Rule 22: definition}
%
%
A one--dimensional cellular automaton CA$(k, r)$ is an array of cells $x_i$ where $i \in \mathbb{Z}$.
Each cell takes on a value from an alphabet $S = \{0, 1,..., k-1\}$ with $k$ symbols.
A chain of cells \{$x_i$\} of finite length $n$ represents a string or global configuration $c$ on $\Sigma$. 
The set of finite configurations is represented as $\Sigma^n$. 
An evolution is a sequence of configurations $\{c_i\}$ given by the mapping $\Phi:\Sigma^n \rightarrow \Sigma^n$ and their global relation is provided by $ \Phi(c^t) \rightarrow c^{t+1}$ where $t$ is a discrete time and every global state of $c$ is a sequence of cell states. 
Cells of each configuration $c^t$ are updated to the next configuration $c^{t+1}$ simultaneously by a local transition function 
$
\varphi: S^{  \, 2 \, r + 1} \rightarrow S
$
as
\[ 
\varphi(x_{i-r}^t, \ldots, x_{i}^t, \ldots, x_{i+r}^t) \rightarrow x_i^{t+1}
\]
acting on a neighborhood of $x_i$ of length $2 \, r + 1$.
For (elementary) ECA$(2,1)$, 
$
\varphi: S^{3} \rightarrow S
$
becomes 
\begin{equation}
\varphi(x_{i-1}^t, x_{i}^t, x_{i+1}^t) \rightarrow x_i^{t+1}
\label{eq: ECA(2;1)}	
\end{equation}
and for Rule 22, its local cell--state transition is given by:
\begin{equation}
\varphi_{R22} = \left\{
	\begin{array}{lcl}
		1 & \mbox{if} & 100, 010, 001 \\
		0 & \mbox{if} & 111, 110, 101, 011, 000
	\end{array} \right. 
\label{eq:ECAR22}	
\end{equation}
\begin{figure}[!tbp]
\begin{center}
\subfigure[]{\includegraphics[width=0.49\textwidth]{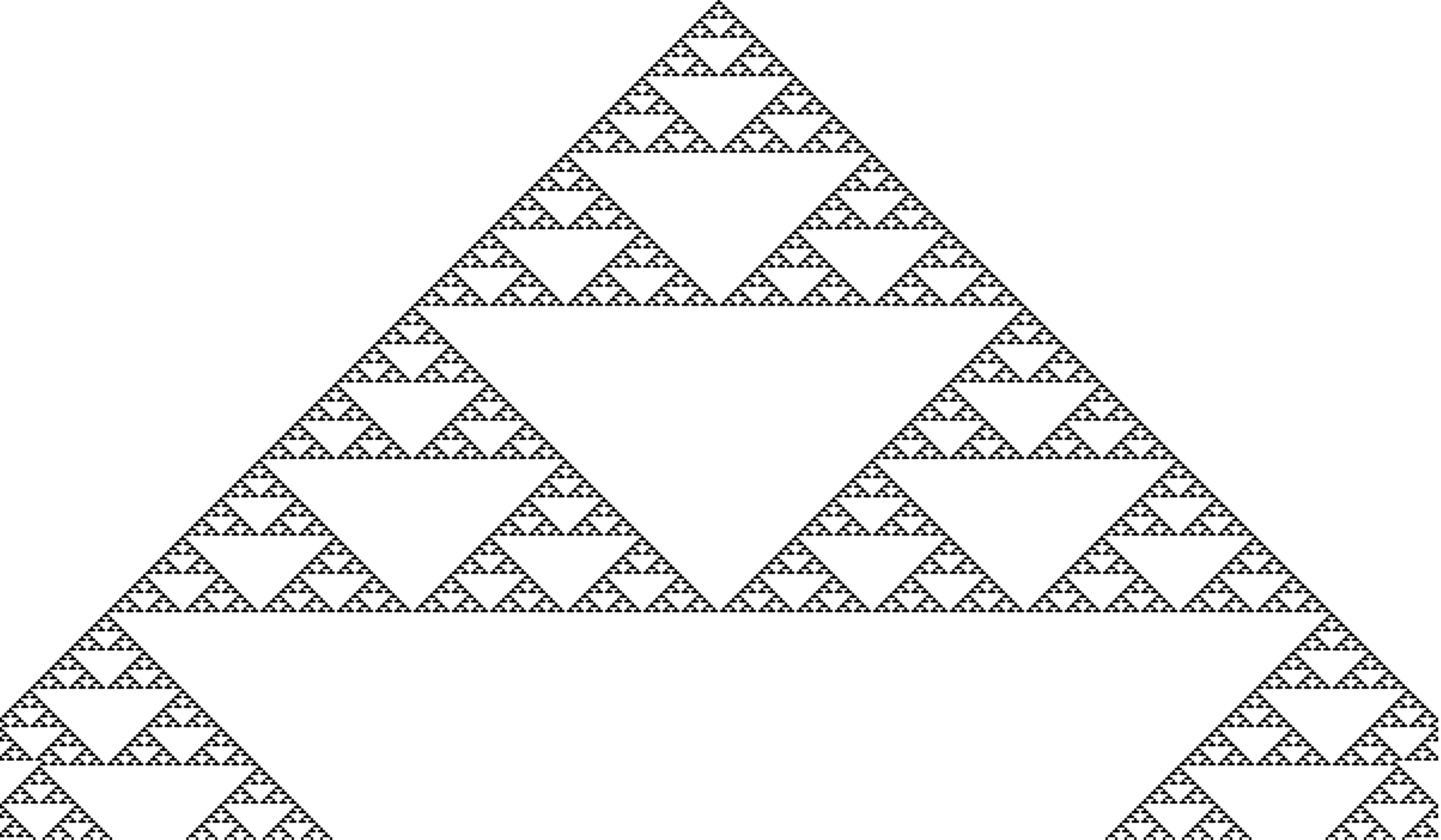}} 
\subfigure[]{\includegraphics[width=0.49\textwidth]{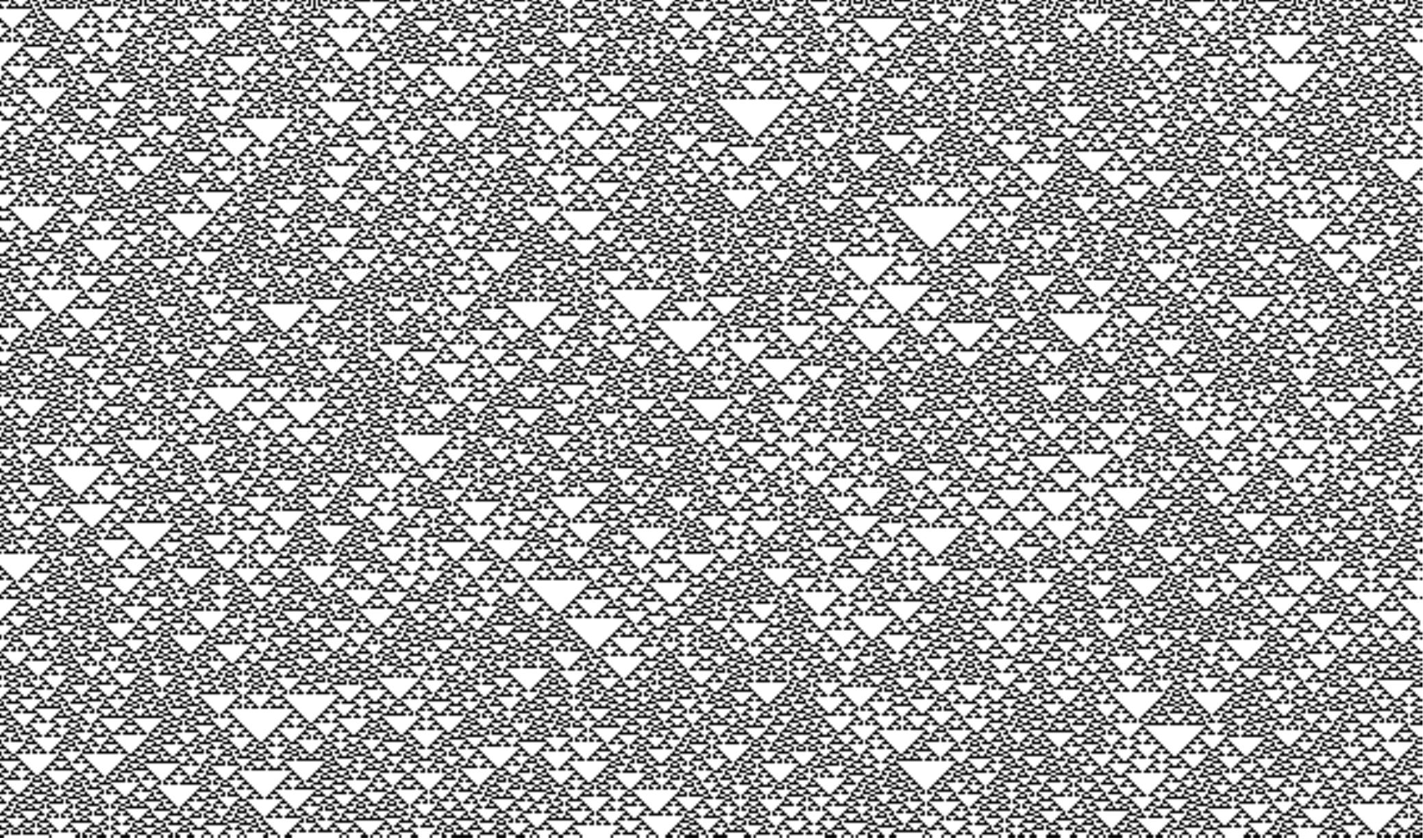}}
\end{center}
\caption{Exemplar dynamics in ECA Rule 22. (a)~Development from a single cell in state 1. (b)~Development from a random configuration with density of 1-cells 0.5. 
Both space-time diagrams evolve on a ring of 600 cells for 350 generations. 
Time evolves from top to bottom.
}
\label{evolR22}
\end{figure}
Rule 22 displays a typical chaotic global behavior from random initial conditions. Figure~\ref{evolR22}a shows the evolution with an initial condition starting with a single cell in state `1'. A pattern growing is a fractal, similar to a Sierpi\'{n}ski gasket. Figure~\ref{evolR22}b shows a development from a random initial configuration with a density 0.5 of cells in state `1'. 
%
%
\subsection{On patterns and dynamics in Rule 22}
%
%
Rule 22 is considered as chaotic because:
\begin{enumerate}
\item future configuration of the automaton is  completely determined from its initial state because of the deterministic rule and synchronous updating,
\item development of the automaton is sensitive to initial conditions (tiny perturbation might lead to dramatic events),
\item global transition graph has dense periodic orbits (attractors),
\item configurations evolved can be characterized as random.
\end{enumerate}
We undertake an extensive and systematic analysis of Rule 22 using different approximations aiming at discovering an emergence of novel non-trivial patterns, periodic patterns, Garden of Eden configurations. 
These configurations are discovered with the help of encoding initial conditions into regular expressions, de Bruijn diagrams, subset diagrams, cycle diagrams, fractals, and jump-graphs. We also show an effect of memory upon dynamics of Rule 22.
%
%
\section{Mean field theory}
\label{section:Mean field theory}
%
%
Mean field theory allows us to describe statistical properties of CA without analyzing evolution spaces of individual rules
\cite{Gutowitz:Victor-1987,McIntosh-1990a}. 
This approximation assumes that elements of a set of states $\Sigma$ are {\em independent} and not correlated with each other in the rule's evolution space. 
One can study probabilities of states in the neighborhood in terms of probability of a single state (the state in which the neighborhood evolves), thus a probability of the neighborhood--state is a product of the probabilities of each cell-state in the neighborhood. 
A polynomial on the probabilities is derived and its curve can be used to classify the rules, as proposed by McIntosh in 
\cite{McIntosh-1990a}. 
%
%
\subsection{Mean field in the Game of Life}
%
%
Using this approach we can construct a mean field polynomial for a two--dimensional CA with a semi--totalistic evolution rule:
\begin{equation}
p_{t+1} = \sum_{v=S_{min}}^{S_{max}} \left (\begin{array}{c}
n-1 \\
v
\end{array} \right ) p_{t}^{v+1}  \, q_{t}^{n-v-1} + \sum_{v=B_{min}}^{B_{max}} \left (\begin{array}{c}
n-1 \\
v
\end{array} \right ) p_{t}^{v}  \, q_{t}^{n-v}
\label{MFp2D}
\end{equation}
\noindent where $n$ represents the number of cells in Moore's neighborhood, $v$ (resp. $n-v$)  the number of occurrences of state `1' (resp. `0'), $p_{t}$  (resp. $q_{t}$) the probability of a cell being in state `1' (resp. `0')  and with $q_{t}=1-p_{t}$. 
$B$ and $S$ are minimum and maximum of an interval for born and survival conditions in Conway's Game of Life (GoL), respectively. 
The GoL's polynomial is the following: 
\begin{equation}
p_{t+1}=84  \, p_t^{3}  \, q_t^{6} + 56  \, p_t^{4}  \, q_t^{5}.
\label{lifeP}
\end{equation}
\noindent The mean field curve 
$
 {\mathcal{F}}
$
 of 
Eq.~\ref{lifeP} 
displayed in 
Fig.~\ref{fig:meanField}a
shows three fixed points $p_{t+1} = p_t$ when crossing the identity.
The first stable fixed point at the origin guarantees its stable state, the second unstable point $ {\mathcal{F}}=0.1986$ relates to areas of densities where the space--time dynamic is unknown. The last stable point in $ {\mathcal{F}}=0.37$ indicates that GoL will converge almost surely  to configurations with small densities of `1'.
%
%
\subsection{Mean field in ECA Rule 22}
%
%
\begin{figure}[!tbp]
\begin{center}
\subfigure[]{\scalebox{0.146}{\includegraphics{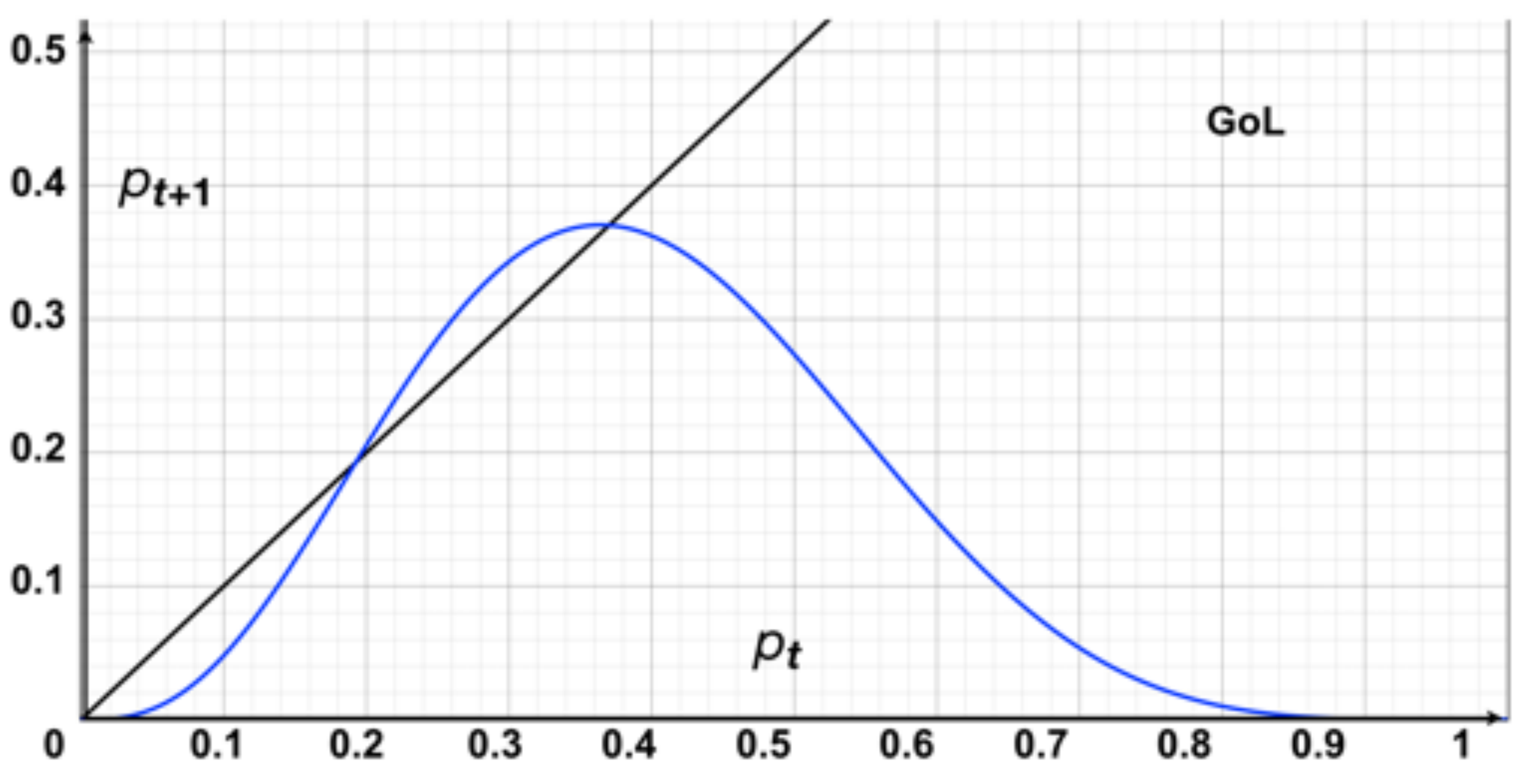}}} 
\subfigure[]{\scalebox{0.146}{\includegraphics{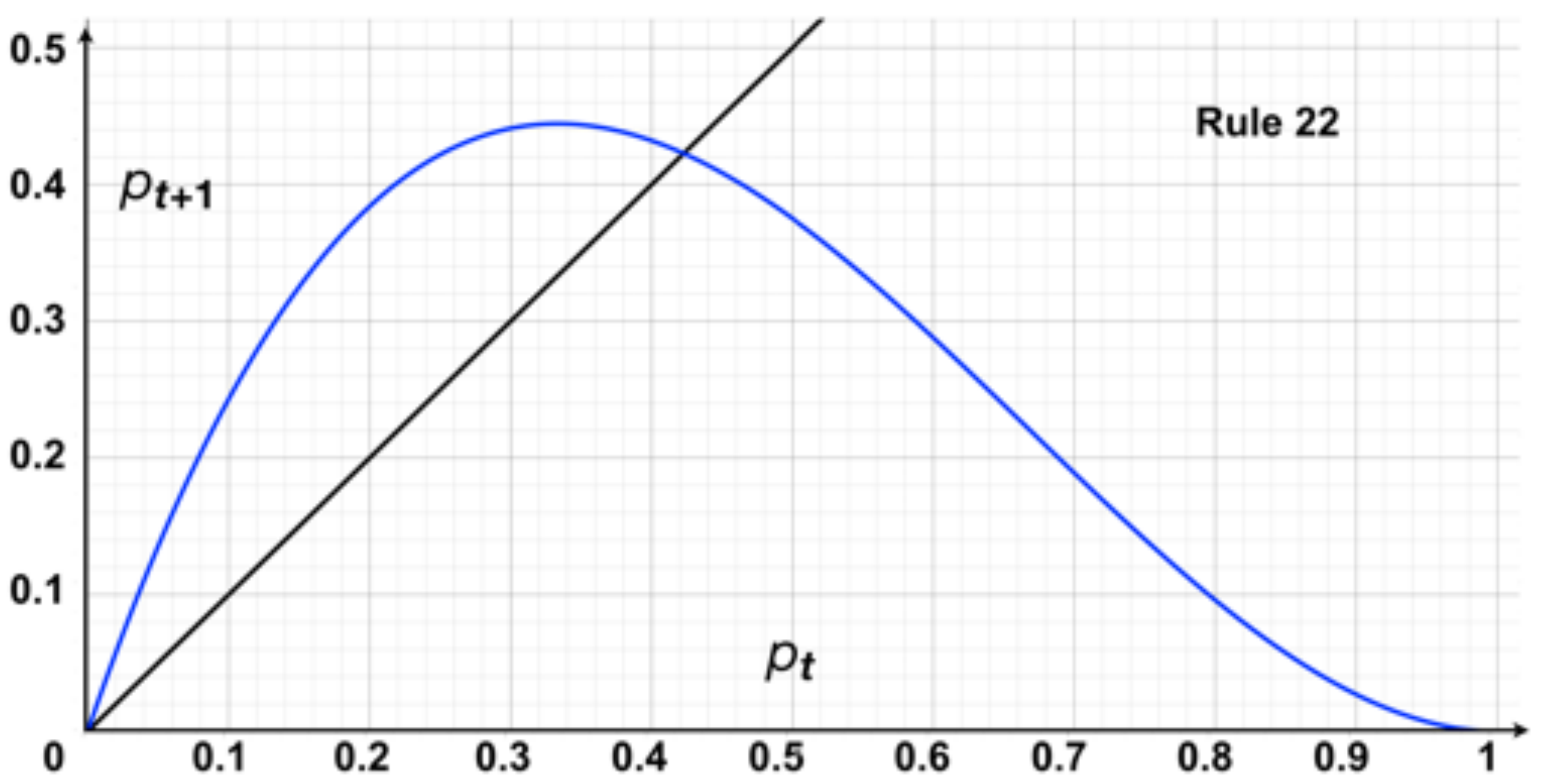}}}
\end{center}
\caption{Mean field curves for (a) GoL and (b) Rule 22.}
\label{fig:meanField}
\end{figure}
For one dimension, we adjust Eq.~\ref{MFp2D} to a full local rule and not only a semi--totalistic one. 
All states of the neighborhood must be considered, thus $p$, $q$, $n$, and $v$ have the same representation as above. But now the product will be with the value of each neighborhood, whence, for a 1$d$ CA$(k, r)$ the mean field polynomial
\[ 
p_{t+1}=\sum_{j=0}^{k^{2r+1}-1}\varphi_{j}(X) \, p_{t}^{v} \, q_{t}^{n-v} 
\]
which gives 
\begin{equation}
p_{t+1} = \sum_{j=0}^{7}\varphi_{j}(X) \, p_{t}^{v}  \, q_{t} ^{3-v} 
\label{MFp1D}
\end{equation}
for ECA$(2,1)$ and where $\varphi_{j}(X)$ denotes the $j$--th transition of $S^3$ in 
Eq.~\ref{eq: ECA(2;1)}.
Finally, the mean field polynomial for Rule 22
\begin{equation}
p_{t+1} = 3 \, p_t \, q_t^{2} 
            = 3 \, p_t \, (1 - p_t)^{2} 
\label{eq:MF-R22}
\end{equation}
is deduced from
(\ref{eq:ECAR22}).

\begin{figure}
\centerline{\includegraphics[width=1\textwidth]{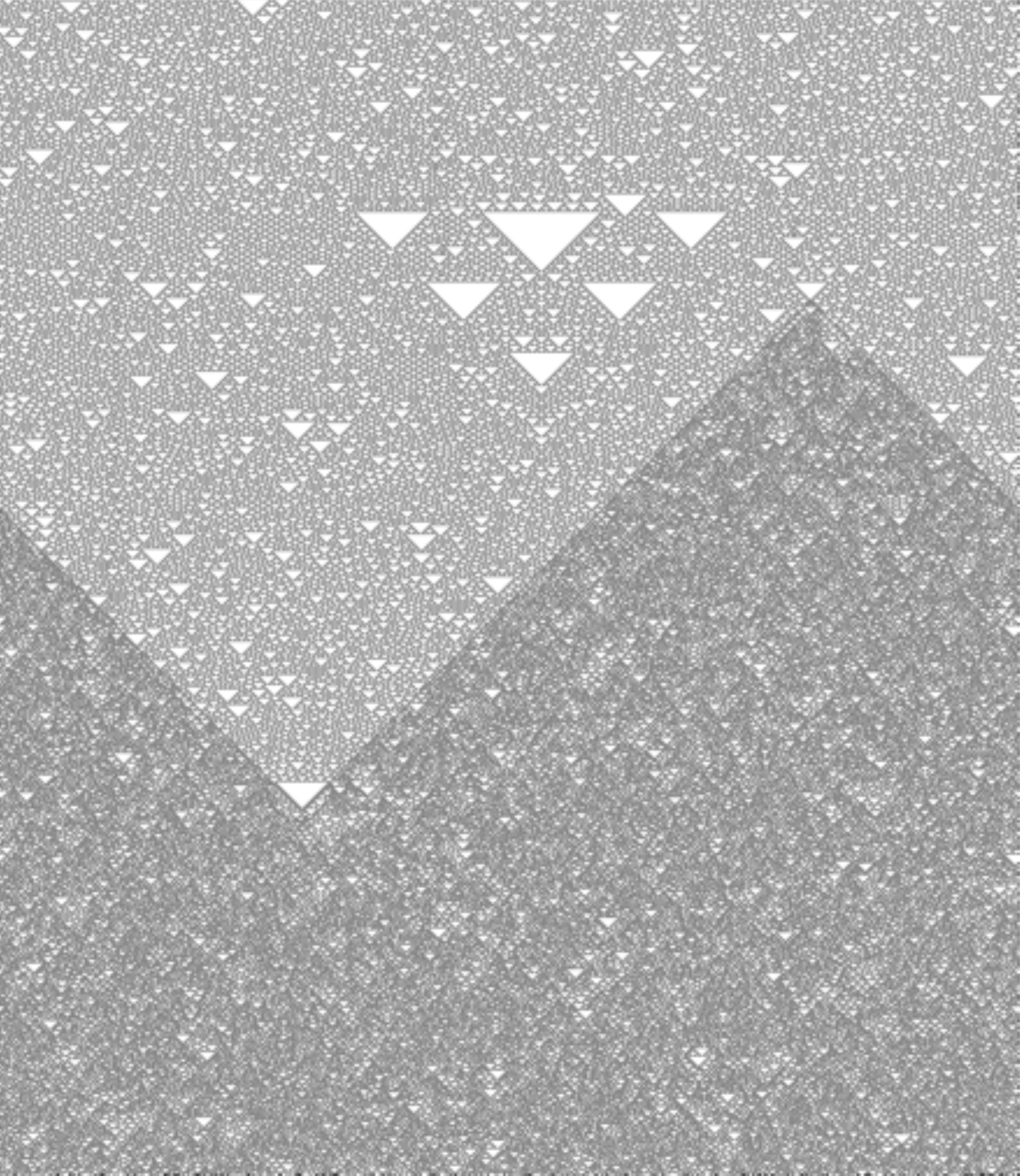}}
\caption{Irreversible phase transition in ECA Rule 22 from a specific initial condition (a regular expression) which after 20,000 generations steps up increases significantly the density of `1', related to the fixed points calculated by mean field theory in Eq.~\ref{eq:MF-R22}. This phase transition merges possible complex dynamics to chaos in ECA Rule 22.}
\label{11101110111011100000-1111111_25284}
\end{figure}

In Rule 22 state `1' appears with probability $  \frac{3}{8}  = 0.375$ (which is close to the fixed stable point 0.37 of GoL). 
\noindent The mean field curve $f$ of 
Eq.~\ref{eq:MF-R22} 
displayed in 
Fig.~\ref{fig:meanField}b
shows a slope 
$
f'(0) = 3
$
at the origin. Density is maximal at 
$
f(1/3) =  \frac{4}{9}  \approx 0.444
$
before reaching the stable fixed point 
$
p_{t+1} = p_t
$
when crossing the identity at 
$
 p_t = 1 - \sqrt{3}/3    \approx  0.423.
$
It then crosses the inflection point 
$
f(2/3) = \frac{1}{2} \cdot f(1/3)
$
with tangential slope
$
f'(2/3) = -1
$
and decreases until 
$
f(1) = f'(1) = 0.
$
Based on the mean field curves classification, Rule 22 is a chaotic ECA 
(Fig.~\ref{11101110111011100000-1111111_25284}).
%
%
\subsection{Mean field behavior of Rule 22}
%
%
Various scenarios of evolution are displayed in 
Fig.~\ref{fig:Evolutions in a ring of 800 cells}:
\begin{figure}
\centerline{\includegraphics[width=12.5cm]{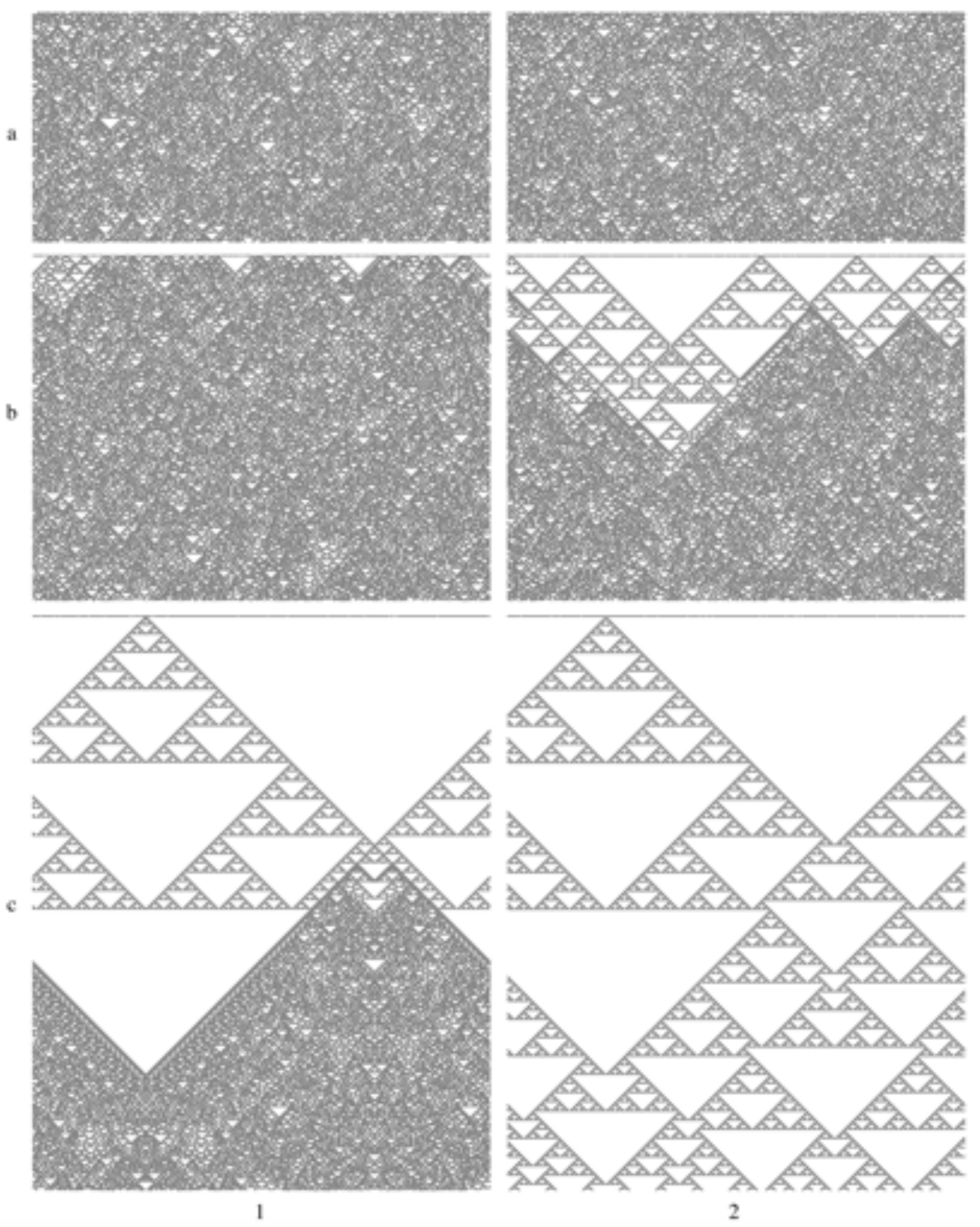}}   
\caption{
Evolutions in a ring of 800 cells:
($a$) 400 generations ($a_1$) from initial density $d_0 = 1/3$ ($a_2$) from density $d_0 = d_F$ reaching spontaneously the fixed point
($b$) 600 generations ($b_1$) from density $d_0 = 0.8$ evolving towards $d_F$ after an early phase transition ($b_2$) from density $d_0 = 0.95$ evolving towards $d_F$ after a later phase transition 
($c$) 1000 generations from density $d_0 = 0.97$ evolving towards ($c_1$) a dense pattern with thin backbones ($c_2$) a sparse pattern with large backbones. 
}  
\label{fig:Evolutions in a ring of 800 cells}
\end{figure}
\begin{itemize}
\item ($a_1$) From initial density $d_0 = 1/3$ reaching spontaneously the maximum before evolving rapidly towards the fixed point with density $d_F = 0.423$. Transition from $d_0$ to $d_F$ is not perceptible.
($a_2$) From initial density $d_0 = d_F$ reaching spontaneously the fixed point. 
\item ($b_1$) From initial density $d_0 = 0.8$ evolving towards the fixed point after an early phase transition.
($b_2$) From initial density $d_0 = 0.95$ evolving towards the fixed point after a later phase transition delimited by a polygonal broken line\footnote{Phase transitions and critical exponents for Rule 22 leading to non--trivial long--range effects were reported in
\cite{Grassberger-1986,Sinha-2006}. Asymptotic properties were described in \cite{Zenil-2013}. 
}.
There exists an interval between two thresholds 
$
d'_0 \approx 0.01
$ 
and 
$
d''_0 \approx 0.92
$ 
such that any (pseudo--)random initial distribution with density 
$
d'_0 < d_0 < d''_0
$
converges almost surely towards fixed point $d_F$.
\item ($c_1$) From initial density $d_0 = 0.97$ evolving towards a {\em  dense} pattern with observable ``backbones''\footnote{Also observable in the ``{\em Exactly 1}'' ECA
\cite{Gravner:Griffeath-2011}
Fig.1d.
}
after crossing the phase transition polygon;
two backbones arise from polygon vertices and their patterns are symmetric from either side.
($c_2$) From the same density $d_0 = 0.97$ and another initial distribution yielding a {\em  sparse} pattern with backbones wherein no phase transition line does appear. 
\end{itemize}
Outside interval 
$
] d'_0, d''_0 [
$
sensitivity to initial conditions is high, with a positive Lyapunov exponent and a chaotic behavior.
Depending on small perturbations from initial configuration, four other evolutions are also possible:
(i) disordered sparse fractals
(ii) convergence towards fixed point  $d_F$ but after a long period
(Fig.~\ref{fig:Disordered pattern evolving towards fixed point})
(iii) initial configuration vanishing at first step
(iv) rare events of periodic patterns
(Fig.~\ref{fig:Rare periodic event}).
That is, six types of evolution altogether. Their estimations of occurrence in interval 
$
 [ d''_0, 1]
$
are displayed in 
Tab.~\ref{table:Statistical estimations of evolutions in interval [d"0, 1]}.
\begin{table}
\centering
\caption{Statistical estimations of evolutions (\%) in interval 
$
 [ d''_0, 1]
$
from samples of 100 initial configurations for each density: 
ergodicity (ERG), disordered sparse fractals (DSF), dense backbones (DBB), sparse backbones (SBB), vanishing (VAN), rare periodic patterns (RPP).
Bolded items reflect an irreversible state.
}
\vspace{0.2cm}
\begin{tabular}{|c||c|c|c|c|c|c|}
\hline	 \rule[-0.1cm]{0cm}{0.5cm} 	
 Density      & \textbf{ERG} & DSF  &   \textbf{DBB}   &   SBB    &   \textbf{VAN}    &   \textbf{RPP}   \\ [0.5ex]
\hline\hline	
$0.920$       & 100    &  0    &  0    &  0    &  0    &  0    \\ \hline 
$0.930$       &   94    &  5    &  1    &  0    &  0    &  0    \\ \hline 
$0.940$       &   93    &  1    &  3    &  1    &  1    &  1    \\  \hline
$0.950$       &   85    &  2    &  7    &  4    &  2    &  0    \\  \hline
$0.960$       &   47    &  4    &  13  &  24  &  12  &  0    \\  \hline
$0.970$       &   21    &  2    &  26  &  26  &  25  &  0    \\  \hline
$0.980$       &    7     &  0    &  18  &  21  &  54  &  0    \\  \hline
$0.990$       &    1     &   0   &   5   &   8   &  86  &  0    \\  \hline
$0.995$       &    0     &   0   &  0    &   0   & 100  &  0    \\  \hline
\end{tabular}
\label{table:Statistical estimations of evolutions in interval [d"0, 1]}
\end{table}
\begin{figure}
\centerline{\includegraphics[width=12.4cm]{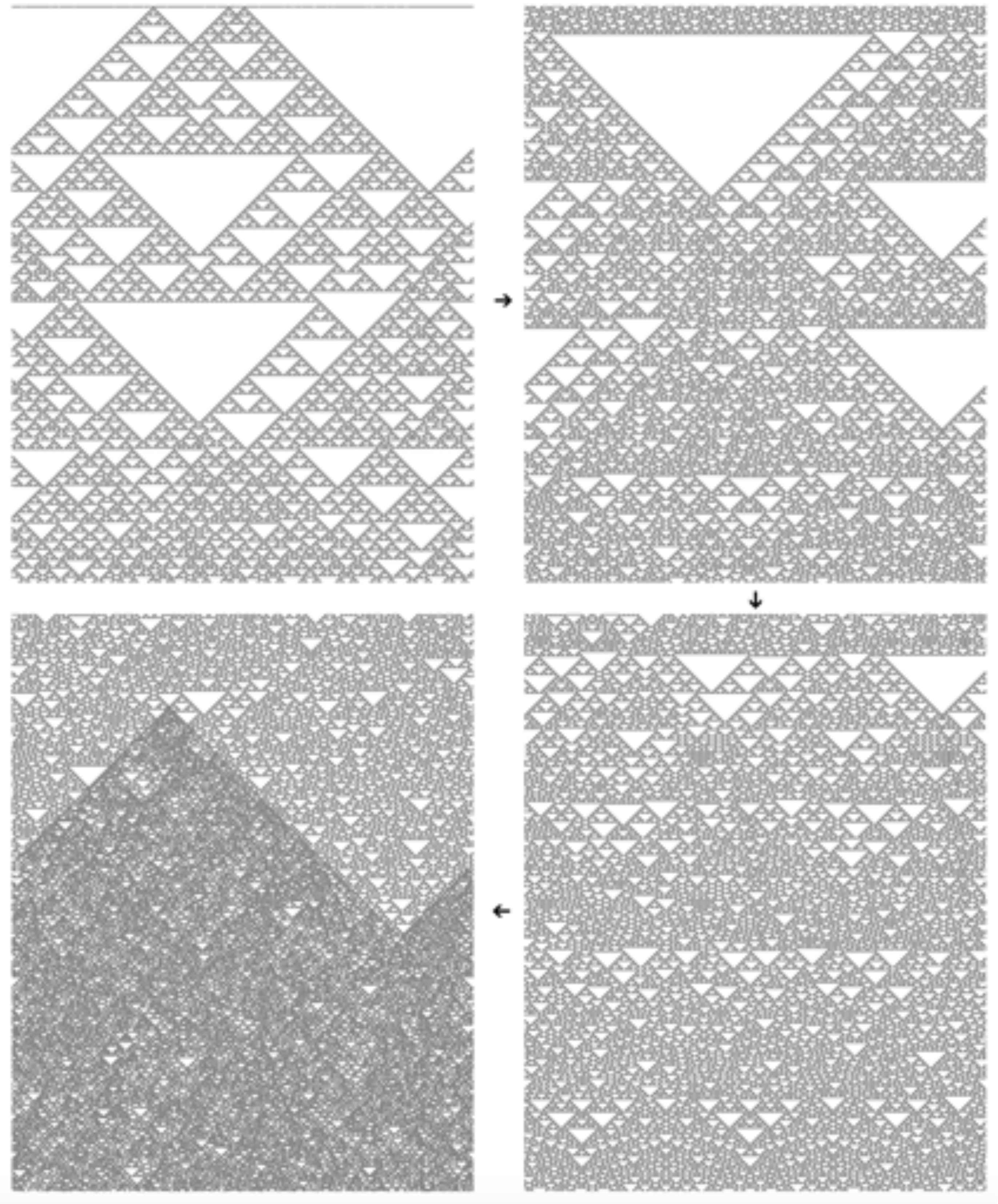}}   
\caption{
Disordered sparse fractal (DSF) evolving from initial density  $d_0 = 0.93$ towards fixed point  $d_F$ after a long period:
$
(0-1000) \rightarrow (2000-3000) \downarrow (3000-4000)  \leftarrow (27000-28000).
$
DSF pattern appears as a long transient state before ergodicity.
Compare with landscape in 
Fig.~\ref{11101110111011100000-1111111_25284}
derived from a regular expression.
}  
\label{fig:Disordered pattern evolving towards fixed point}
\end{figure}
\begin{figure}
\centerline{\includegraphics[width=12.3cm]{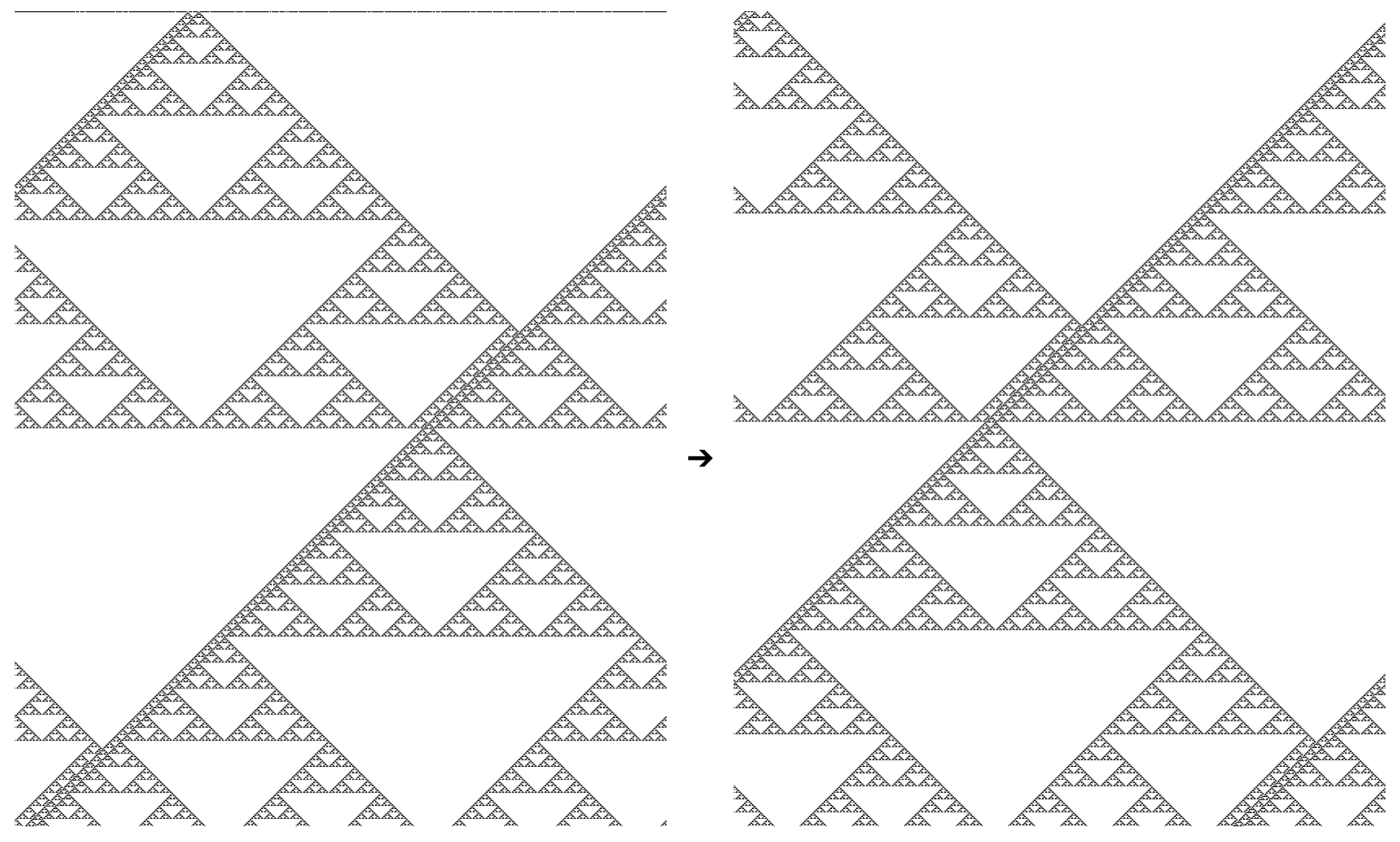}}   
\caption{
Rare periodic event occurring from initial density $ d_0 = 0.94 $ with a complete pattern moving leftward like a glider:
$
(0-1000) \rightarrow (200000-201000).
$
Compare with the patchwork of periodic patterns in 
Fig.~\ref{deBruijn10gen}.
The probability of occurrence of such a pattern from a random initial distribution is about $10^{-3}$.
}  
\label{fig:Rare periodic event}
\end{figure}

Beyond  the phase transition polygon in case of convergence towards fixed point  $d_F$, the evolution becomes ergodic, in the sense that the system has the same behavior either averaged over time or averaged over space
\cite{Halmos-1956}.
The process is stationary and homogeneous at mesoscopic scale. In other words, their exists a smallest macro--cell 
$
 {\mathcal{C}}
$
 of size 
$
   \xi \times \xi,
$
where $\xi$ is the correlation length, as representative (or statistical) volume element such that density 
$
  d_{\mathcal{C}}
$ 
in the macro--cell is close to the mean density averaged within the whole system
\cite{Lesne-1998,Deserable-2011}.
Thus
$
   d_{\mathcal{C}} \approx \frac{3}{8} = 0.375
$
that is, the exact ratio of `1' filling 
$
\varphi_{R22}.
$

It should be observed that a disordered sparse fractal pattern (DSF) may evolve towards ergodicity (ERG) as in 
Fig.~\ref{fig:Disordered pattern evolving towards fixed point}
but sometimes after a long, unpredictable time. We denote as ``DSF'' such evolution remaining in this state at least within a time window of arbitrary length $10^3$. In the same way, sparse backbone patterns  (SBB) may evolve towards a dense backbone landscape (DBB).

A simple way to check (not to prove) whether unstable evolutions become eventually ergodic or to get a more global overview upon evolution is to skip some timesteps with skip time--lengths $\Delta t$. This transformation yields a projective view upon the $(x,t)$--landscape with angle
$
 \arctan{1/\Delta t}.
$
Various skipped scenarios of evolution from initial critical density $d_0 = 0.97$ in a ring of 800 cells and within a time window of length 
$
  10^3\cdot\Delta t
$ 
are displayed in 
Fig.~\ref{fig: Skipped evolutions in a ring of 800 cells}:
\begin{figure}
\centerline{\includegraphics[width=12.4cm]{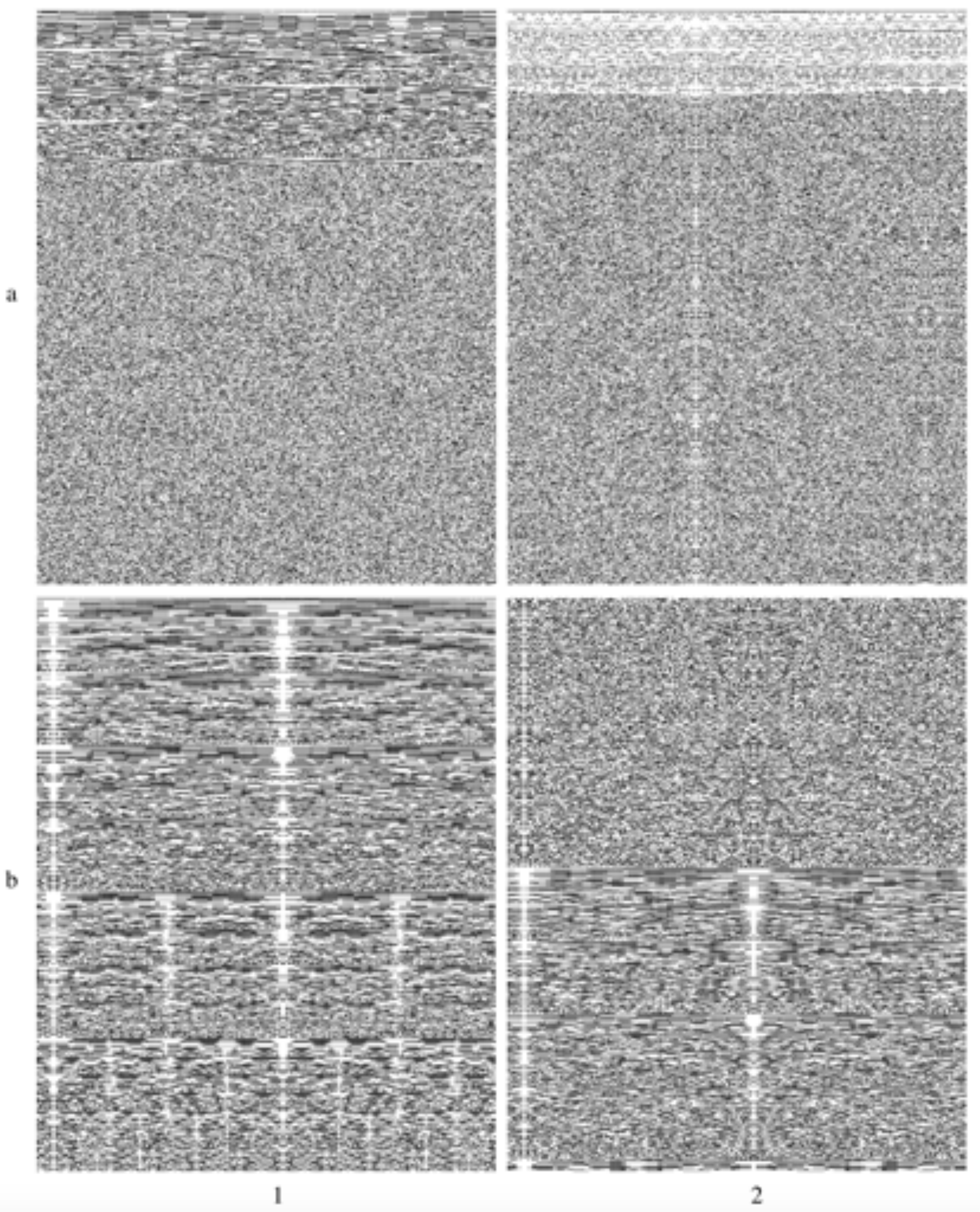}}   
\caption{
Skipped scenarios in a ring of 800 cells with $d_0 = 0.97$ within a time window of length $10^3\cdot\Delta t$.
($a$) Phase transitions
($a_1$) with $\Delta t = 32$, DSF $\to$ ERG from disordered sparse fractal to ergodicity 
($a_2$) with $\Delta t = 31$, SBB $\to$ DBB from sparse to dense backbones.
($b$) Stratified landscapes SBB with $\Delta t = 64$  
($b_1$) up to 64000 generations with observable backbones... and sub--backbones evolving like a Cantor dust 
($b_2$) up to $ 4 \cdot 10^6 $ generations with perpetual phase transitions: a transition line separates a dense backbone (DBB) regime from a sparse backbone (SBB) regime; evolution remains still unstable.
}
\label{fig: Skipped evolutions in a ring of 800 cells}
\end{figure}
\begin{itemize}
\item ($a$) Phase transitions
($a_1$) DSF $\to$ ERG with $\Delta t = 32$ from disordered sparse fractal to ergodicity -- 
up to 32000 generations, transition occurs after about 8000 timesteps
($a_2$) SBB $\to$ DBB  with $\Delta t = 31$ from sparse to dense backbones -- 
up to 31000 generations; transition occurs after about 4000 timesteps.
Note that apparent discrepancies between densities in ($a_1$) and ($a_2$) before phase transition is no more than a side effect resulting from even or odd skip length parity.
\item 
($b$) Stratified landscapes SBB with $\Delta t = 64$  
($b_1$) up to 64000 generations with observable backbones... and sub--backbones evolving like a Cantor dust 
($b_2$) up to $ 4 \cdot 10^6 $ generations with perpetual phase transitions.
\end{itemize}
Failing to prove the existence or not of ergodic evolution, this skipping approach nevertheless emphasizes several chaotic behavior with long--range correlations. As well as in statistical physics, renormalization methods
\cite{Lesne-1998}
overcome the weakness of mean field approximations that may fail or, at least, produce insufficient information. Mean field theory is a rough approximation which assumes {\em independence} between neighboring sites. On the contrary, other deterministic approaches like de Bruijn diagrams assume {\em dependence}. They will be discussed thereafter. 
%
\section{Attractors}
\label{section:Attractors}
%
%
\subsection{Basins of attraction}
\label{subsection:Basins of attraction}
%
%
\begin{figure}[!tbp]
\centerline{\includegraphics[width=1.03\textwidth]{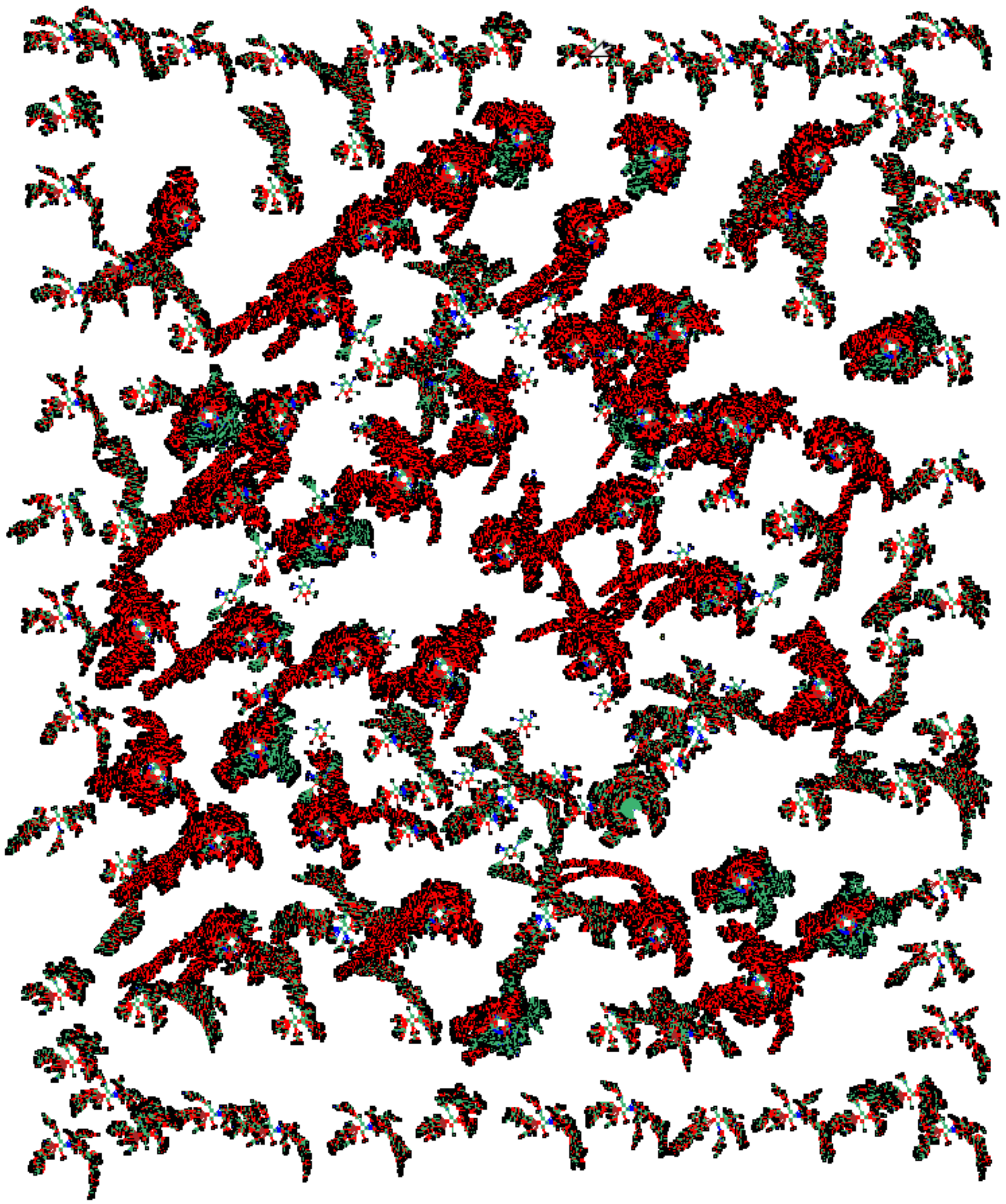}}
\caption{Basin of attractors in ECA Rule 22 for rings of size 20. The number of attractors are 108 with 12 non--equivalent types. Based in attractors characterization, Rule 22 displays chaotic behavior with highly dense, not long transients, and several symmetric trees.}
\label{attractors-r20ECA22}
\end{figure}
\begin{figure}[!tbp]
\centerline{\includegraphics[width=1\textwidth]{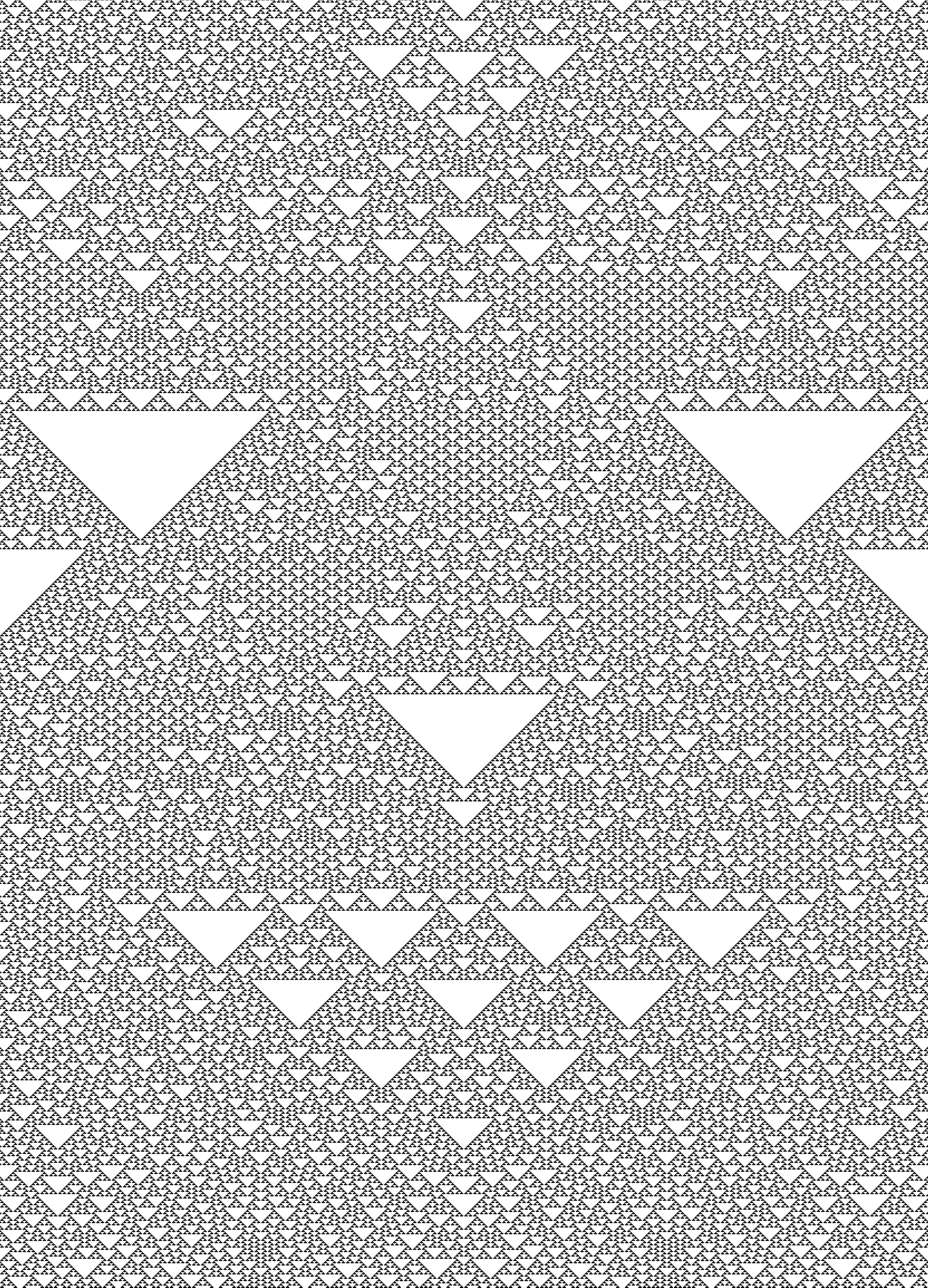}}
\caption{Discovering non--trivial patterns emerging in ECA Rule 22 displaying a family of tilings of different sizes from a string of a basin of length 20  (Fig.~\ref{attractors-r20ECA22}). A lot of these patterns can be reached with concatenation of the string 00000001000100000000. }
\label{complexDynamics22syme-874x8092}
\end{figure}
Basins of attraction have been studied by Andy Wuensche in the framework of ECA and random Boolean networks
\cite{Wuensche:Lesser-1992,Wuensche-2016,McIntosh-1993}. 
A string of cell--states $x_i^t$ is a configuration $c$. An evolution is represented by a sequence of configurations $\{c_0, c_1, c_2, \ldots, c_{m-1}\}$, such that $\Phi:\Sigma^n \rightarrow \Sigma^n$, and the global transition can be represented as $\Phi(c^t) \rightarrow c^{t+1}$. A number of all global states of $c$ is determined by the length of a string $m^n$ (where $n$ is the length and $m$ the number of symbols). The structure of an attractor (Fig.~\ref{attractors-r20ECA22}) is given in three parts. Leaves represent Garden of Eden, i.e. unreachable in the evolution but only as initial global states (these states have no ancestors). Branches are configurations that have at least one ancestor and just one successor. Height of branches determines a number of generations to reach the attractor. An attractor is the final state of a string of length $n$. Numbers labelling vertices represent the decimal values of the strings.

Wuensche
\cite{Wuensche:Lesser-1992} 
proposed that Wolfram's classes can be represented as a {\it basin classification}. In this classification complex behavior is characterized by moderate transients, moderate--length periodic attractors, moderate in--degree and small density of leaves. This way, Fig.~\ref{complexDynamics22syme-874x8092} displays a type of non--trivial behavior later of thousands of generations starting with a concatenation of one of these strings calculated by one attractor of length 20:

\begin{quote}
00000001000100000000 $\rightarrow$ 00000011101110000000 \\ $\rightarrow$ 00000100000001000000 $\rightarrow$ 00001110000011100000 \\ $\rightarrow$ 00010001000100010000 $\rightarrow$ 00111011101110111000 \\ $\rightarrow$ 01000000000000000100 $\rightarrow$ 11100000000000001110 \\ $\rightarrow$ 00010000000000010000 $\rightarrow$ 00111000000000111000 \\ $\rightarrow$ 01000100000001000100 $\rightarrow$ 11101110000011101110.
\end{quote}

\noindent Particularly, the average density for this evolution space is 4/15. 
This density lives exactly between the density of the rule itself  (Eq. \ref{eq:ECAR22}) and the stable fixed point in mean field theory  (Eq. \ref{eq:MF-R22} and Fig. \ref{fig:meanField}b). 
We will note that this value is not reachable from the statistical analysis done for ECA Rule 22. Also, it is the average where we report non--trivial behavior in ECA Rule 22. 
Of course, when an evolution is evolving to this value and later switches to the density of the stable fixed point hence the phase transition is irreversible 
(see Fig. \ref{11101110111011100000-1111111_25284}).

By calculating large attractors, we can discover landscapes of complexity in basins featured with non--symmetric, high and dense ramifications: these kind of ramifications are indicators of `unpredictable' behavior on most large configurations. No rarely chaotic rules tend to have symmetric basins. Basins of attraction can be connected into a meta--network, called the `jump-graph'
\cite{Wuensche-2016}. 
Jump--graphs determine the next level of CA complexity by showing a probability to jump from one attractor to another attractor given a mutation on the same domain of strings
\cite{Martinez:Adamatzky:Chen:Chen:Mora-2017}.
\begin{figure}[!tbp]
\centerline{\includegraphics[width=1.05\textwidth]{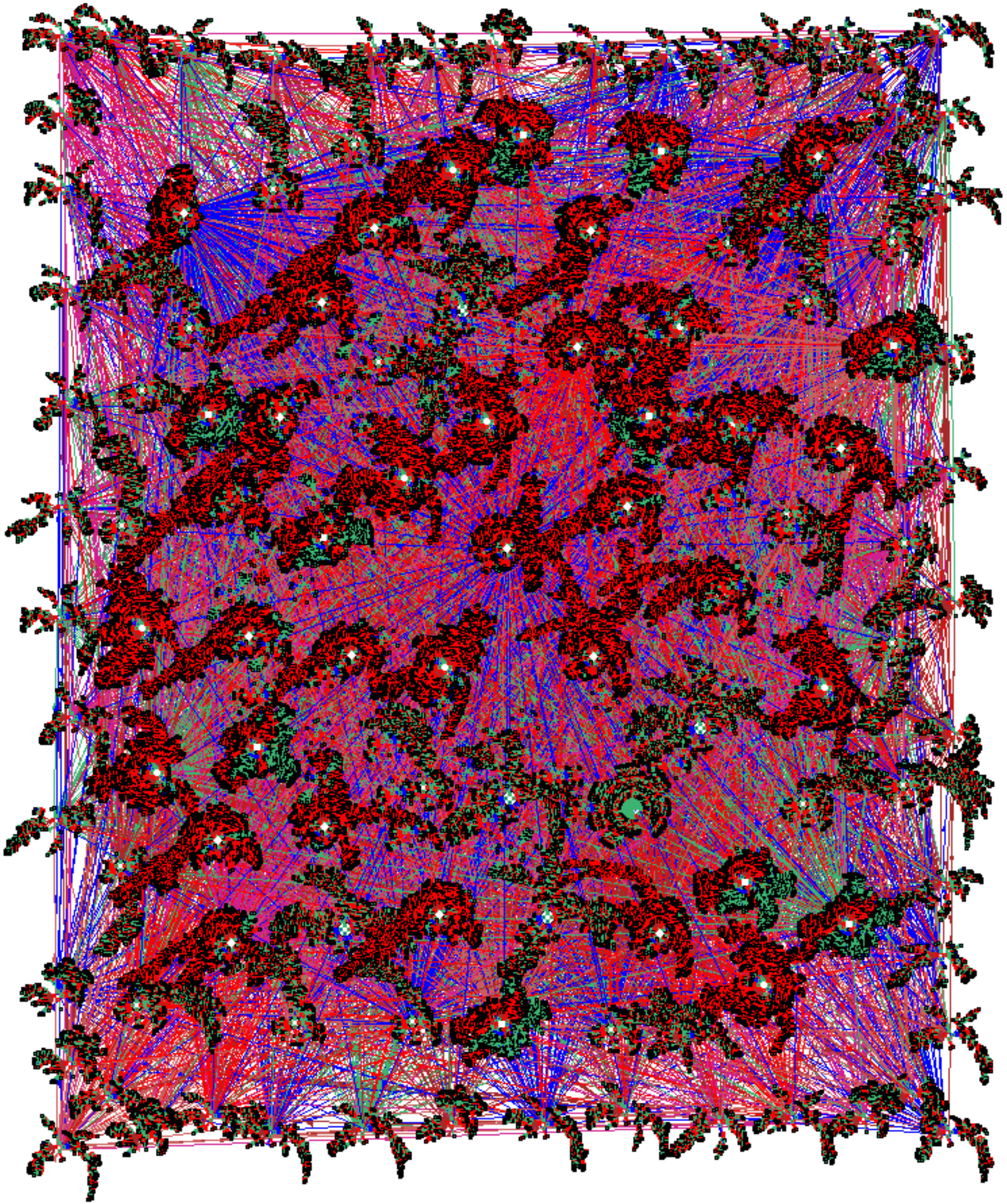}}
\caption{Jump-graph in ECA Rule 22 constructed with a base of attractors of length 20. The connection is determined by mutation of one bit in the strings. The chaotic behavior from jump-graphs is characterized to the high connectivity between all attractors (for details see
\cite{Martinez:Adamatzky:Chen:Chen:Mora-2017}).}
\label{jump-r20ECA22}
\end{figure}

Let us consider a one--bit value mutation $\Psi(\Phi(c_i)) \rightarrow \Phi(c_j)$
\cite{Wuensche-2016}. 
A configuration $c_i$ expressed as a string $w_i = a_0 a_1 \ldots a_{n-1}$, such than it can jump into other configuration $c_j$ expressed as a string $w_j = b_0 b_1 \ldots b_{n-1}$. Hence $a_i$ can mutate to one $b_i$, where each configuration $c$ belongs at the same field of attractors $\Psi$. Also, the mutation represents a loop in the same basin if $a_i = b_i$.  Figure \ref{jump-r20ECA22} shows a jump--graph from the basins of attraction of length 20 (Fig. \ref{attractors-r20ECA22}). This way, a chaotic system presents a high density of connectivity with all the attractors in the jump--graph.
%
%
\subsection{Longest paths and representative cycles}
\label{subsection:Longest paths and representative cycles}
%
%
In this section we will use the term \textit{state} as an alias for global state $c_{i}$, configuration, or 1$d$ string. CA($n$) denotes an ECA Rule 22 with $n$ cells and cyclic boundary. 
The aim of this section is to study 
\begin{itemize}
\item What is the length of the longest path until the zero-state $(00...0)$ is reached, and how does a related initial Garden of Eden state look like?
\item How many cycles exist, how long are they, and what is a representative state for each cycle? 
Cycles that belong to the same class (cycles' states are equivalent under shift and mirroring) shall be listed only once as \textit{representative cycle}, the kind of cycles we are interested in here. 
\item Are there similar states (cyclically shifted) that appear periodically within a cycle?
\item What is the length of the longest path until a certain cycle is reached, and how does a related initial state look like?
\end{itemize}
The following terms and functions are used here:
\begin{itemize}
\item \textit{path(A, C)} is a sequence of states (from state $A$ to state $C$). 
\item \textit{length(path)} gives the number of states of a path or cycle. 
\item \textit{prefix(C)} is a path($A$, $B$) where $B$ is direct predecessor of $C$.
\item \textit{maxprefix(C)} is a prefix($C$) of maximum length. 
\item $\alpha$ \textit{= length(maxprefix(0))} where (0) is the zero-state (00...0), the longest prefix(0).
\item \textit{cycle} denotes a periodic attractor, a cyclic path.
\item \textit{k-cycle} is a cycle of length $k$.
\item $\omega(cycle)$ gives the length of a cycle.
\item \textit{similar(S)} is a state that can be derived from state $S$ by cyclic shift and optional mirroring. 
\item $\epsilon$ is called \textit{intra-cycle-period}.
\item In some cycles similar states appear again after $\epsilon$ time-steps. 
\item \textit{k/e-cycle} is a $k$-cycle where $e=\epsilon$($k$-cycle).
We may call $k/e$-cycles \textit{strong} if $k=e$, and \textit{weak} if $k>e$.
\item \textit{cycle-prefix} is a prefix($D$) where $D$ belongs to a cycle.
\item $\lambda(cycle)$ gives the length of the longest cycle-prefix.
\end{itemize}
%
\textbf{First method and results.}
%
The CA($n$), $n = 3 \ldots 20$, were simulated for all possible initial states and then analyzed by special programs and manual inspection. In order to avoid unnecessary simulations, similar initial states were excluded. Similar states are states which are equivalent under cyclic shift and mirroring. For instance the number of different cases to be simulated for $n=18$ is only 7685, which is significantly lower than $2^{18}$.
It should be noted that the cycles for $n$ up to 34 were already computed by McIntosh 
\cite{McIntosh-1990b}.
%
\begin{table}[!tbp]
\vspace{0.1cm}
\caption{Longest path to the zero state.}
\begin{center}
\begin{tabular}{| c | c | r |}
\hline	
  & $\alpha$: length           &   normalized    \\
$n$ & max prefix(0)              &  initial state \\
\hline 

3  & 2  & 001\\\hline
4  & 2  & 0001\\\hline
5  & 6  & 01011\\\hline
6  & 5  & 010111\\\hline
7  & 6  & 0001111 \\\hline
8  & 8  & 00101011\\\hline
9  & 2  & 000100101\\\hline
10 & 7  & 0010111101\\\hline
11 & 6  & 00101010101\\\hline
12 & 23 & 001001100111\\\hline
13 & 20 & 0001001100111\\\hline
14 & 24 & 00010101100111\\\hline
15 & 32 & 001001010100111\\\hline
16 & 41 & 0000000101010011\\\hline
17 & 53 & 00101010101110011\\\hline
18 & 8  & 000001011000010011\\\hline
19 & 17 & 0000101010111010101\\\hline
20 & 18 & 00001000101010111101\\\hline
\end{tabular}
\normalsize
\end{center} 
\label{table-max-prefix}
\end{table}
%
\begin{table}[!tbp]
\vspace{0.1cm}
\caption{Representative cycles for $n=3 \ldots 20$. $\omega$: cycle length, 
$\epsilon$: intra-cycle-period, $\lambda$: length of longest cycle prefix.
Not all leading zeroes are displayed.}   
\vspace{0.2cm}
\scriptsize 
\begin{tabular}{| c | c | c | c | c | c| c |}

\hline	
    &           & representative &            &    repetition  &           & initial state       \\
$n$ &  $\omega$ & cycle state   &  $\epsilon$ &    within      & $\lambda$ & of longest     \\
    &           & (normalised)  &             &    cycle       &           & cycle-prefix        \\
\hline 

3, 5, 6  &       &   no cycle $\omega > 1$  &       &      &       &    \\\hline
4   & 2   & (0011)*  & 1  & $c^{t+1}$=shl2($c^t$) & 0  &   no prefix \\\hline

7  & 7   & (0001011)* & 1  & $c^{t+1}$=shl4($c^t$)  & 0  &   no prefix \\\hline

   & 2   & (0011)*  & 1  & $c^{t+1}$=shl2($c^t$) & 0  &   no prefix \\
8  & 4   & 00000101 & 2  & $c^{t+2}$=shl4($c^t$) & 2  & 00100111 \\
   & 6   & 00000011 & 3  & $c^{t+3}$=shl4($c^t$) & 2  & 00101101 \\
\hline

9  & 4  & 000000101 & ~  &                   & 9  & 000100111 \\\hline

   & 4  & 0000001001 & 2  & $c^{t+2}$=shl5($c^t$)   & 7  & 0011010101 \\ 
10 & 4  & 0000000101 & ~  &                         & 5  & 0010110011 \\
   & 6  & 0000010011 & 3  & $c^{t+3}$=shl3m($c^t$)  & 1  & 0000010011 \\\hline

   & 4  & 00000001001 & ~  &                   & 7  & 00000100111 \\
11 & 5  & 00000001111 & ~  &                   & 7  & 00101010011 \\
   & 11 & 00001001111 & 1  & $c^{t+1}$=shl7($c^t$) & 9  & 00010110011 \\\hline

12               & 2  & (0011)*      & 1  & $c^{t+1}$=shl2($c^t$) & 0  &    no prefix \\
$=3\times 4$     & 4  & 000000010001 & 2  & $c^{t+2}$=shl6($c^t$) & 6  & 000100110011 \\
                 & 5  & 000000001111 & ~  &                   & 9  & 001010101101 \\\hline

13 & 5  & 0000000001111 & ~  &    & 13  & 0000101001101 \\\hline

14            & 7   & (0001011)*     & 1  & $c^{t+1}$=shl3($c^t$)  & 0  &    no prefix \\
$=2\times 7$  & 12  & 00000010001111 & 6  & $c^{t+6}$=shl9m($c^t$) & 4  & 00000101010011 \\\hline

15            & 20  & 000000010000011 & 4  &  $c^{t+4}$=shl4($c^t$)  & 20  & 001010100101111 \\\hline

              & 2   & (0011)*          & 1  & $c^{t+1}$=shl2($c^t$) & 0   &   no prefix \\
16						& 4   & (00000101)*      & 2  & $c^{t+2}$=shl4($c^t$) & 2   & (00100111)* \\
$=2\times8$	  & 6   & (00000011)*      & 3  & $c^{t+3}$=shl4($c^t$) & 2   & (00101101)* \\
$=4\times4$   & 7   & 000000000000011   & ~  &                       & 11  & 0000000010010101 \\
              & 12  & 000000000000101  & 6  & $c^{t+6}$=shl8($c^t$) & 29  & 0000001001010111 \\
              & 12  & 000000000100001   & ~  &                       & 10  & 0010101010101011\\\hline 
							
              & 4   & 000010100000101   & ~   &                          & 2   & 00100101100100111 \\
17            & 12  & 000000000000101   & ~   &                          & 21  & 00101010101101011 \\
              & 26  & 000000000010011   & 13  & $c^{t+13}$=shl15m($c^t$) & 26  & 00101010010110011 \\\hline

              & 4   & 000100100000101 & 2  & $c^{t+2}$=shl4($c^t$)  & 2   & 00101010100101101 \\
              & 4   & (000000101)*   & ~  &                        & 45  & 00111001010101111 \\
18            & 4   & 000010100000101 & ~  &                        & 5   & 00101010011001111 \\
$=2\times 9$  & 12  & 000000000001001 & 6  & $c^{t+6}$=shl9($c^t$)  & 35  & 00000001000011101 \\
              & 12  & 000000000000101 & ~  &                        & 55  & 00010010100110011 \\
              & 18  & 000001101001011 & 9  & $c^{t+9}$=shl9($c^t$)  & 7   & 01111001100110011 \\\hline

              & 4   & 000100100000101 &    &    											& 21          & 00001011100010111 \\
              & 4   & 000101000000101 &    &    											& 6           & 10100101010101011 \\%
19            & 4   & 001001000000101 &    &    											& 1           & 00110111000110111 \\%
              & 12  & 000000000001001 &    &    											& 78          & 00001100010010111 \\
              & 28  & 000001000011101 & 14 &  $c^{t+14}$=shl4m($c^t$) &  9          & 01001100110011101 \\\hline
	                                                              
              & 2   & (0011)*         & 1  & $c^{t+1}$=shl2($c^t$)  & 0  &   no prefix \\		
												
              & 4   & (0000001001)*   & 2  & $c^{t+2}$=shl5($c^t$)  & 44 & 00101100011010011\\	
              & 4   & 001000100000101 & 2  & $c^{t+2}$=shl4($c^t$)  & 5 & 00110110011001000\\	%
																				
	     & 4   & (0000000101)*   & ~  &                        & 12 & 11010001110101111 \\ 							
20            & 4   & 001001000001001 & ~  &                        & 62 & 01111001010101111 \\ 
$=2\times 10$ & 4   & 001001000000101 & ~  &                        & 10 & 00010000100001111 \\ %
							
$=4 \times 5$ & 6   & 100110000101111 & 3  & $c^{t+3}$=shl7m($c^t$)      & 1  & 01011010100101111 \\	%
              & 6   & (0000010011)*   & 3  & $c^{t+3}$=shl7m($c^t$)   & 8  & 00001010100010111 \\				
              & 8   & 000000001000001 & 4  & $c^{t+4}$=shl10($c^t$) & 42 & 00000010001110101 \\              
							& 12  & 000000000010001 & 6  & $c^{t+6}$=shl10($c^t$) & 24 & 00100101100100011 \\ 
							& 24  & 000000000100001 & ~  &                        & 98 & 01010101110101101 \\            
\hline
\end{tabular}
\normalsize
\label{table-cycles-3-20}
\end{table}
%

Table \ref{table-max-prefix} shows $\alpha$ (the length of the \textit{maxprefix}(0)), and the related \textit{normalized} initial state.
A normalized state is a representative of all states which are equivalent under cyclic shift and mirroring. 
It is found by selecting the state with the smallest binary number among all equivalents.
E.g.  \textit{normalize}(100011) = 000111, and \textit{normalize}(110110100) = 001011011 (by mirror and shift).
The $\alpha$ values are much smaller than $2^n$, and not monotonically increasing with $n$.

Table \ref{table-cycles-3-20} shows the results obtained by analyzing all the simulations. 
The operator $shlP(c)$ means shift $c$ to the left by $P$ positions,
and $shlPm(c)$  means  that first the mirror operator is applied before shifting. 

We find always the trivial $(00..0) \rightarrow (00..0)$ cycle of length 1.
For even $n$ there always exists the fixed point (01)* $\rightarrow$ (01)*, a lonely 1--cycle with no prefix.
We will not further mention or pay special attention to these basic 1--cycles.
 
For $n=4$, and multiples of 4, we get the 2/1--cycle (0011)* $\leftrightarrow$ (1100)*.
The two strings are similar under shift of two positions, so the inherent pattern is the same.

For $n=7$, and multiples of 7, we get a 7--cycle.
In order to characterize this cycle, one representative is chosen, it is the one with the smallest normalized value, i.e. (0001011)*. 

For $n=8$, we get three cycles with length $\omega / \epsilon= 2/1, 4/2, 6/3$.

For $n=10$, there exists a 6/3--cycle. After every 3 time-steps, the same string appears in mirrored form and shifted 3 positions to the left. 

For $n=12$, the cycles of CA(4) form a subset (to be included if not detected) which is the 2/1-cycle (0011 0011 0011) $\leftrightarrow$ (110 1100 1100).

For $n=14$, the cycles of CA(7) form a subset which is shown as 7/1--cycle.

Fig. \ref{attractors-r20ECA22} 
shows all possible cycles where many of them are similar (equivalent under shift and mirroring). From that figure we may anticipate a very complex attractor structure, but we should realize that the number of representative cycles in CA(20) is only 11.

In general, because of the cyclic boundary, if $k$ is a factor of $n$, then the CA($k$) cycles are a subset of the CA($n$) cycles. For example, the cycles of CA($k=4,5,10$) form a subset of the CA($n=20$) cycles. However there is a difference: the strings of CA($n$) are cyclic repetitions of the CA($k$) strings, and the original intra-cycle period $\epsilon$ may not appear in CA($n$). 

\vspace{4pt}
%
\noindent\textbf{Second method and results.}
%
For larger $n$, the first method cannot further be applied due to extensive computational costs. Therefore, now only a relatively small random subset of all possible $2^n$ initial states is used in order to find a subset of all cycles and paths which are not necessarily the longest ones. 

10,0000 random initial states were generated for $n=25, 30, 35, ... , 60$.
For 5,000 of the states the probability 0.125 was used for each cell to generate a cell state 1.
For the other 5,000 of the initial states, at first a probability $p$ between 0 and 1 (in steps of 1/1000) was randomly selected. Then $p$ was used for each cell to generate a cell state 1, otherwise 0.
This technique of randomizing gave better results in experiments for Rule 22 compared to the usage of a fixed probability of 0.5. CAs were simulated and cycles and path lengths were computed and processed in a semi-automatic mode. In addition, a genetic algorithm was used to find near-optimal $\alpha$-values. The results are presented in Table \ref{table-cycles-25-60}. Note that because of the statistical approach, the listed cycles are not complete and the true maximum path lengths could be longer, e.g. for CA(60) all cycles already found for the factors 
CA(4, 5, 10, 12, 15, 20, 30)  have to be included.

We can summarize that for  $n\leq 60$, the longest paths are much smaller than $2^n - 1$ (which is achievable with other ECA rules 
\cite{Adak:Mukherjee:Das-2018}). 
Further work remains to find a general formula or at least boundaries for the longest paths and the cycle distribution, depending on a number of cells. 
%
\begin{table}[!tbp]
\caption{Representative cycles for $n=25 \ldots 60$. $\omega$: cycle length, $\lambda$: length of longest cycle prefix, 
$\omega+\lambda$: length of longest path detected. 
The values were obtained by simulation of 10,000 random initial states.} 
\begin{center}
\small
\begin{tabular}{ |c | c | c | c |}
\hline	
    & $\omega$          &  $\omega+\lambda$  &   $\alpha$       \\
$n$ &  length of cycles &  longest path      &     longest      \\
    &   detected         & ending in a cycle   &    prefix(0)     \\
\hline 
25 & $4,5,26,28,50,55,150$ &$150 + 57 = 207 $ & 152 \\
\hline
30&$1,4,6,14,20,40,70,86,120,240,1070$ &$ 1070+153=1223 $&  419 \\
\hline
35  &$4,5,12,28,64,1015 $&$ 1015+302=1317  $& 179  \\ 
\hline
40  &$ 1,4,8,12,16,24,52,80,124,206,320 $&$ 124+2551=2675  $& 303  \\ 
\hline
45  &$4,8,16,19,2295 $&$ 4+4815=4819  $& 540  \\ 
\hline
50  &$ 1,12,28,31,55,56,100,117,150, $&$ 252+13956=14208 $& 750   \\ 
    &$                  252,700,3150 $&$                 $&       \\
\hline
55  &$12,28,30,56,60,330,440,660, $&$ 148225+7124=155349 $& 13904   \\ 
    &$         990,4620,36190,148225 $&$                    $&     \\ 
\hline
60  &$ 1,12,60,120,138,395,476,  $&$ 40980+1004=41984  $& 35579  \\ 
    &$ 480,22740,40980             $&$                   $&    \\ 
\hline
\end{tabular}
\normalsize
\end{center}
\label{table-cycles-25-60}
\end{table}
%
%
\section{De Bruijn diagrams}
\label{section:De Bruijn diagrams}
%
%
De Bruijn diagrams 
\cite{deBruijn-1946}
were originally proposed in shift--register theory
\cite{Golomb-1967}. 
For a 1$d$ CA$(k,r)$ the de Bruijn diagram is defined as a directed graph with $k^{2r}$ vertices and $k^{2r+1}$ edges. 
Vertices are labelled with the elements of symbols of length $2r$. 
An edge is directed from vertex $i$ to vertex $j$, if and only if, the $2r-1$ final symbols of $i$ are the same as the $2r-1$ initial ones in $j$ forming a neighborhood of $2r+1$ states represented by $i \diamond j$. In this case, the edge connecting $i$ to $j$ is labelled with $\varphi(i \diamond j)$ (the value of the neighborhood defined by the local function) 
\cite{Voorhees-1996}, 
as shown in Fig.~\ref{deBruijnGenerico} for ECA(2,1) and for Rule 22 in Figs.~\ref{deBruijnR22}--\ref{deBruijnR22-2}.
%
%
\subsection{Basic de Bruijn diagram}
\label{subsection:Basic de Bruijn diagram}
%
%
The connection matrix $M$ corresponding to the de Bruijn diagram
\cite{McIntosh-1991}
 is as follows:
\begin{equation}
	M_{i,j} = \left\{\begin{array}{ll}
			          1 & \mbox{if } j \in \{ k \,  i, k \,  i+1, \hdots, k \,  i+k-1 \mbox{ (mod } k^{2\cdot r}) \} \\
		           	 0 & \mbox{otherwise} \\
		       \end{array}
			\right.
\label{eq-Bruijn}
\end{equation}
\begin{figure}[!tbp]
\centerline{\includegraphics[width=1.5in]{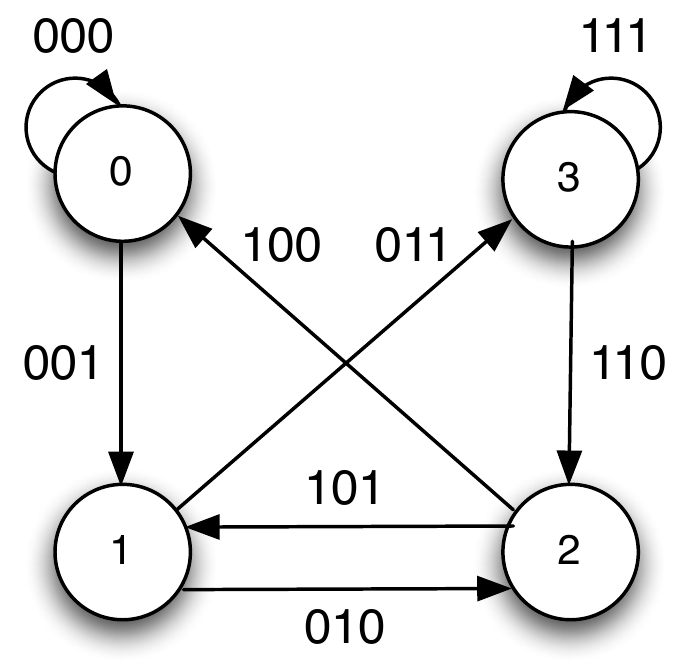}}
\caption{Generic de Bruijn diagram for ECA (2,1).}
\label{deBruijnGenerico}
\end{figure}
\begin{figure}[!tbp]
\centerline{\includegraphics[width=3.8in]{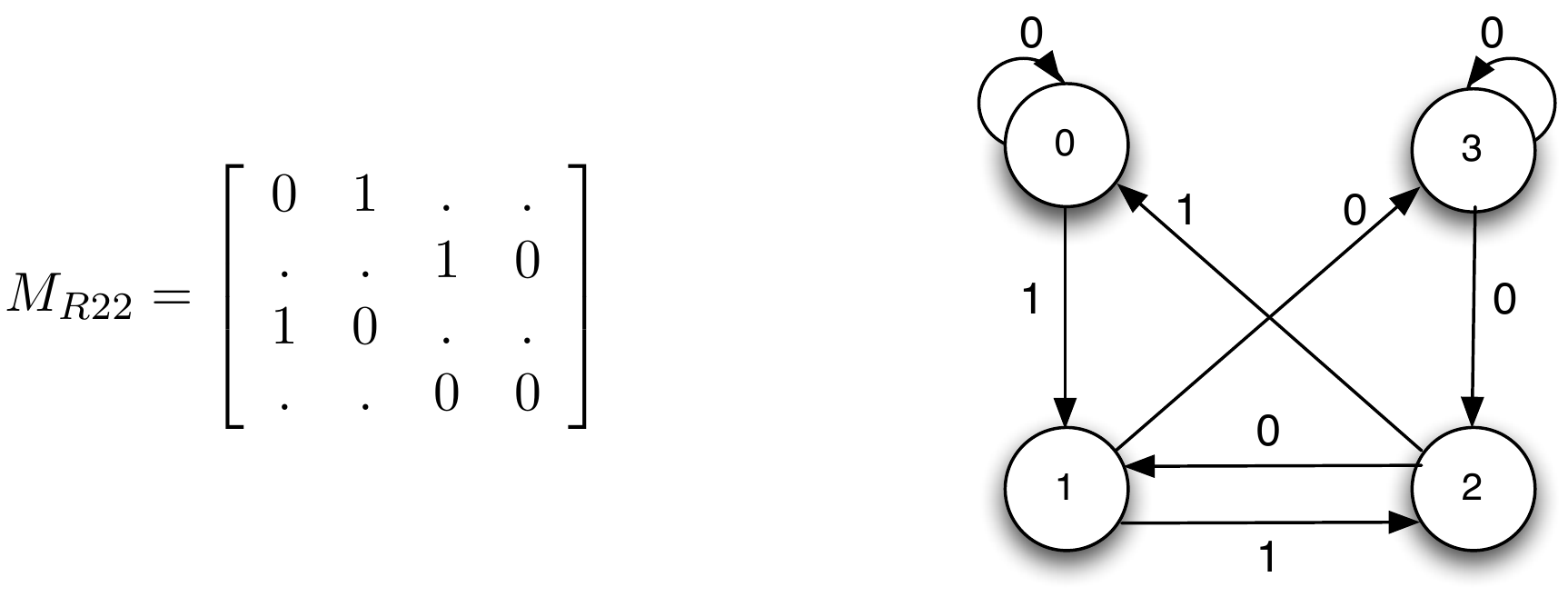}}
\caption{Connection matrix and de Bruijn diagram for ECA Rule 22.}
\label{deBruijnR22}
\end{figure}
\noindent wherein module $k^{2\cdot r}$ represents the number of vertices and $j$ takes on values in 
$
\{ k \,  i, k \,  i+1, \hdots, k \,  i+k-1 \mbox{ (mod } k^{2\cdot r}) \}.
$
Hence for ECA(2,1)
\begin{equation}
	M_{i,j} = \left\{\begin{array}{ll}
			          1 & \mbox{if } j \in \{ 2 \,  i, 2 \, i+1 \mbox{ (mod } 4) \} \\
		           	 0 & \mbox{otherwise} \\
		       \end{array}
			\right.
\label{equation:de Bruijn ECA(2,1)}
\end{equation}
and vertices are labelled by fractions of the overlapping of neighborhoods originated by 00, 01, 10 and 11, and the overlaps of the full neighborhood determine each connection:
$$
\begin{array}{cc}
(0,{\bf 0}) \diamond ({\bf 0},0) \rightarrow 0{\bf 0}0 & (1,{\bf 0}) \diamond ({\bf 0},0) \rightarrow 1{\bf 0}0 \\
(0,{\bf 0}) \diamond ({\bf 0},1) \rightarrow 0{\bf 0}1 & (1,{\bf 0}) \diamond ({\bf 0},1) \rightarrow 1{\bf 0}1 \\
(0,{\bf 1}) \diamond ({\bf 1},0) \rightarrow 0{\bf 1}0 & (1,{\bf 1}) \diamond ({\bf 1},0) \rightarrow 1{\bf 1}0 \\
(0,{\bf 1}) \diamond ({\bf 1},1) \rightarrow 0{\bf 1}1 & (1,{\bf 1}) \diamond ({\bf 1},1) \rightarrow 1{\bf 1}1
\end{array}
$$
They are the edges of the generic de Bruijn diagram in Fig.~\ref{deBruijnGenerico}. 
The de Bruijn diagram has four vertices which can be renamed as $\{0,1,2,3\}$ corresponding to four partial neighborhoods of two cells $\{00,01,10,11\}$, and eight edges representing neighborhoods of size 
$2 \, r+1$.
De Bruijn diagram for Rule 22 is derived from the generic one (Fig.~\ref{deBruijnGenerico}) and it is calculated in Fig.~\ref{deBruijnR22}, where the edges are labelled by the next state.
\begin{figure}[!tbp]
\centerline{\includegraphics[width=4in]{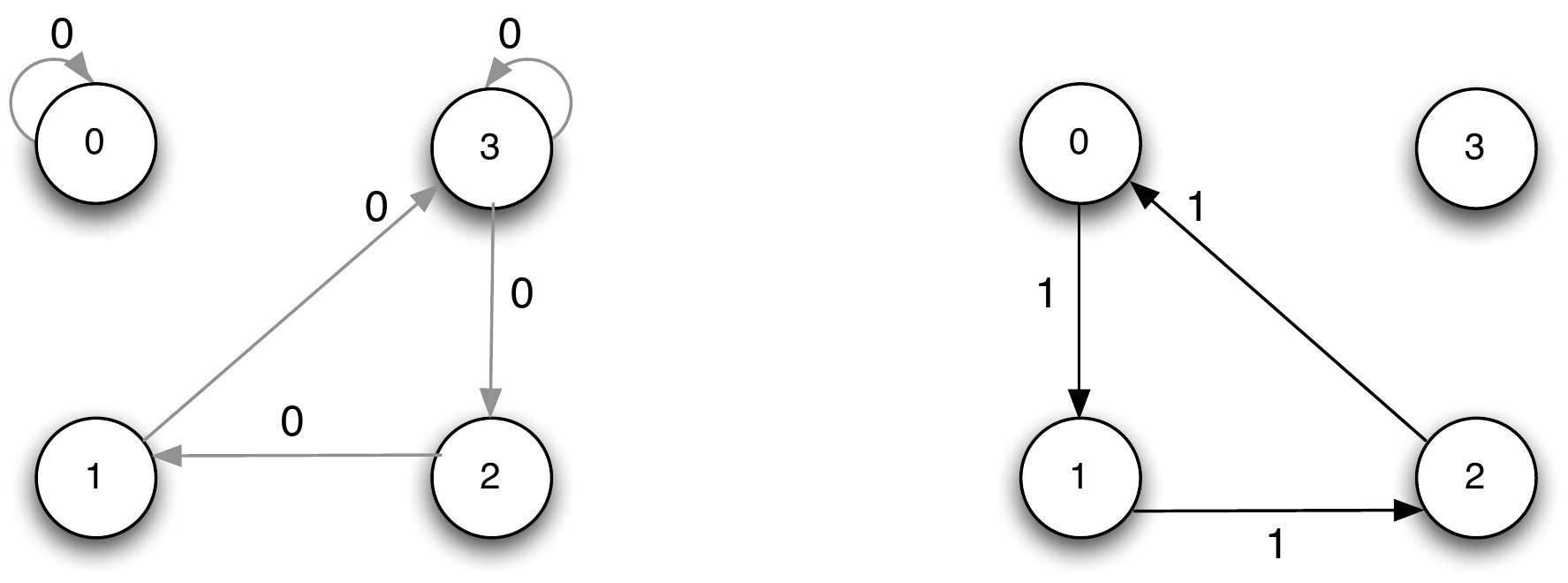}}             
\caption{De Bruijn subdiagrams showing unreachable states.}
\label{deBruijnR22-2}
\end{figure}

Paths in the de Bruijn diagram may represent chains, configurations or classes of configurations in the evolution space. Vertices are sequences of symbols in the set of states and the strings are sequences of vertices in the diagram. The edges represent overlapping of the sequences. Different intersection degrees evoke different de Bruijn diagrams (Fig.~\ref{deBruijnR22-2}). Thus, the connection takes place between an initial symbol, the overlapping symbols and a terminal one. For practical reasons we can use colors, thus the color of an edge represents the next state to which each neighborhood, as shown in Fig. \ref{deBruijnGenerico}, evolves.

\begin{figure}[!tbp]
\centerline{\includegraphics[width=0.74\textwidth]{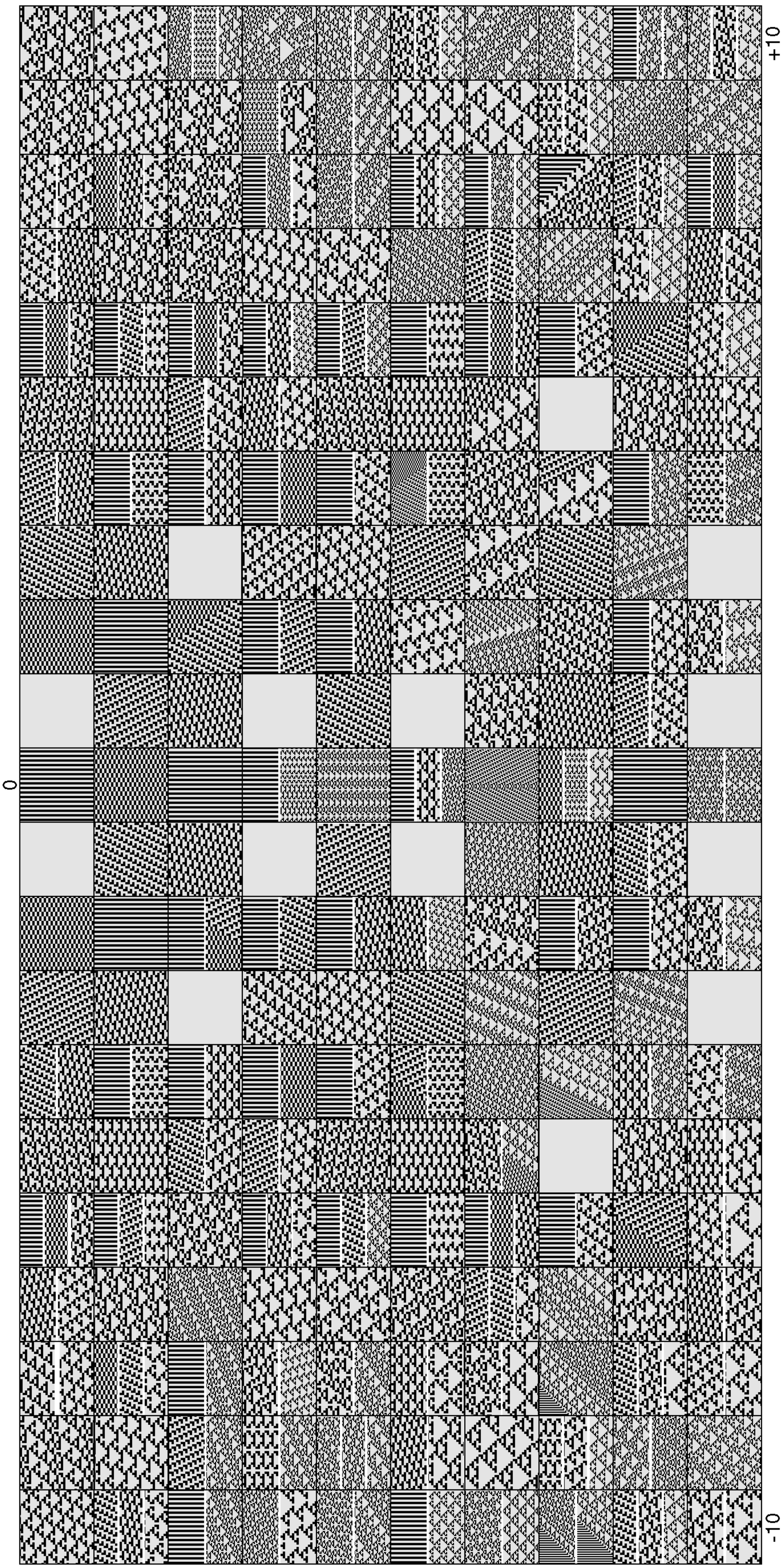}}
\caption{The whole set of periodic patterns yielded from extended de Bruijn diagrams to 10 generations with positive and negative shifts to 10 cells.}
\label{deBruijn10gen}
\end{figure}

\begin{figure}[!tbp]
\centerline{\includegraphics[width=0.92\textwidth]{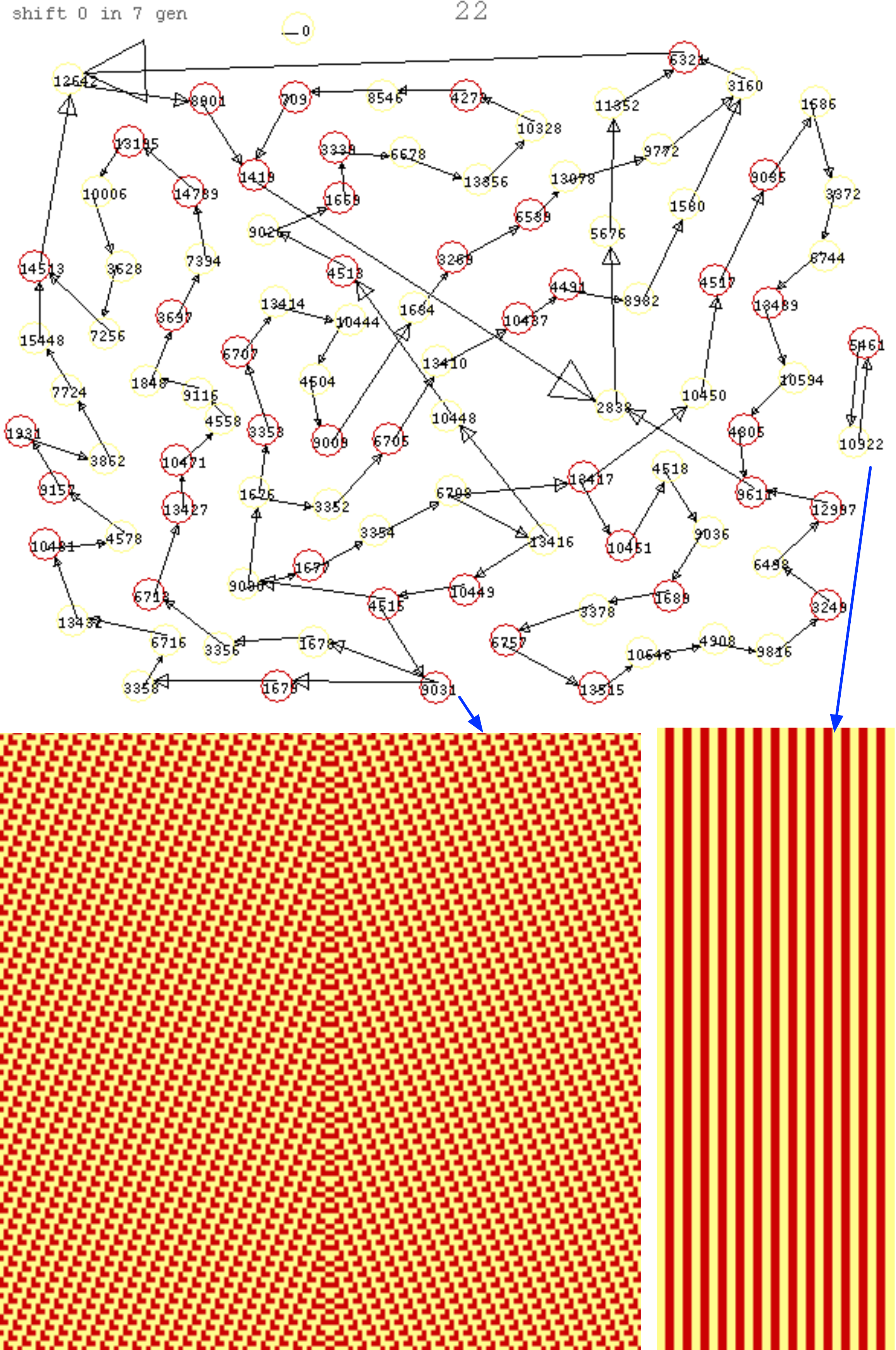}}
\caption{The extended de Bruijn diagram (0,7) calculating a pattern emitting  mobile self-localizations. To reproduce this pattern we concatenate the expressions $(1010001)^n$-111000101100010-$(1100010)^n$, where $n>0$ is the number of copies. The small cycle represents still life patterns.}
\label{deBruijn0s7g}
\end{figure}

\begin{figure}[!tbp]
\centerline{\includegraphics[width=1\textwidth]{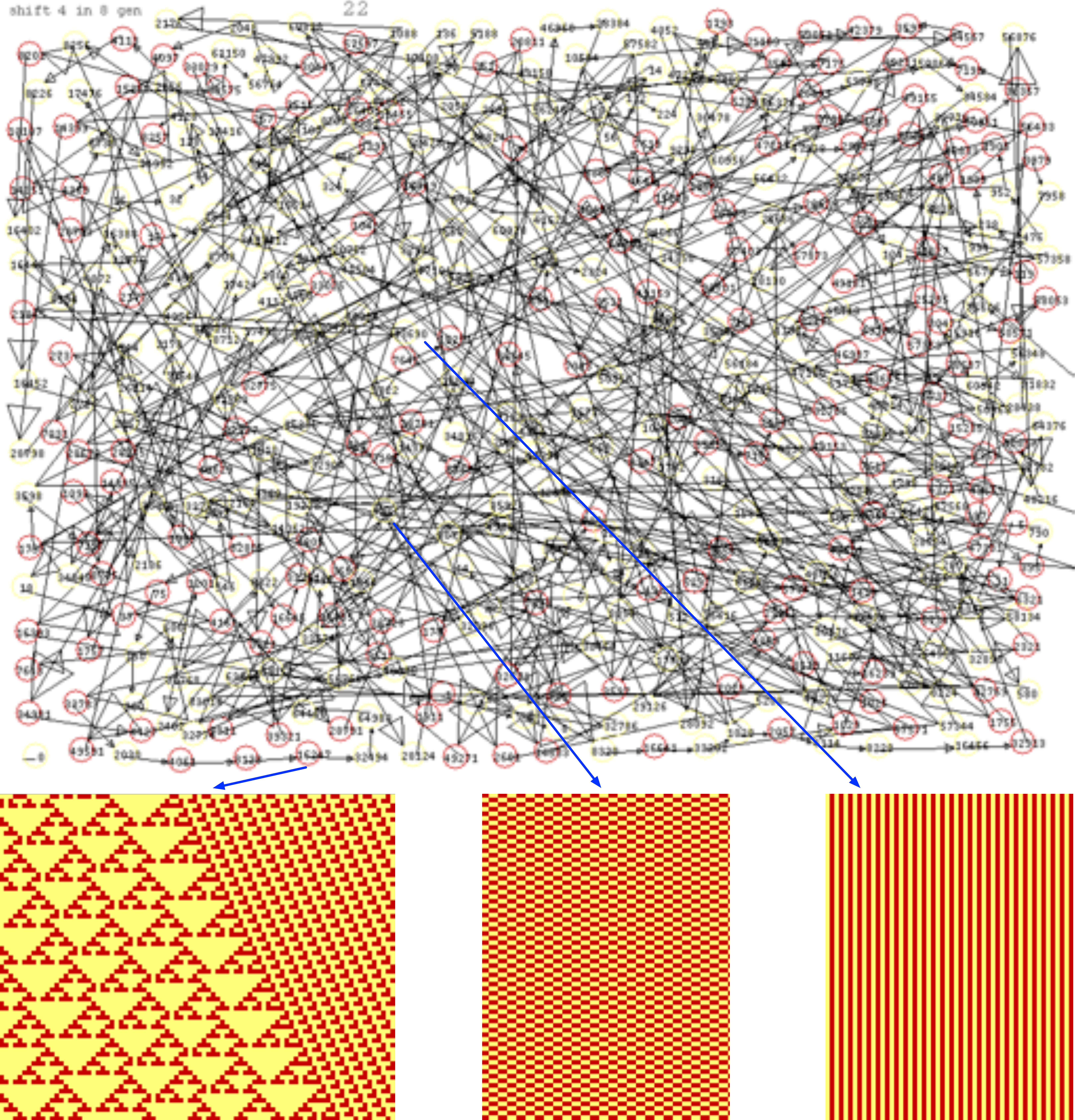}}
\caption{The extended de Bruijn diagram (4,8) calculating mobile self-localizations, small tilings and meshes. The first large cycle calculates a configuration known as fuse 
\cite{McIntosh-2009}
because two periodic patterns with different densities can evolve together without perturbing each others boundaries.}
\label{deBruijn4s8g}
\end{figure}

\begin{figure}[!tbp]
\centerline{\includegraphics[width=0.97\textwidth]{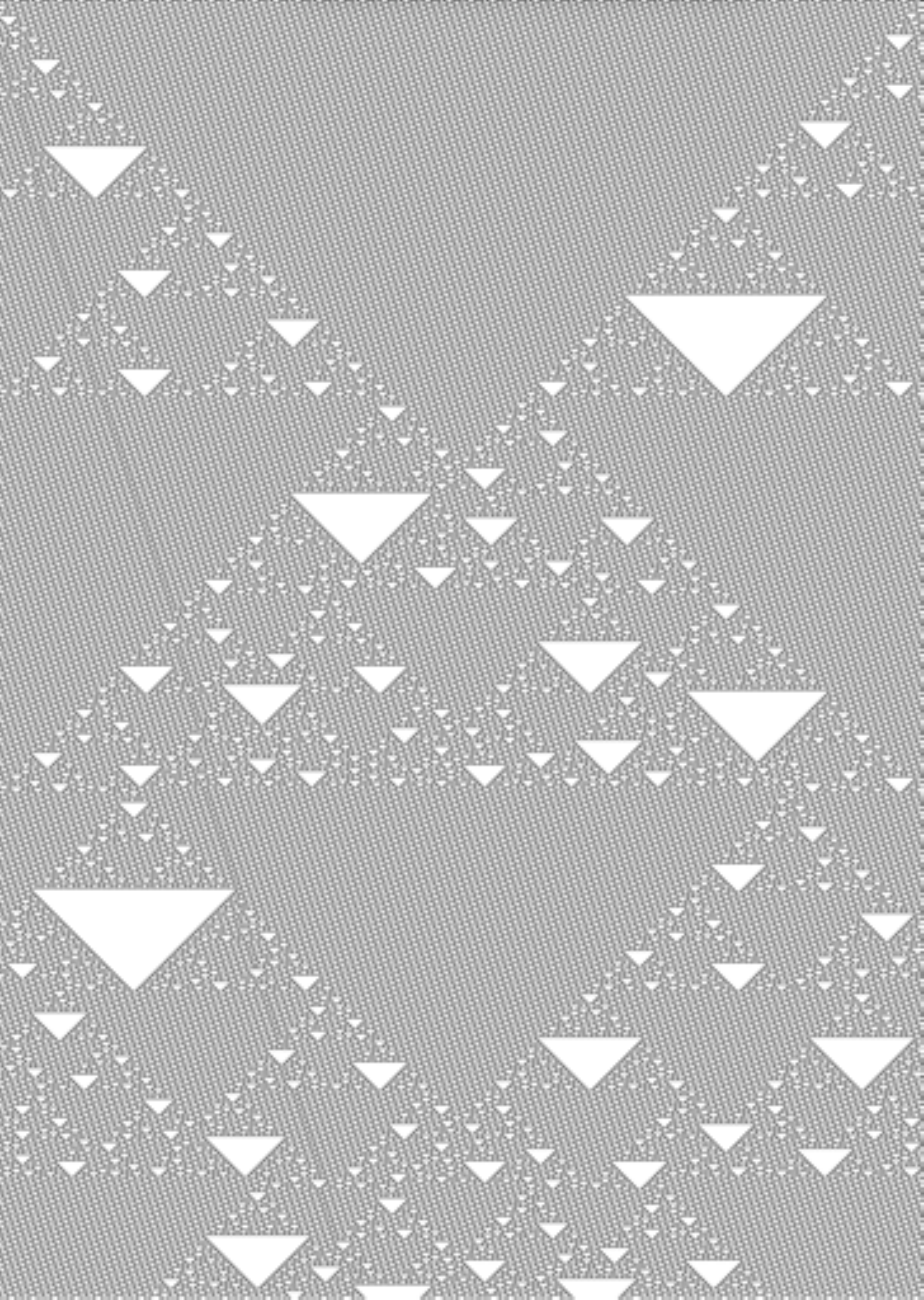}}
\caption{Non-trivial dynamics emerging in ECA Rule 22 on a ring of 1,198 cells during 1,679 generations. Mobile localizations emerge as triangular polygons traveling in a mobile periodic background. The localization conserve their shape when collide with each other. This dynamics was discovered with the extended de Bruijn diagram order (10,2), see Fig. \ref{deBruijnR22}. These configurations cannot be reached from a random initial condition.}
\label{complexDynamics22b-3358}
\end{figure}
%
%
\subsection{Extended de Bruijn diagram}
\label{subsection:Extended de Bruijn diagram}
%
%
An extended de Bruijn diagram
\cite{McIntosh-1991,McIntosh-2009}
 takes into account wide overlapping of neighborhoods. 
 We represent $M^{(2)}_{R22}$ by indexes $i=j=2  \, r  \, n$, where $n \in \mathbb Z^{+}$, $M^{(3)}_{R22}$ and $i=j=3  \, r \, n$, $M^{(4)}_{R22}$ and $i=j=4  \, r  \, n$, and so up to $M^{(m)}_{R22}$ with $i=j=m  \, r  \, n$; consequently basic de Bruijn diagram is obtained when $m=1$. 
The regular expressions derived from the de Bruijn diagram for Rule 22 (Tab.~\ref{regularexpressions}) can be linked to space--time dynamics phenomena exhibited by the rules. These includes symmetric complex behavior, chaos, stable periodic behavior. 
Figure \ref{deBruijn10gen} shows in detail every periodic pattern yielded from extended de Bruijn diagrams. To read the diagram we use notation $(i,j)$, where $i$ is a displacement (left or right) and $j$ is  a number of generations.  Thus the pattern in position $(0,0)$ (upper center) displays a periodic pattern without both displacement and period, the expression reproducing this pattern is $(01)^*$ 
(de Bruijn subdiagram in Fig. \ref{deBruijn10gen} and Eq. 2 in Table \ref{regularexpressions}).
\begin{figure}
\centerline{\includegraphics[width=1\textwidth]{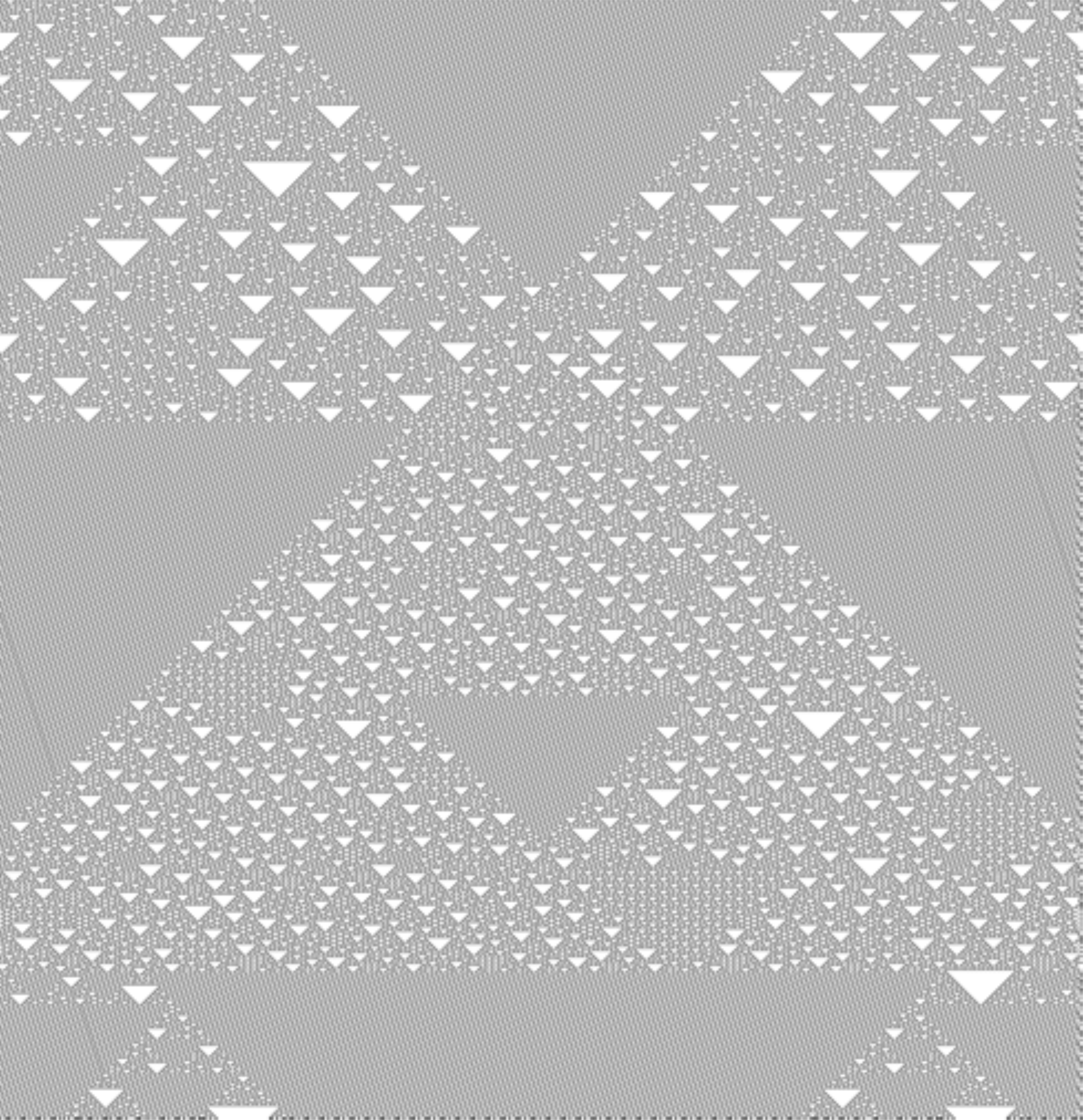}}
\caption{Non-trivial dynamics emerging in ECA Rule 22 with the mobile periodic background (10,2) evolving with a high density of small tilings. Different large triangular polygons can be constructed from the interactions of other polygons.}
\label{complexDynamicsR22a}
\end{figure}
\begin{figure}[!tbp]
\centerline{\includegraphics[width=0.95\textwidth]{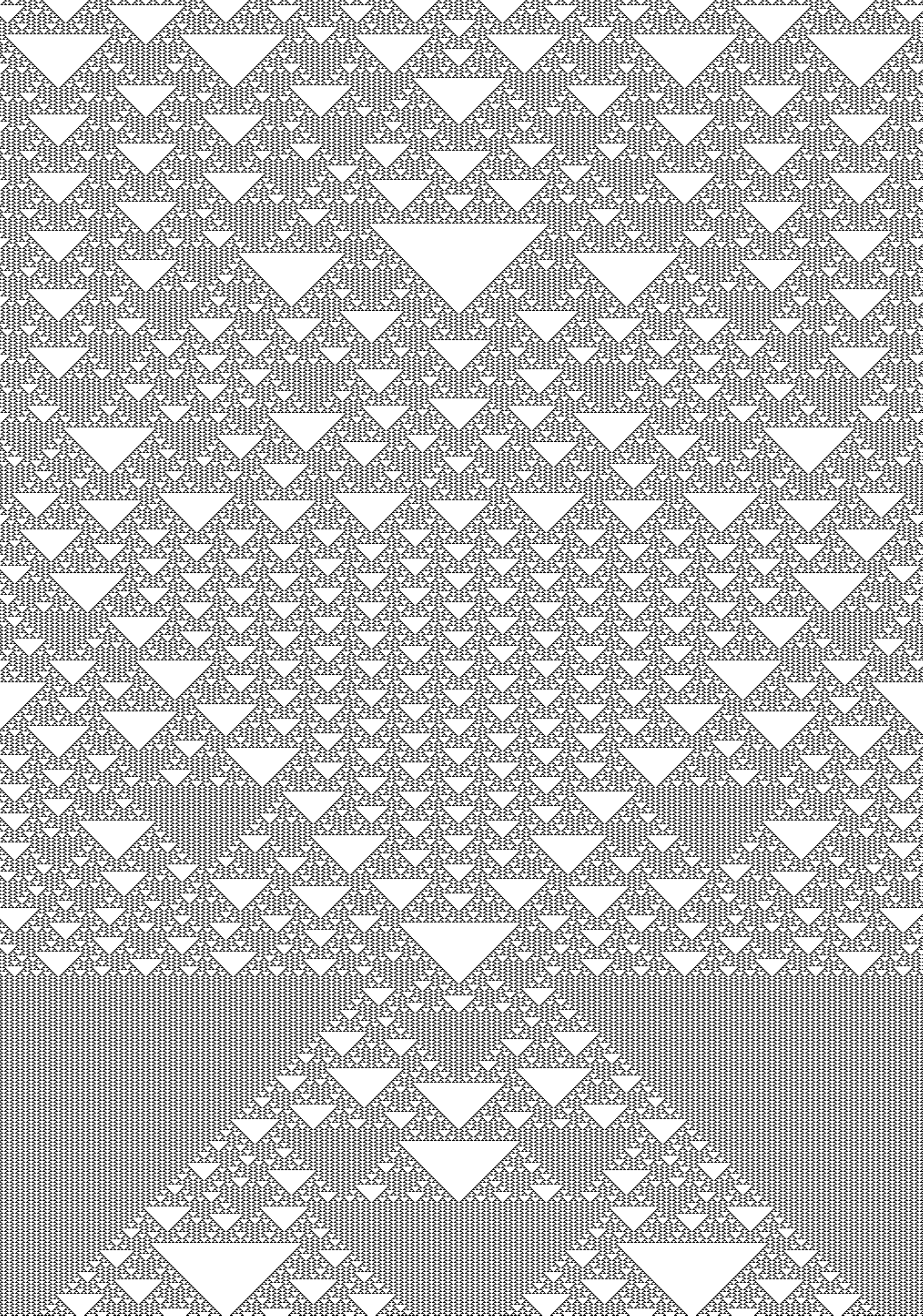}}
\caption{Non-trivial behavior emerging in a fix periodic background, i.e., this background does not move. In this periodic background complex large triangular polygons can emerge as well, including several types of small tiles. This fixed periodic background can be calculated from the extended de Bruijn diagram (0,4), see Fig. \ref{deBruijnR22}.}
\label{complexDynamics23-294984}
\end{figure}
In this way, a lot of  non-trivial patterns can be extracted. Let us consider few examples. 
\begin{itemize}
\item $(0,7)$. The graph is characterized by two cycles, the small cycle yielding simple still life patterns and the large cycle represents  configurations emitting traveling localizations, ro particles. These configurations can not  evolve naturally, i.e. from an initial random conditions,  because they are destroyed when a certain limit of their size is reached (Fig. \ref{deBruijn0s7g}). They configurations are  expressed for the following cycle--ways: 

$4515 \rightarrow 9080 \rightarrow 167 \rightarrow 3354 \rightarrow 6708 \rightarrow 13416 \rightarrow 10449 \rightarrow 4515 \rightarrow 9031 \rightarrow 1679 \rightarrow 3358 \rightarrow 6716 \rightarrow 13432 \rightarrow 10481 \rightarrow 4578 \rightarrow 9157 \rightarrow 1931 \rightarrow 3862 \rightarrow 7724 \rightarrow 15448 \rightarrow 14513 \rightarrow 12642 \rightarrow 8901 \rightarrow 1419 \rightarrow 2838 \rightarrow 5676 \rightarrow 11352 \rightarrow 6321 \rightarrow 12642$ 

The regular expression to reproduce the same pattern is calculated as Eq. 8 in Table \ref{regularexpressions}.
\item $(4,8)$. The graph has several paths between different cycles. Some of these cycles calculate trivial patterns. Other cycles represent configurations developed from a `fusion' of two periodic regions competing for the space (Fig. \ref{deBruijn4s8g}). Other fused configurations can be found in coordinates (6,9), $(-4,6)$, $(-4,8)$, $(-5,7)$, $(-6,9)$, $(-8,7)$ and $(-10,8)$.
\item $(10,2)$. Here we observe composed triangular polygons. The polygons sustain in a periodic mobile background. The background features small particles crossing the space (Fig. \ref{complexDynamics22b-3358}). This complex behavior emerges with probability between 1/7 and 3/7. This complex behavior was discovered with the help of expressions derived from the basins of attraction (see Fig. \ref{complexDynamics22syme-874x8092}). One more example of interaction of large triangular domains evolving in the periodic background is shown in Fig.~\ref{complexDynamicsR22a}.
\item $(0,4)$.  The periodic background is stationary. Several types of triangular domains and several families of small tiles emerge (Fig. \ref{complexDynamics23-294984}). Other coordinates leading to similar configurations are $(0,8)$, $(-6,2)$, $(-6,6)$, $(6,2)$ and $(6,6)$.
\end{itemize}
During the analysis of configurations derived with the de Bruijn diagrams and basin of attractors we have referred to some specific regular expressions. 
Regular expressions can be calculated recursively by following paths on a graph with 
\cite{Hopcroft:Ullman-1979}:
\begin{equation}
R_{i,j}^k = R_{i,j}^{k-1} + R_{i,k}^{k-1} (R_{k,k}^{k-1})^* R_{k,j}^{k-1}
\label{reeq}
\end{equation}
\noindent where $i$ is the initial state and $j$ the final state. Base case when $k=0$ is the direct path to every node. This way, by using the basic de Bruijn diagram in Fig. \ref{deBruijnR22}, we have calculated the whole set of regular expressions, summarized in Table \ref{regularexpressions}. The first column shows an equation number, the second column the regular expression, and the third column the kind of behavior that emerges when we codify configurations by these regular expressions. This way, we could evaluate these equations and explore an unlimited number of configurations.
\begin{table}[th]
\centering
\caption{Regular expressions derived in ECA Rule 22. The set of equations is calculated using the recursive function $R_{i,j}^{k}$ (Eq.~\ref{reeq}) to recognise $k$ paths between nodes $i$ to $j$ in the de Bruijn diagram (Fig.~\ref{deBruijnR22}).}
\vspace{0.1cm}
\begin{tabular}{ | c | p{4.5cm} | p{4.6cm} |}
\hline
Eq & expression & evolution \\
\hline \hline
1 & $(0+1)^*$ & stable state \\ \hline
2 & $(01)^*$ & stable periodic \\ \hline
3 & $(001)^*$ & stable state \\ \hline
4 & $11(01)^*00$ & still life \& symmetric complex behavior \\ \hline
5 & $(01)^*(0+1)$ & stable periodic \& symmetric complex behavior \\ \hline 
6 & $(000111)^*$ & stable state \\ \hline
7 & $(0+0(01)^*00)^*$ & stable periodic, chaos \& big gaps \\ \hline
8 & $(((01)^*+0)00^*1)^*$ & chaos, complex behavior \\ \hline
9 & $(0+1)+11(01)^*(0+1)$ & complex behavior \\ \hline
10 & $((01)^*00)(0+0(01)^*00)^*$ & chaos, stable periodic, chaos \& big gaps \\ \hline
11 & $(11(01)^*00)(0+0(01)^*00)^*$ & chaos, still life \& symmetric complex behavior \\ \hline
12 & $(0+0(01)^*00)^*0(01)^*(0+1)$ & stable state, stable periodic, chaos \& big gaps \\ \hline
13 & $(0+1(01)^*00)(0+0(01)^*00)^*$ & chaos, stable periodic, chaos \& big gaps \\ \hline
14 & $((0+1)+11(01)^*1)+(11(01)^*00)(0+0(01)^*00)^*(0(01)^*(0+1))$ & -- \\ \hline
15 & $(0^*10^*)(10^*10+(10^*+10^*10)0^*)^*$ & -- \\ \hline
\end{tabular}
\label{regularexpressions}
\end{table}
%
%
\section{Garden of Eden}
\label{section:Garden of Eden}
%
%
The question ``does a complex CA contains a universal constructor?'' is a classic problem appearing in the CA literature since von Neumann works
\cite{von Neumann-1966}. 
A configuration of a universal constructor in the Game of Life CA is proposed by Goucher in 2010
\cite{Goucher-2010}. 
In this context, our aim is to know if ECA Rule 22 is able to construct any string. Previously this problem was studied by McIntosh
\cite{McIntosh-2009} 
who found that Rule 22 has a global injective relation and therefore configurations without ancestors exist. 

We  use a subset diagram to calculate {\it Garden of Eden} configurations
\cite{McIntosh-2009}, 
i.e., the configurations without ancestors
\cite{Amoroso:Cooper-1970}. 
A subset diagram has $2^{k^{2r}}$ vertices with $k$ states and $r$ neighbors. If all the configurations of the certain length have ancestors then all the configurations with extensions both to the left and to the right with the same equivalence must have ancestors. But if this is not the case, then the vertices represent Garden of Eden configurations.

We can define the subset diagram as the power set of $2^{k^{2r}}$. Such that each subset $S \in U_{S}$ (where $U_{S}$ is a power set) and one symbol $a \in \Sigma$:
\begin{equation}
	\alpha(S,a) = \bigcup_{q_i \in S} \mbox{ } \varphi(q_i,a).
\label{eqSubsetR22}
\end{equation}
Vertices of the subset diagram are formed by the combination of each subset formed from the states of the de Bruijn diagram. Symbolic de Bruijn matrices $M_{k,s}$ or $M_{s}$ are characterized by $k$ states and $s$ number of states in the partial neighborhood. Thus, for Rule 22 we can obtain symbolic matrices, derived from the de Bruijn sub--diagrams shown in Fig.~\ref{deBruijnR22-2}. For any ECA we have four sequences of states in the Bruijn diagram enumerated as 0, 1, 2 and 3 (see Fig.~\ref{deBruijnGenerico}).

\begin{figure}[!tbp]
\centering
\includegraphics[width=0.35\textwidth]{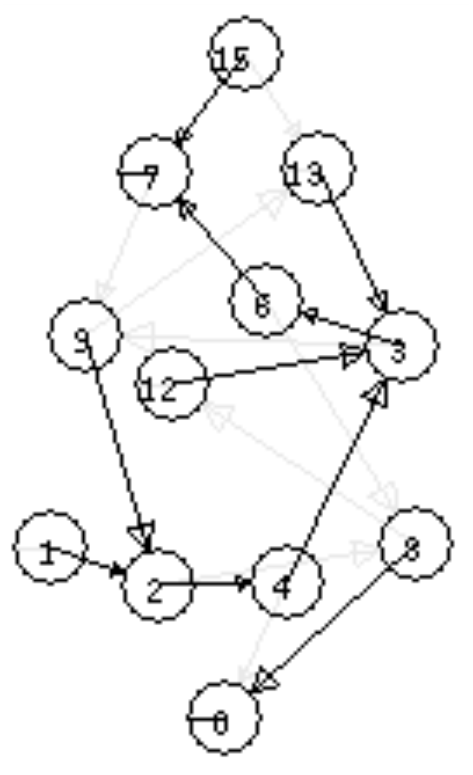}
\caption{The simplified subset diagram for ECA Rule 22.}
\label{subSetR22}
\end{figure}

Union between subsets is represented by the state in which each sequence evolves and is assigned to the states (subsets that form it) as governed by Eq.~\ref{eqSubsetR22}. 
Relations between subsets for Rule 22 are constructed in Table~\ref{relationsubsetR22}. 
Figure~\ref{subSetR22} shows the full scalar subset diagram. Each class of edges defines a function on $\Sigma_{0}$ or $\Sigma_{1}$. The subset diagram describes the union $\Sigma_{0} \cup \Sigma_{1}$ that by itself is not functional 
\cite{McIntosh-1991}.

We must distinguish four types of subsets, where it is possible to make a transition between its four unit classes. Also, we should observe that a residual of the de Bruijn diagram can be found in the subset diagram. This is because a unit class is precisely defined by the nodes of the original diagram. At first instance, we can see some relations are more frequent than others. Also there are nodes without inputs, or nodes with most connections including loops. Most important are cycles of different lengths. They are used to infer words, or sequences, that a CA could recognize. 
Thus, the subset diagram can be used as a general machine to recognize the universe of words in which a CA could evolve.
\begin{table}[th]
\caption{Relations between states of the subset diagram in Rule 22.}
\centering
\vspace{0.2cm}
\begin{tabular}{|c|c|c|c|}
\hline
$S$ & label & 0 & 1 \\
\hline
 	$\phi$ & 0 & 0 & 0 \\
 	\{0\} & 1 & 1 & 2 \\
 	\{1\} & 2 & 0 & 12 \\
 	\{2\} & 4 & 1 & 2 \\
 	\{3\} & 8 & 8 & 4 \\
 	\{0,1\} & 3 & 1 & 14 \\
 	\{0,2\} & 5 & 1 & 2 \\
	\{0,3\} & 9 & 9 & 6 \\
 	\{1,2\} & 6 & 1 & 14 \\
 	\{1,3\} & 10 & 8 & 12 \\
  	\{2,3\} & 12 & 9 & 6 \\
 	\{0,1,2\} & 7 & 1 & 14 \\
 	\{0,1,3\} & 11 & 9 & 6 \\
 	\{0,2,3\} & 13 & 9 & 6 \\
 	\{1,2,3\} & 14 & 9 & 14 \\
 	\{0,1,2,3\} & 15 & 9 & 14  \\
\hline
\end{tabular}
\label{relationsubsetR22}
\end{table}

By analyzing the full diagram we can derive a small subset diagram which is deduced from the original diagram (de Bruijn diagram). This diagram includes only vertexes with cycles, the universal and empty set and the subset with one element, yielding a new diagram that will be more practical for us. The reduction gives yet a more small diagram to read quickly strings belonging to Garden of Eden configurations. The reduction is also useful to calculate the degree of Welch indices for reversible CAs
\cite{Seck-Tuoh-Mora:Medina-Marin:Hernandez-Romero:Martinez:Barragan-Vite-2017}.
%
The expressions that determine Garden of Eden configurations in ECA Rule 22 are listed below:
\begin{itemize}
\item 10110
\item 01111*01101*
\item 11*(011111)*1*0110
\end{itemize}
%
%
\section{Fractals}
%
%
\subsection{Iterated functions in Rule 22}
%
%
 A fractal is constructed recursively from a self--replication of a pattern  
 \cite{Eppstein-Fractals}. 
 Chaotic systems often bear properties of fractals.

\begin{figure}[!tbp]
\begin{center}
\subfigure[]{\scalebox{0.12}{\includegraphics{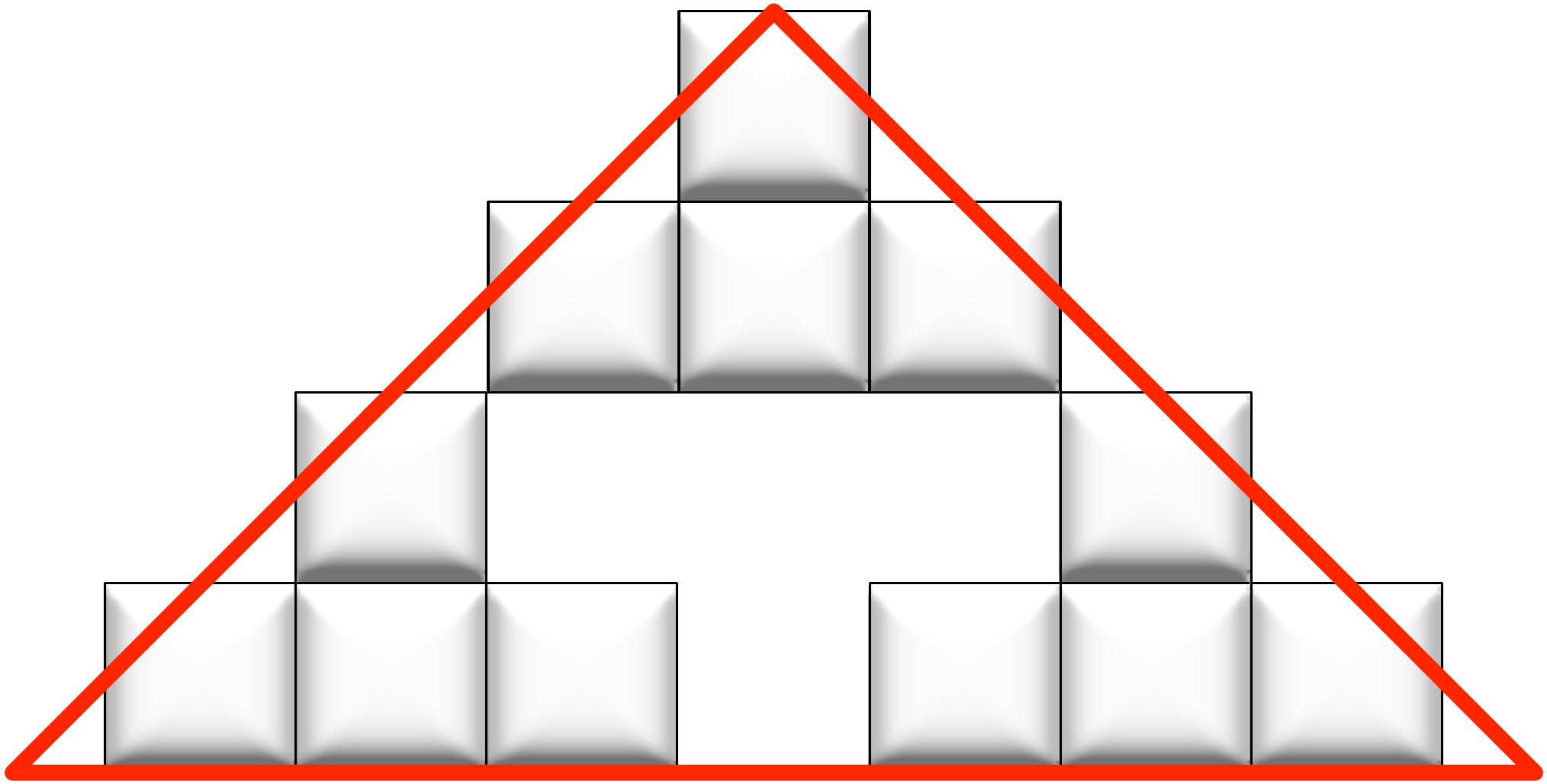}}} \hspace{0.5cm}
\subfigure[]{\scalebox{0.06}{\includegraphics{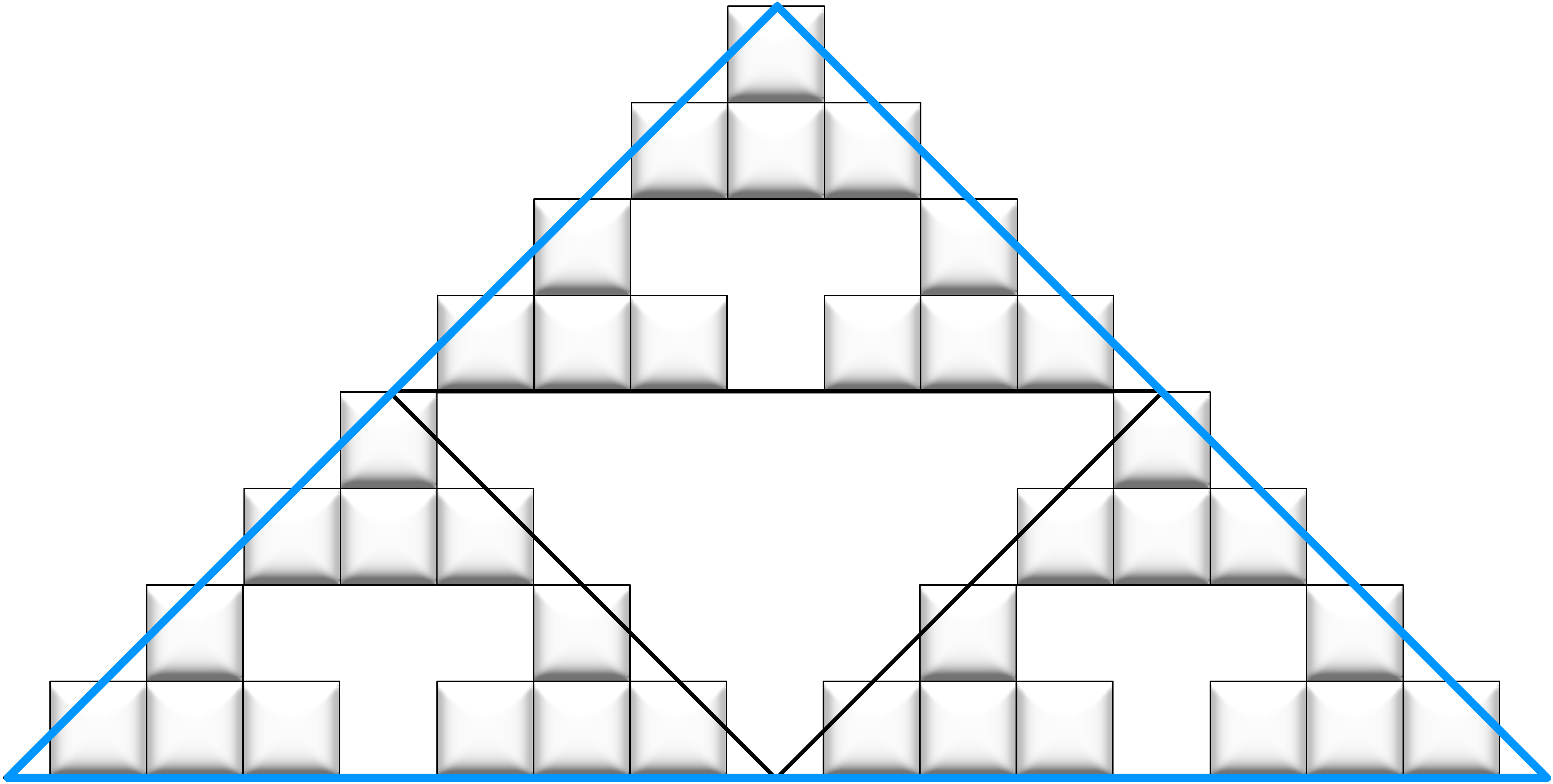}}} \hspace{1.8cm}
\subfigure[]{\scalebox{0.074}{\includegraphics{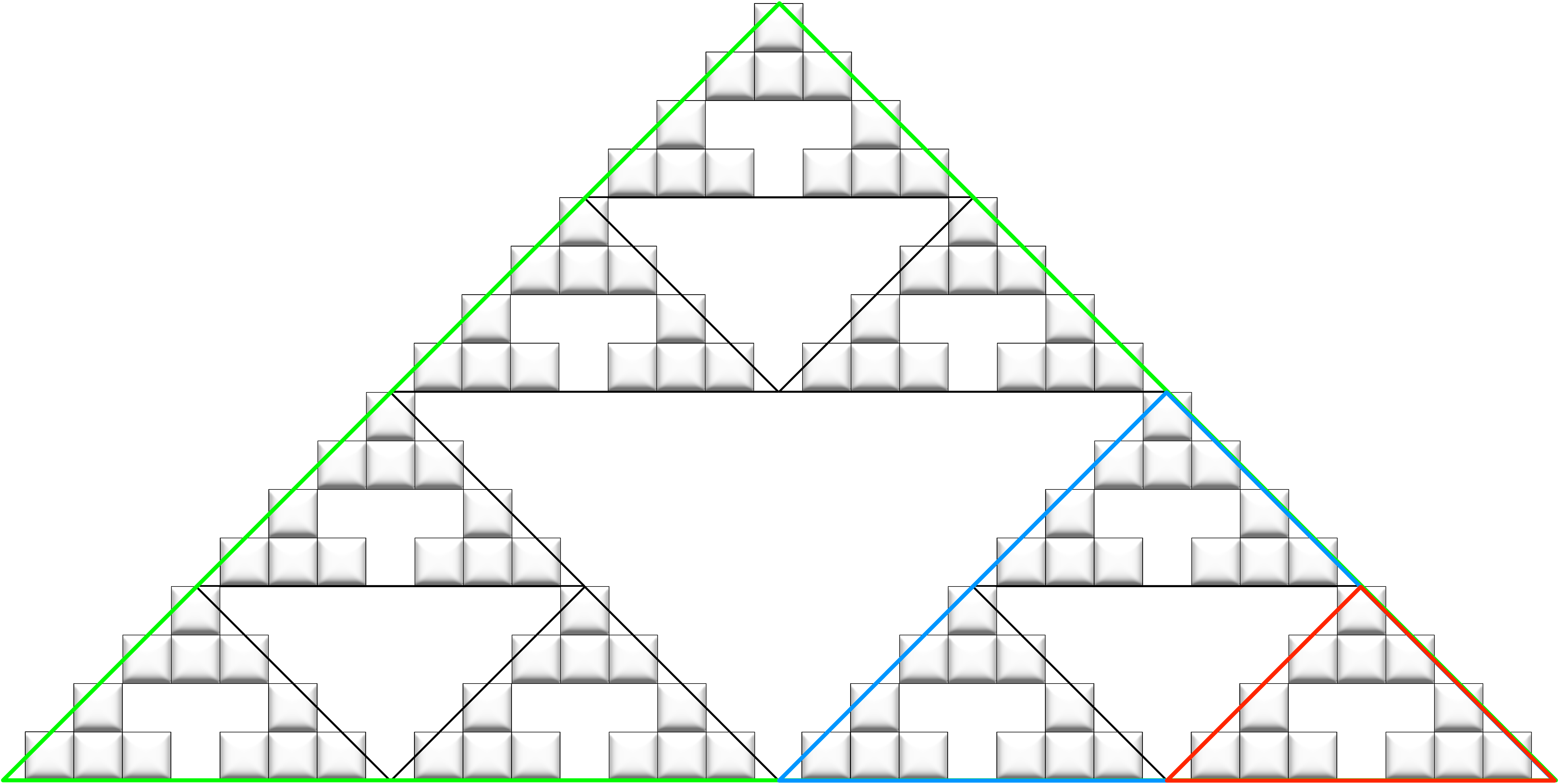}}}  
\end{center}
\caption{Iterated function determines a fractal defining a Sierpi\'{n}ski triangle in ECA Rule 22 from a composition of three tiles starting with a 1.}
\label{IteratedFunction}
\end{figure}

\begin{figure}[!tbp]
\begin{center}
\subfigure[]{{\includegraphics[width=1.02\textwidth]{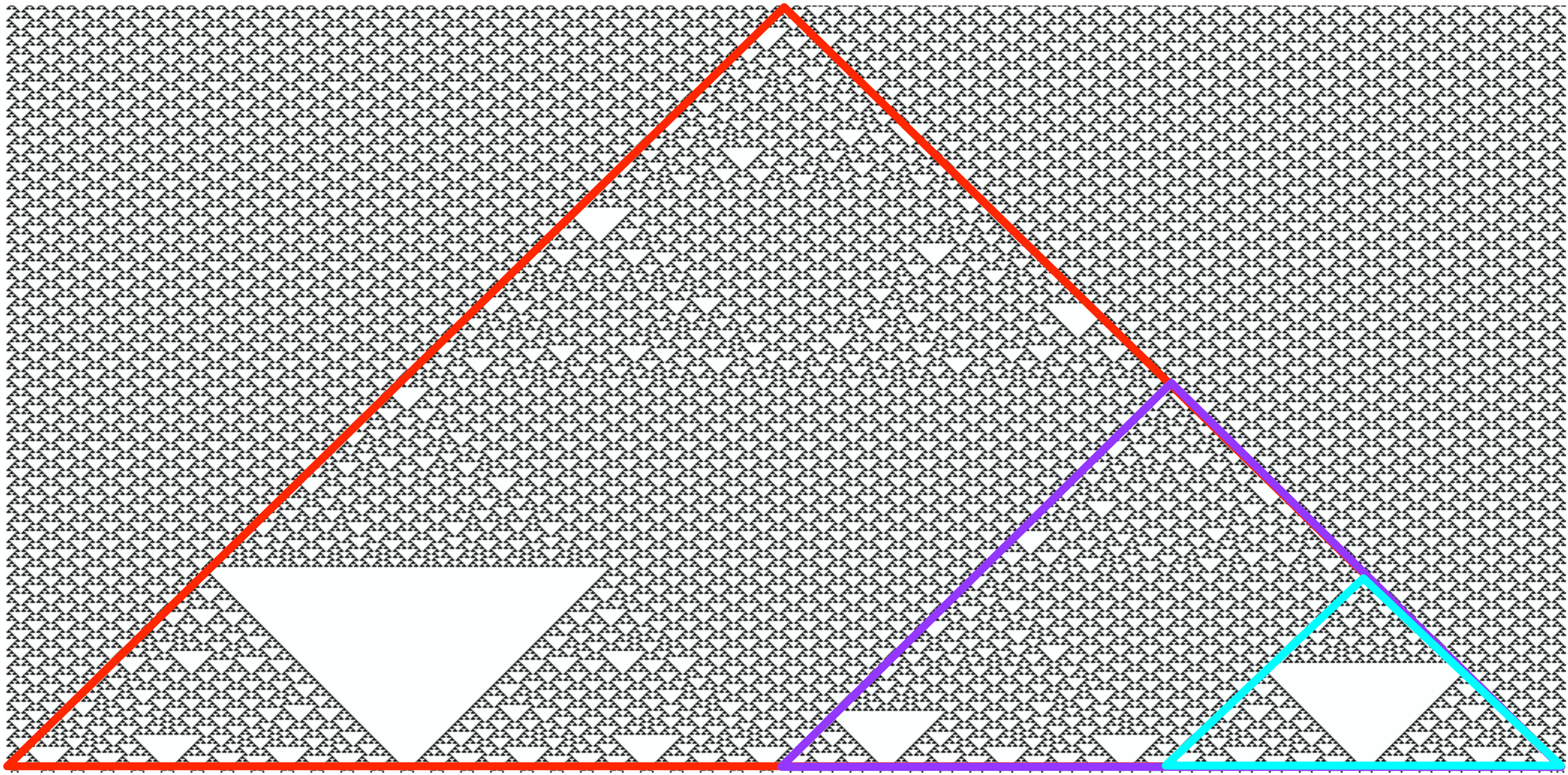}}}  
\subfigure[]{{\includegraphics[width=1.02\textwidth]{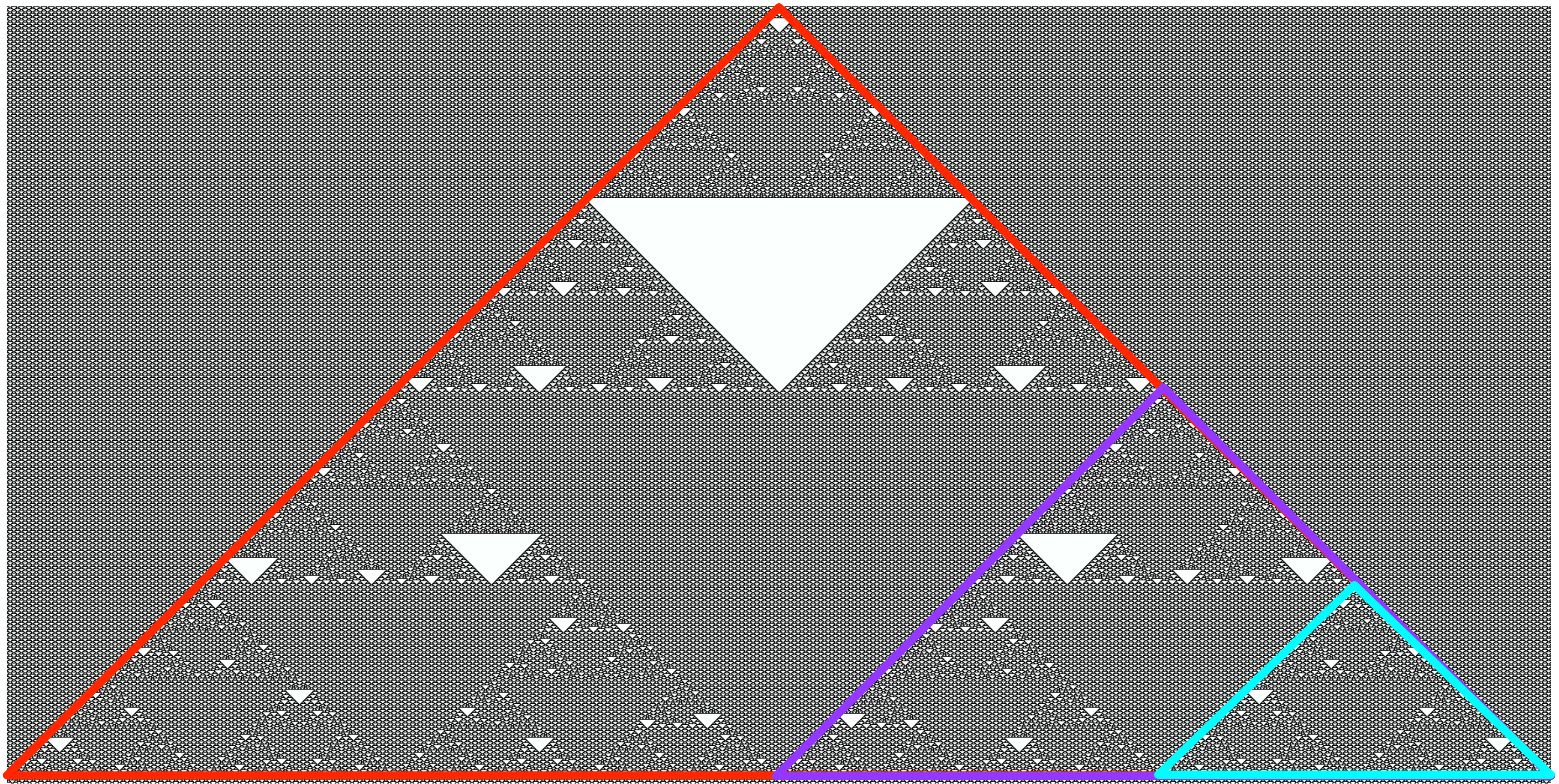}}}
\end{center}
\caption{Composition of non-trivial fractals emerging in ECA Rule 22 after thousands of generations. The iterated function preserves its fractal dimension. These fractals evolve on a periodic background without displacement.}
\label{IteratedFunctionef}
\end{figure}

\begin{figure}[!tbp]
\begin{center}
\subfigure[]{\scalebox{0.345}{\includegraphics{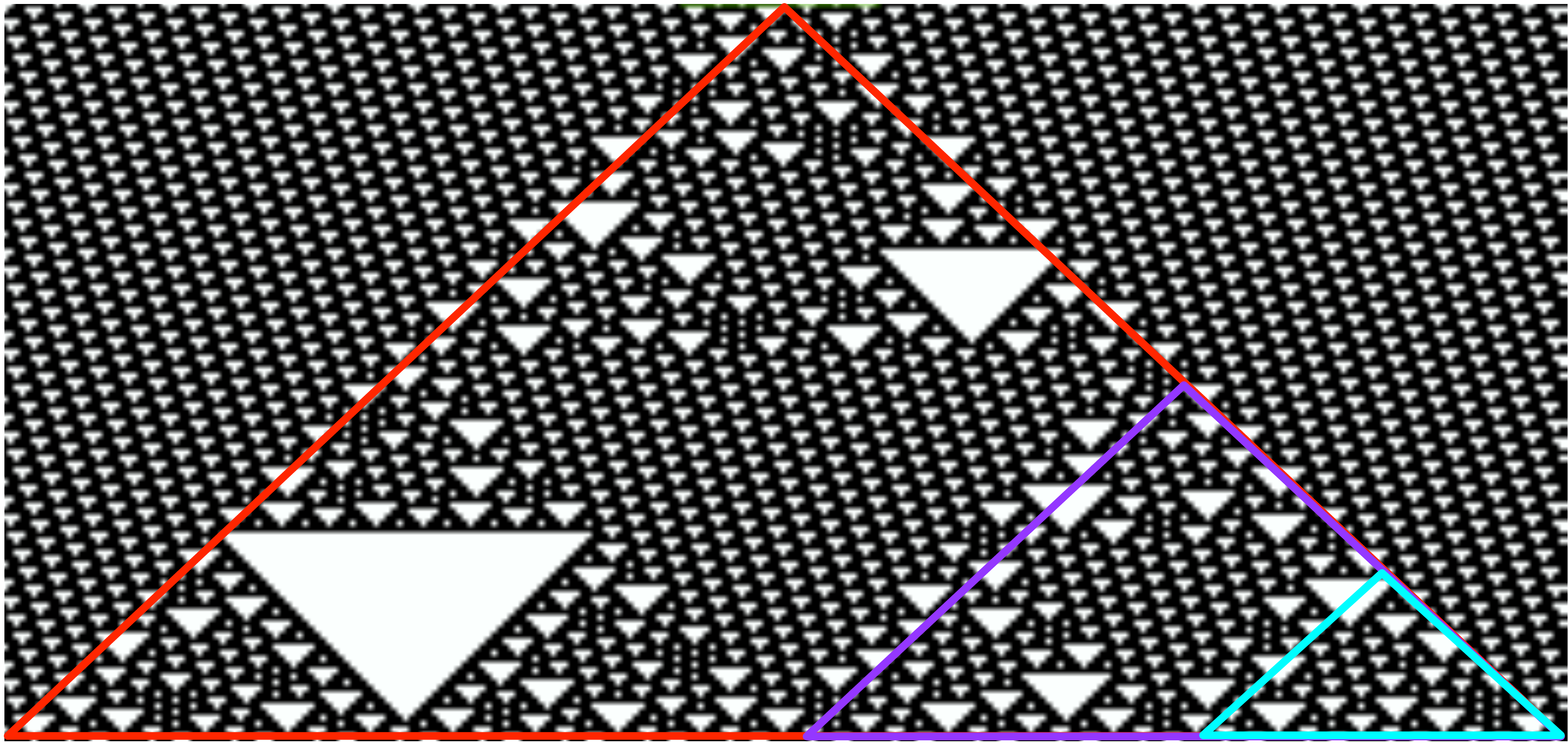}}}  
\subfigure[]{\scalebox{0.364}{\includegraphics{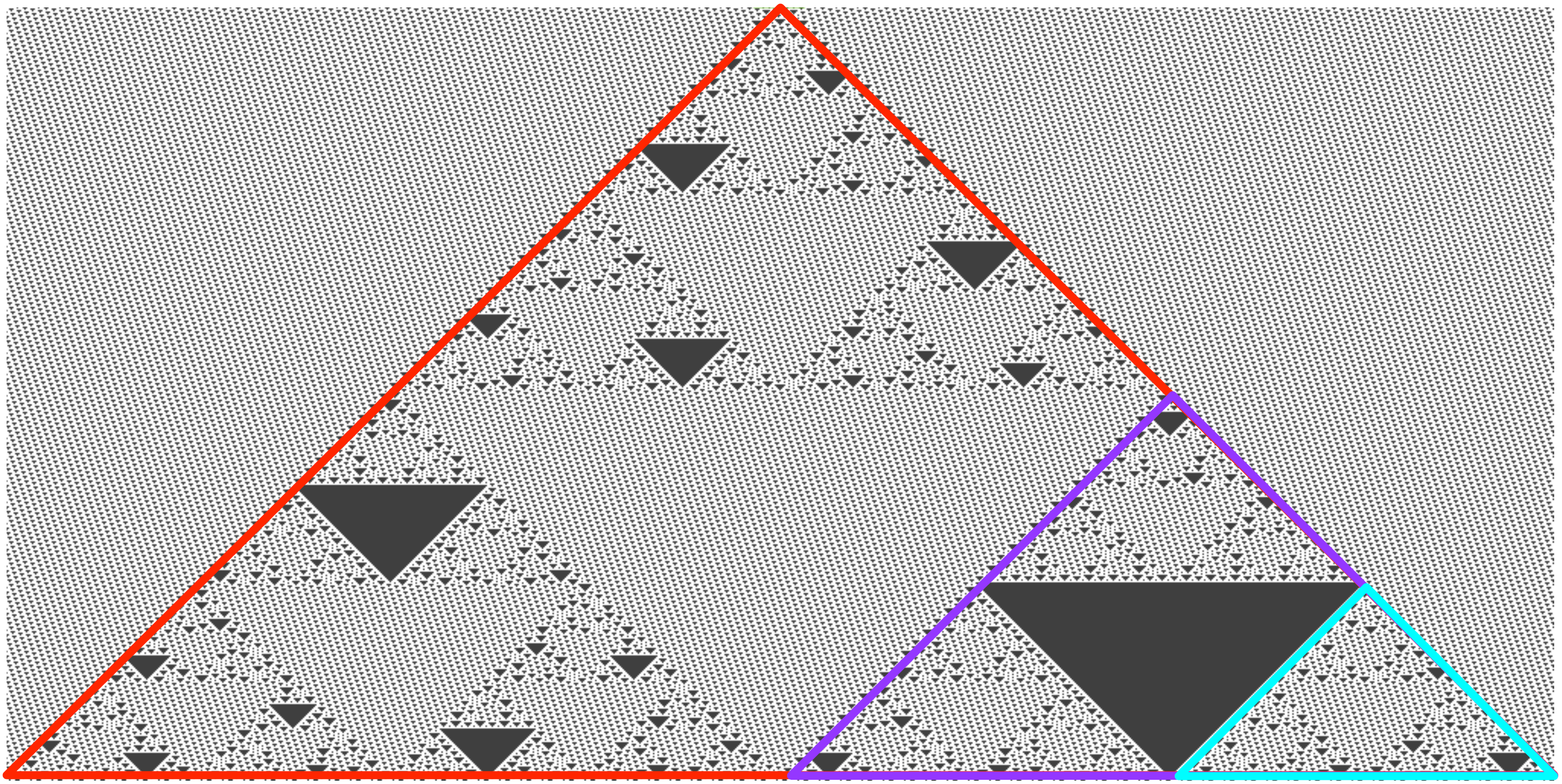}}}
\end{center}
\caption{Composition of non-trivial fractals emerging in ECA Rule 22 after thousands of generations. The iterated function preserves its fractal dimension. These fractals evolve on a periodic background without displacement.}
\label{IteratedFunctiongh}
\end{figure}

ECA Rule 22 produces a fractal pattern, known as Sierpi\'{n}ski triangle, starting from a single cell in state 1  (Fig.~\ref{evolR22}). Figure~\ref{IteratedFunction} shows a triangle constructed with three small tiles derived in Rule 22, this triangle grows in power of two with respect the number of cells. The main triangle (Fig.~\ref{IteratedFunction}a) has three replicas in the next iteration (Fig.~\ref{IteratedFunction}b), and following iteration produces nine base replicas (Fig.~\ref{IteratedFunction}c). The fractal dimension $D$ can be calculated given the number of replicates $N$ and the scaling factor $m$ 
\cite{Chen:Dong-1998}. 
The fractal dimension of the patterns generated by Rule 22 is the following:
\begin{equation}
D = \frac{log(N)}{log(m)} = \frac{log(3)}{log(2)} = 1.5849625.
\label{equation:Frac dim Rule 22}
\end{equation}

Also ECA Rule 22 displays non trivial behavior via fractals where they emerge in different stages during the evolution. 
These fractals are combined with other fractals constructed from Rule 22 over thousands of generations.

Using regular expressions we found two different periodic backgrounds emerging in ECA Rule 22, as discussed in Sections 5, 6 and 7. 
Let us illustrate two fractals growing in intervals of other fractals with different composition of tiles. Figure~\ref{IteratedFunctionef}a shows the initial state of fractals growing in a periodic background without displacement conserving the same fractal dimension.  Figure~\ref{IteratedFunctionef}b shows the same iterated function over thousands of generations. The same behavior is tested on a periodic background with displacement in Figs.~\ref{IteratedFunctiongh}ab. Composed fractals emerging in periodic backgrounds with or without displacement are disjoint.
%
%
\subsection{Rule 18, mutations, gaskets and seashells}
\begin{figure}
\centerline{\includegraphics[width=10.5cm]{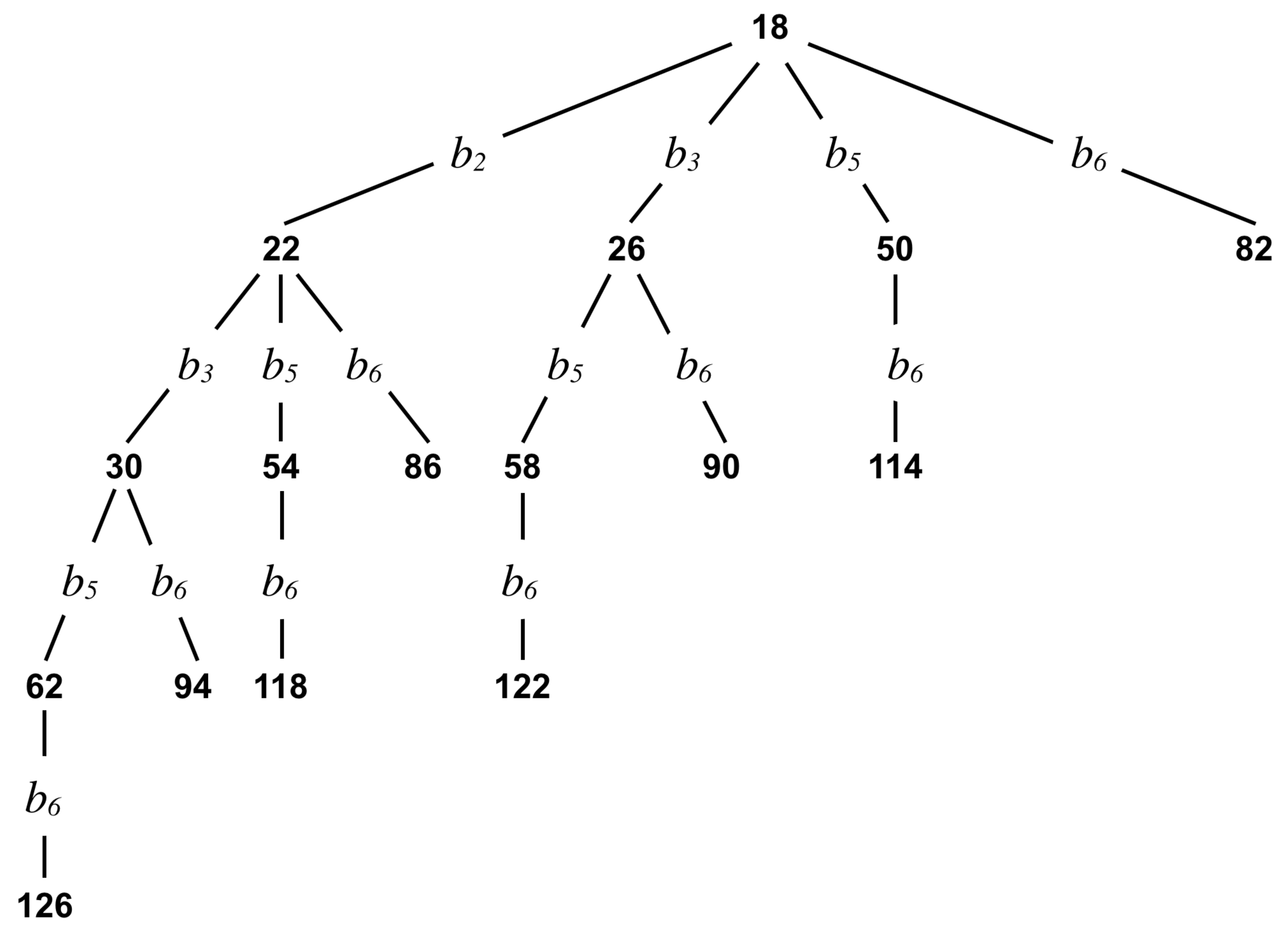}}   
\caption{Mutation tree from Rule 18 derived from 
Tab.~\ref{table:Mutation table from Rule 18}.
A branch from Rule $R$ to Rule $R'$ 
$
   (R|b_i \leadsto R')
$
bears exactly the 1--bit genetic change $b_i$ and 
$
  d_H(R,R') = 1
$
where $d_H$ is the Hamming distance.
}  
\label{fig:Mutation tree from Rule 18}
\end{figure}

%
%
\begin{table}
\centering
\caption{Mutation table from Rule 18 in the ECA subset 
$
 (b_7 b_6 b_5 b_4 b_3 b_2 b_1 b_0) =  ( \textbf{0} \ b_6  \ b_5 \  \textbf{1} \ b_3  \ b_2 \  \textbf{1} \  \textbf{0}).
$
A rule $R$  {\em mutates} into Rule $R'$ through bit $b_i$ 
$
   (R|b_i \leadsto R')
$
with exactly a 1--bit change.
}
\vspace{0.2cm}
{\setlength{\tabcolsep}{0.1mm}}
\begin{tabular}{|c||c|c|c|c|c|c|c|c||c|}
\hline	 \rule[-0.1cm]{0cm}{0.5cm} 	
 \footnotesize Rule & \footnotesize\textbf{111}&\footnotesize\textbf{110}&\footnotesize\textbf{101}&\footnotesize\textbf{100}&\footnotesize\textbf{011}&\footnotesize\textbf{010}&\footnotesize\textbf{001}&\footnotesize\textbf{000}&\footnotesize Mutation\\ [0.5ex]
\hline\hline	
\footnotesize\textbf{18} &\small\textbf{0}  & \small 0  &  \small 0  &  \small \textbf{1} & \small 0  & \small 0  &  \small\textbf{1} & \small\textbf{0}  &  \\ \hline 
\footnotesize\textbf{22} &\small\textbf{0}   & \small 0  & \small 0   &  \small\textbf{1} &  \small 0  &  \small 1 & \small\textbf{1} & \small\textbf{0}  & {\small $18|b_2 \leadsto 22$}   \\ \hline 
\footnotesize\textbf{26} & \small\textbf{0}  &  \small 0  & \small 0  &  \small\textbf{1} &  \small 1  & \small 0  & \small \textbf{1} & \small\textbf{0}  & {\small $18|b_3 \leadsto 26$}   \\ \hline 
\footnotesize\textbf{30} & \small\textbf{0}  &  \small 0  &  \small 0 &  \small\textbf{1} &  \small 1  & \small 1  &  \small\textbf{1} & \small\textbf{0}  & {\small $22|b_3 \leadsto 30$}  \\ \hline\hline 
\footnotesize\textbf{50} & \small\textbf{0}  &  \small 0  &  \small 1 & \small \textbf{1} & \small  0  &\small  0  &  \small\textbf{1} & \small\textbf{0}  & {\small $18|b_5 \leadsto 50$}    \\ \hline 
\footnotesize\textbf{54} & \small\textbf{0}  &  \small 0  &  \small 1 & \small \textbf{1} &  \small 0  & \small  1  &  \small\textbf{1} & \small\textbf{0}   & {\small $22|b_5 \leadsto 54$}  \\ \hline 
\footnotesize\textbf{58} & \small\textbf{0}  &  \small 0  &  \small 1 & \small\textbf{1}  &  \small 1  & \small  0  &  \small\textbf{1} & \small\textbf{0}  & {\small $26|b_5 \leadsto 58$}   \\ \hline 
\footnotesize\textbf{62} & \small\textbf{0}  &  \small 0  &  \small 1 & \small \textbf{1} &  \small 1  &  \small 1  & \small \textbf{1} & \small\textbf{0}  & {\small $30|b_5 \leadsto 62$}   \\ \hline\hline 
\footnotesize\textbf{82} & \small\textbf{0}  &  \small 1  &  \small 0  & \small \textbf{1} &  \small 0  & \small 0  &  \small\textbf{1} & \small\textbf{0}  & {\small $18|b_6 \leadsto 82$}    \\ \hline 
\footnotesize\textbf{86} & \small\textbf{0}  &  \small 1  &  \small 0  & \small \textbf{1} &  \small 0  &  \small 1  &  \small\textbf{1} & \small\textbf{0}  & {\small $22|b_6 \leadsto 86$}     \\ \hline 
\footnotesize\textbf{90} & \small\textbf{0}  &  \small 1  &  \small 0  & \small \textbf{1} &  \small 1  & \small  0  &  \small\textbf{1} &\small \textbf{0}  & {\small $26|b_6 \leadsto 90$}   \\ \hline 
\footnotesize\textbf{94} &\small \textbf{0}  &  \small 1  &  \small 0  & \small \textbf{1} &  \small 1  &  \small 1  &  \small\textbf{1} & \small\textbf{0}  & {\small $30|b_6 \leadsto 94$}   \\ \hline\hline 
\footnotesize\textbf{114} & \small\textbf{0}  &  \small 1 &  \small 1  &\small \textbf{1} &  \small 0  &  \small 0  &  \small\textbf{1} & \small\textbf{0}  & {\small $50|b_6 \leadsto 114$}   \\ \hline 
\footnotesize\textbf{118} & \small\textbf{0}  &  \small 1 &  \small 1  & \small \textbf{1} &  \small 0  &  \small 1 &  \small\textbf{1}  & \small\textbf{0} & {\small $54|b_6 \leadsto 118$}  \\ \hline 
\footnotesize\textbf{122} & \small\textbf{0}  &  \small 1 &  \small 1  & \small \textbf{1} &  \small 1  & \small 0  &  \small\textbf{1} & \small\textbf{0}  & {\small $58|b_6 \leadsto 122$}   \\ \hline 
\footnotesize\textbf{126} & \small\textbf{0}  &  \small 1 &  \small 1  & \small \textbf{1} &  \small 1  & \small 1  &  \small\textbf{1} & \small\textbf{0} & {\small $62|b_6 \leadsto 126$}   \\ \hline 
\end{tabular}
\label{table:Mutation table from Rule 18}
\end{table}
Rule 18 and Rule 22 are complex rules widely reported in ECA literature
\cite{Wolfram-1983b,Wolfram-1984,McIntosh-1990b,Sinha-2006,Gravner:Griffeath-2011}.
Let us redefine them as follows: a cell takes state  `1' if exactly one
\begin{itemize}
\item (R18):  of its neighbors is in state `1': $(100,001) \rightarrow 1 $
\item (R22): in its neighborhood is in state `1': $(100,010,001) \rightarrow 1 $
\end{itemize}
and consider the subset of the 256 ECA
\begin{equation}
\Psi_{\widehat{R18}} = \left\{
	\begin{array}{lcl}
		1 & \mbox{if} & 100, 001 \\
		0 & \mbox{if} & 111, 000
	\end{array} \right. 
\label{equation:R18 subset}	
\end{equation}
that defines the sixteen rules displayed in
Tab.~\ref{table:Mutation table from Rule 18}.
Referring to the genotype paradigm in
\cite{Li:Packard-1990}
with a rule defined by the sequence 
$
 (b_7 b_6... b_1 b_0),
$
a rule $R$  {\em mutates} into Rule $R'$ through bit (or ``{\em gene}'') $b_i$, or 
$
   (R|b_i \leadsto R')
$
with exactly a 1--bit change, that yields the mutation (or inheritance) tree in
Fig.~\ref{fig:Mutation tree from Rule 18}.

This tree contains a set of rules with complex behavior inherited from Rule 18 and where Sierpi\'{n}ski patterns often appear.
Rules 18, 22, 122, 126 reveal a similar behavior towards ergodic regime and unstable areas, as well as Rule 90 but that  displays a multilayered pattern, whereas Rule 30 and, by reflection, Rule 86 exhibit small striped patterns near the polygonal border.
Rule 26 and, by reflection, Rule 82 exhibit large striped patterns at equilibrium, that may evolve towards sparse backbones at low or high densities, or periodic patterns similar to those in 
Fig.~\ref{fig:Rare periodic event}.
Rule 94 displays a Sierpi\'{n}ski gasket, but evanescent and quickly entering a uniform polygon with vertical stripes.
Finally it should be emphasized that Rules 50, 54, 58 and by reflection 114, 62 and by reflection 118, display patterns somewhat far from the sieve but nevertheless all of them enter the phase transition polygon\footnote{Rules 30, 54, 90, 126 are examined elsewhere 
\cite{Martinez:Adamatzky:Alonso-Sanz:Seck-Tuoh-Mora-2009,Martinez:Adamatzky:McIntosh-2014,Martin:Odlyzko:Wolfram-1984,Martinez:Adamatzky:Seck-Tuoh-Mora:Alonso-Sanz-2010}.
}.

\begin{figure}
\centerline{\includegraphics[width=11.6cm]{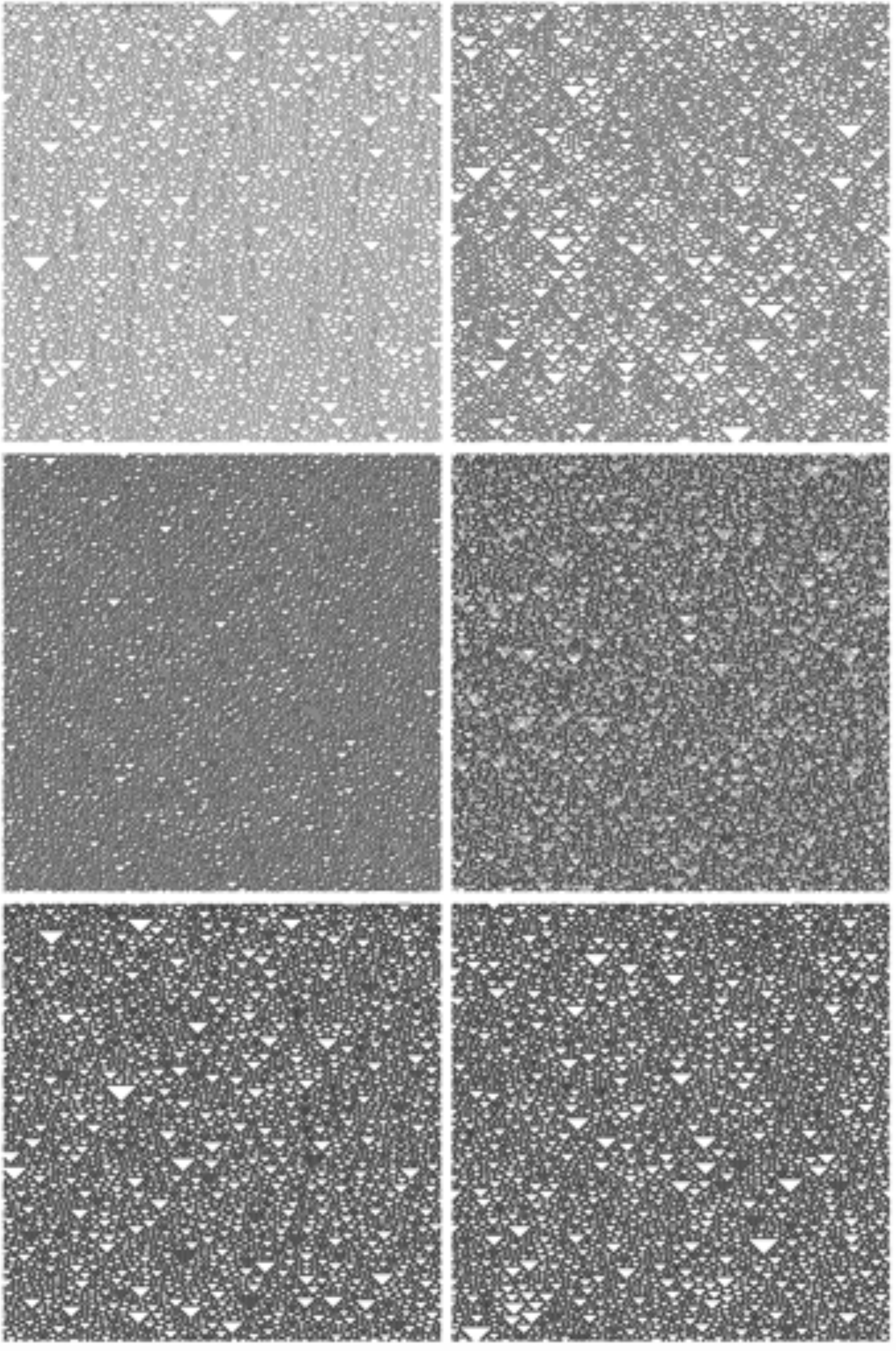}}   
\caption{
Evolutions in a ring of 400 cells, 400 generations from initial density $d_0 = 0.50$. 
Left to right, top to bottom: Rules 18, 22, 30, 90, 122, 126 in ergodic regime.
All patterns have the Hausdorff dimension $\log_2(3)$ of 
Eq.~\ref{equation:Frac dim Rule 22}.
}  
\label{fig:Evolutions Rules 18-22-30-90-122-126}
\end{figure}

This tree does not contain the subset of rules associated either by conjugation or by conjugation-reflection. Beginning from Rule 129 --conjugate with Rule 126-- this subset would display a large collection of Sierpi\'{n}ski--like patterns but this time in white on a black background.

Even in the synchronous ``1nCA'' 
\cite{Bandini:Bonomi:Vizzari-2012}
with a minimal neighborhood of two cells --the cell itself and the either left or right adjacent cell alternating at either odd or even timesteps-- Sierpi\'{n}ski patterns appears in Rules 6$\Delta$ and 9$\Delta$, namely for the only symmetric rules with an equal number of black and white cells, and for the only rules fulfilling the maximal ``sensitivity parameter''.
\begin{figure}
\centerline{\includegraphics[width=6.2cm]{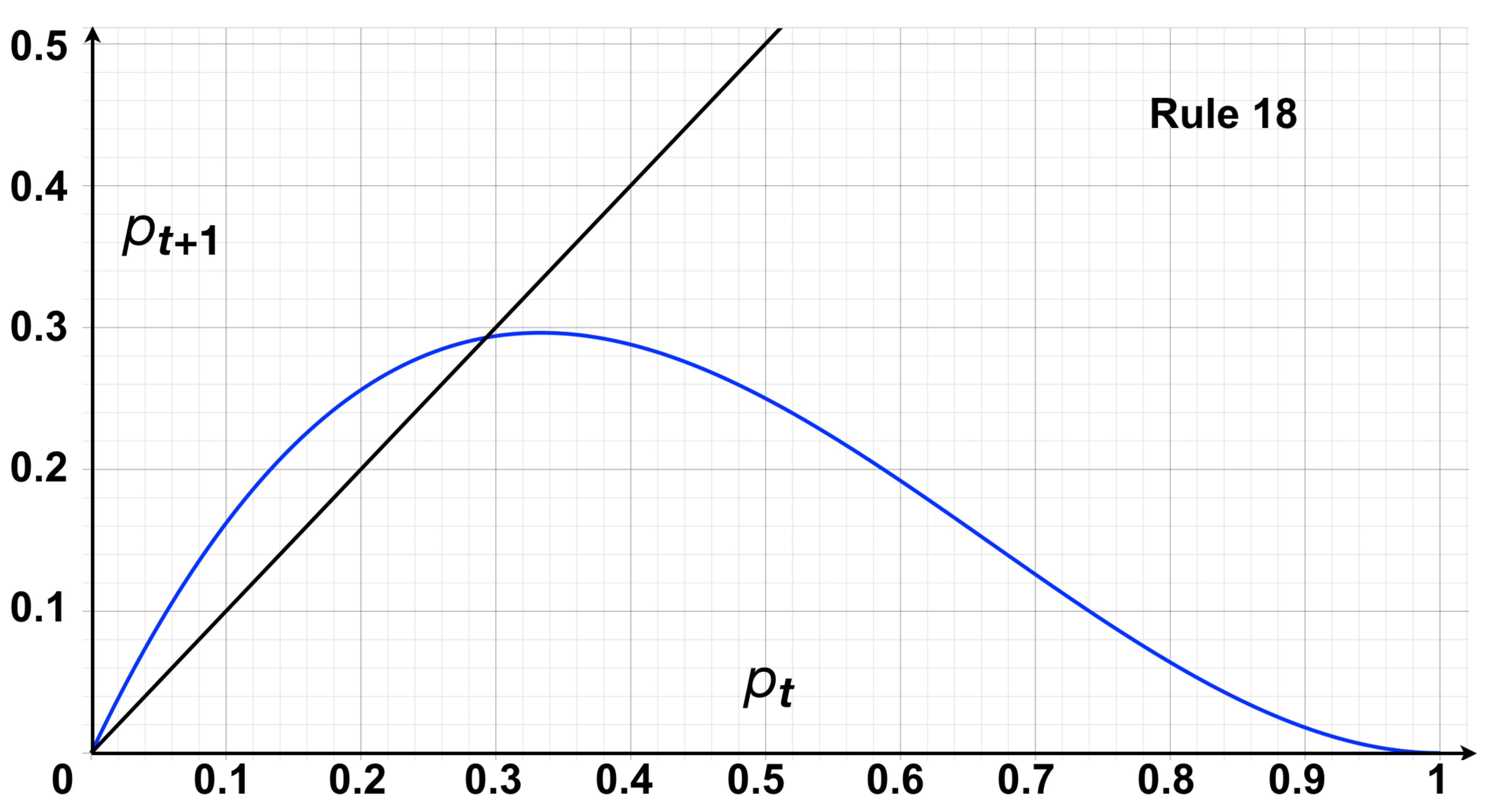}\hspace{0.2cm}\includegraphics[width=5.7cm]{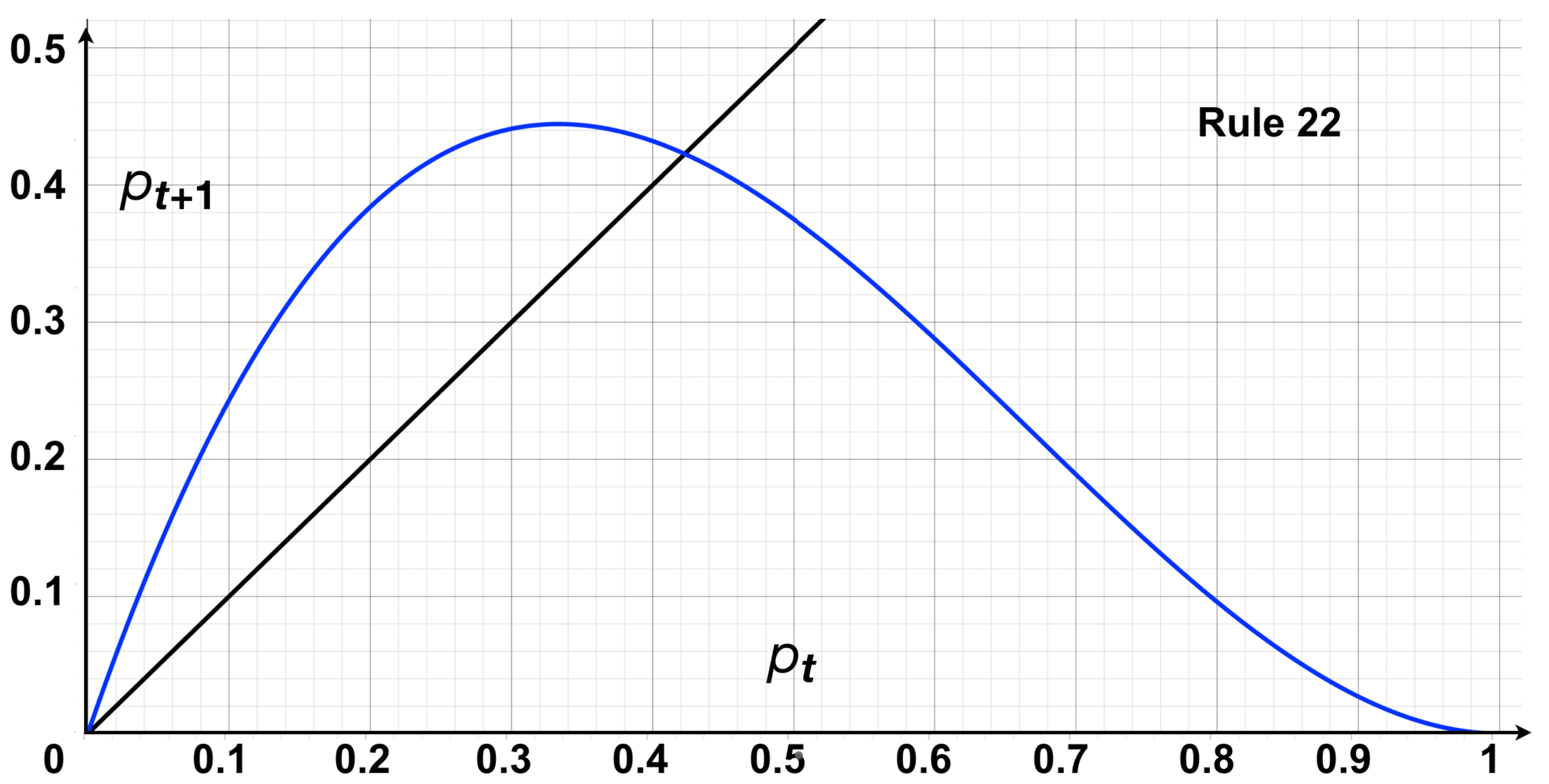}}   
\caption{
Rule 18  
($p_{t+1}=2 p_t q_t^{2}$)
and Rule 22
($p_{t+1}=3 p_t q_t^{2}$)
with their mean field curves.
}  
\label{fig:Mean field curves Rules 18-22}
\end{figure}

Six rules (18, 22, 30, 90, 122, 126) evolving in their ergodic regime are displayed in
Fig.~\ref{fig:Evolutions Rules 18-22-30-90-122-126}.
All patterns have the Hausdorff dimension $\log_2(3)$ of 
Eq.~\ref{equation:Frac dim Rule 22}.
They only differ from their average density 
$
  d_{\mathcal{C}}
$ 
of their mesoscopic minimal macrocell
$
 {\mathcal{C}}.
$

Referring again to the mean field curves in
Sect.~\ref{section:Mean field theory}
now displayed for Rule 18 and Rule 22 in 
Fig.~\ref{fig:Mean field curves Rules 18-22}
we observe that Rule 18 curve reaches its stable fixed point
$
p_{t+1} = p_t
$
when crossing the identity at 
$
 p_t  \approx  0.293
$
whereas for Rule 22 the fixed point is got at 
$
 p_t  \approx  0.423
$
whence the discrepancies between densities in ergodic regime, observable in
Fig.~\ref{fig:Evolutions Rules 18-22-30-90-122-126}.
Moreover, Rule 18 curve shows a slope
$
f'_{R18}(0) = 2
$
whereas Rule 22 shows a slope
$
f'_{R22}(0) = 3
$
at the origin. That comes from the fact that these rules induce from a single source their following evolution:
\begin{itemize}
\item (R18): $ 0^*10^* \rightarrow 0^*(101)0^* $
\item (R22): $ 0^*10^* \rightarrow 0^*(111)0^* $
\end{itemize}
and that their density ratio at timesteps $2^p - 1$ ($p>0$) remains $2/3$.

The  Sierpi\'{n}ski gasket 
\cite{Sierpinski-1916}
appears in a wide variety of situations
\cite{Barnsley-1988}.
The binomial coefficients can be arranged to form Pascal's triangle and Pascal's triangle turns into Sierpi\'{n}ski gasket  with coefficients modulo two.
It may turn into something like a natural tree by some diffeomorphism.
This tree is embeddable into the 2$d$ diffusion graphs embedded into the triangulate lattice
\cite{Deserable-1999,Deserable-2014}:
its vertex dust forms the Sierpi\'{n}ski gasket  patterns.
Sierpi\'{n}ski gasket is often known as a Banach fixed point from some contractive affine transformation into three elements.

But the most fascinating is the formation of patterns from random initial distributions of pigmentations on certain varieties of seashells 
\cite{Wolfram-1983b,Meinhardt-2009}.
This phenomenon can be explained from the Gierer--Meinhardt reaction-diffusion model of the activator--inhibitor type, arising in various situations of pattern formation in morphogenesis
\cite{Gierer:Meinhardt-1972,Turing-1952}.
%
%
\section{Discussion}
\label{section:Discussion}

%
%
\subsection{Chaos or determinism?}
%
%
Despite the fact that Rule 22 is based on simple and determined interactions, the behavior generated by such system is visibly complex and seems to be non--deterministic. To test the kind of the rule's behavior tests chaos 0--1 
\cite{Gottwald:Melbourne-2009} 
(classifier returning value near 1 if series is chaotic and near to 0 if deterministic) was used to determine whether behavior is chaotic/random or deterministic. It is applied directly to the time series data and does not require phase space reconstruction. These two kinds of behavior are significantly different. Physically, randomness has a stochastic nature, while deterministic chaos is generated by even simple system that does not contain any source of randomness, as commonly understood by public. If executed on a PC, then algorithms simulating such behavior generate only pseudo-random/chaotic behavior and series generated in such a way, are essential deterministic and periodic but with very long period. The period is usually long enough to simulate randomness/chaos.

The term \textit{chaos} covers a rather broad class of phenomena whose behavior may seem erratic, chaotic at first glance. Till now chaos was observed in many systems (including evolutionary ones) and, in the last few years, it has been also used to replace pseudo-random number generators (PRGNs) in evolutionary algorithms (EAs). Let us mention for example research papers like 
\cite{IZelinka:Celikovsky:Richter:Chen-2010} 
(a comprehensive overview of mutual intersection between EAs and chaos is discussed in this paper), one of the first use of chaos inside EAs 
\cite{Caponetto:Fortuna:Fazzino:Xibilia-2003}, 
\cite{Pluhacek:Senkerik:Davendra:Oplatkova-2013,Pluhacek:Budikova:Senkerik:Oplatkova:Zelinka-2012} 
discussing the using of deterministic chaos inside particle swarm algorithm instead of PRGNs, 
\cite{Lozi-2012,Wang:Yang-2012} 
investigating relations between chaos and randomness or the latest one 
\cite{Sun:Zhang:Gu-2012},
and
\cite{Hong:Dong:Zhang:Chen:Panigrahi-2013,Senkerik:Davendra:Zelinka:Oplatkova:Pluhacek-2013,Davendra:Zelinka:Senkerik-2010}, 
using chaos with EAs in applications, amongst others.

Another research joining deterministic chaos and pseudorandom number generator has been done for example in 
\cite{Lozi-2012}. 
The possibility of generation of random or pseudorandom numbers by use of the ultra weak multidimensional coupling of 1-dimensional dynamical systems is discussed there. Another paper 
\cite{Persohn:Povinelli-2012} 
deeply investigates a logistic map as a possible pseudo-random number generator and it is compared with contemporary pseudo-random number generators. A comparison of logistic map results is made with conventional methods of generating pseudo-random numbers. The approach is used to determine the number, delay, and period of the orbits of the logistic map at varying degrees of precision (3 to 23 bits). Logistic map, we are using here, was also used in 
\cite{Drutarovsky:Galajda-2007} 
like chaos-based true random number generator embedded in reconfigurable switched-capacitor hardware. Another paper 
\cite{Wang:Qin-2012} 
proposed an algorithm of generating a pseudorandom number generator, which is called (couple map lattice based on discrete chaotic iteration) and combine the couple map lattice and chaotic iteration. Authors also tested this algorithm in NIST 800-22 statistical test suits and is used in image encryption.

In 
\cite{Pareek:Patidar:Sud-2010} 
authors exploit interesting properties of chaotic systems to design a random bit generator, called CCCBG, in which two chaotic systems are cross-coupled with each other. For evaluation of the bit streams generated by the CCCBG, the four basic tests are performed: monobit test, serial test, auto-correlation, Poker test. Also the most stringent tests of randomness: the NIST suite tests have been used. 
\begin{figure}[!tbp]
\centering
\subfigure[]{
\includegraphics[width=0.476\textwidth]{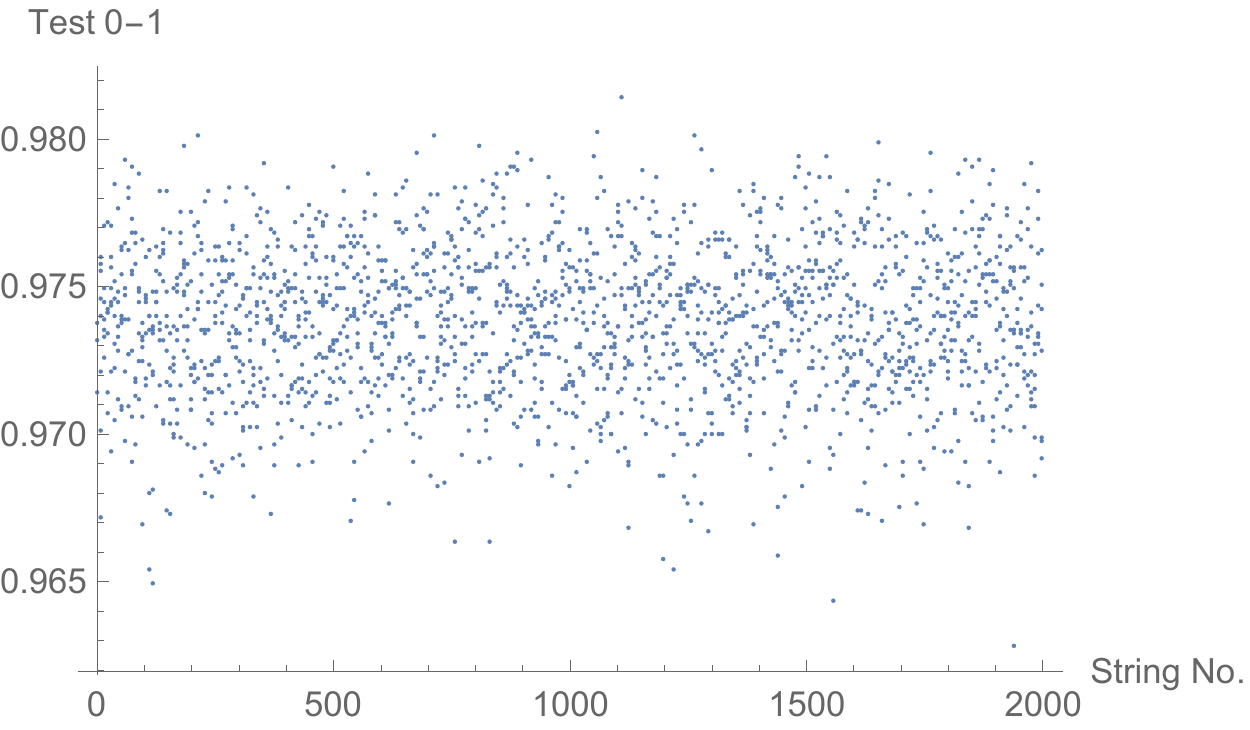}\label{le01}
}
\subfigure[]{
\includegraphics[width=0.476\textwidth]{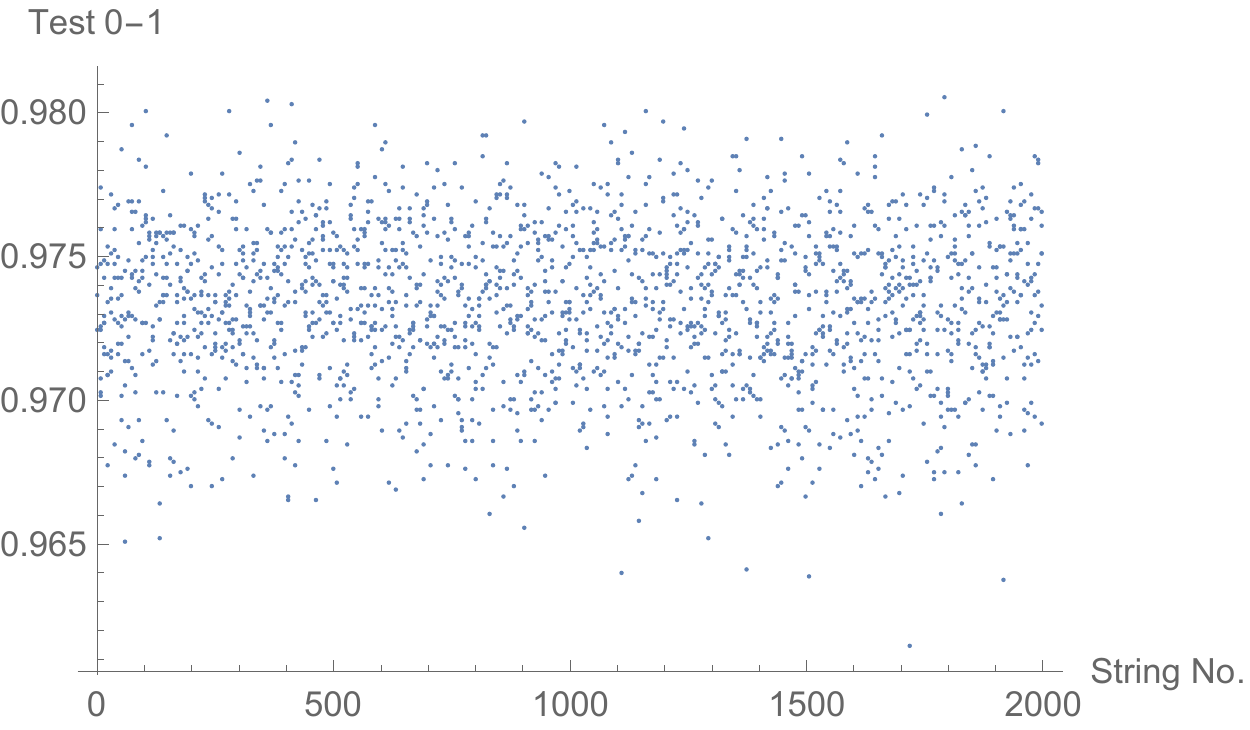}\label{figHistEntr22}
}
\subfigure[]{
\includegraphics[width=0.476\textwidth]{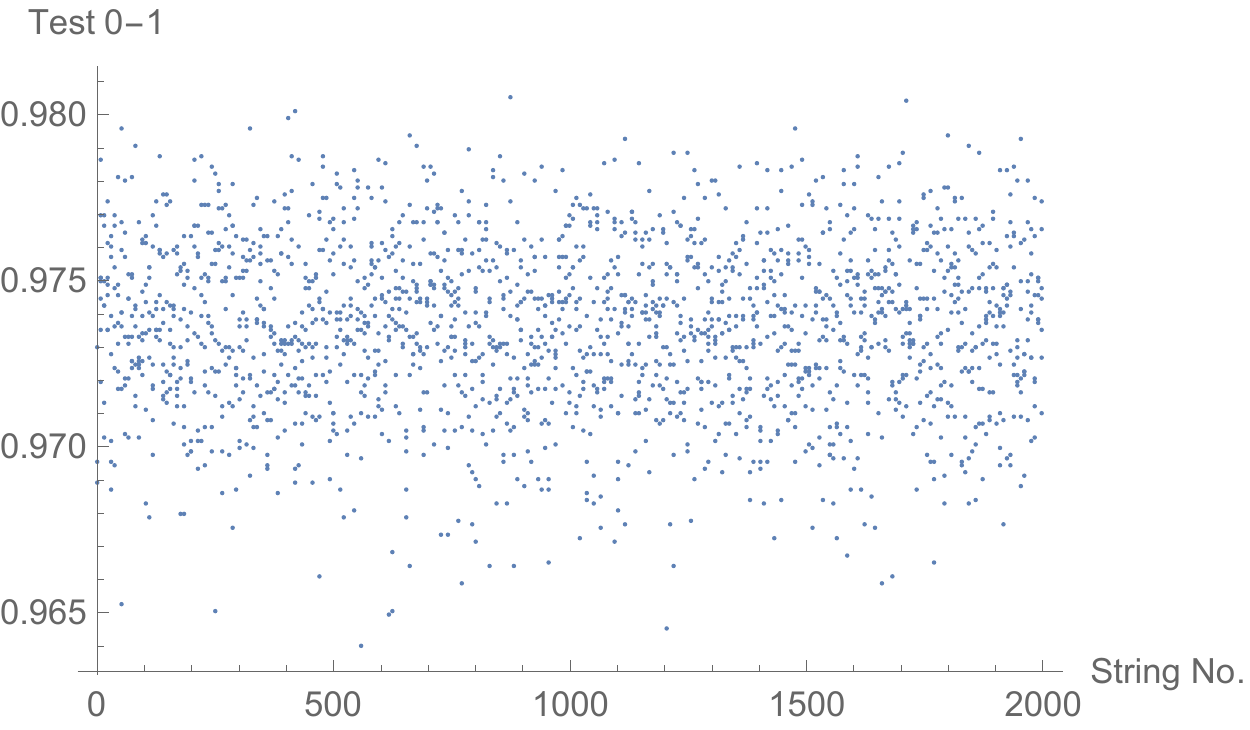}\label{PRNG}
}
\subfigure[]{
\includegraphics[width=0.476\textwidth]{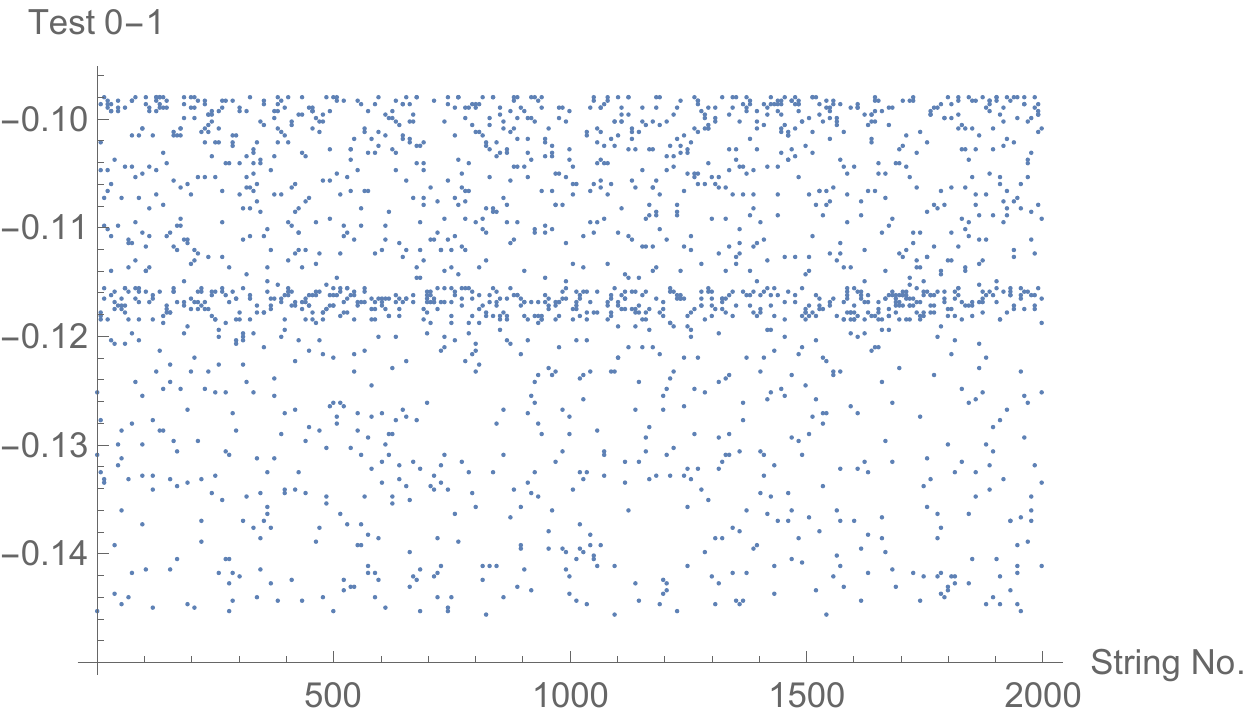}\label{sin}
}
\subfigure[]{
\includegraphics[width=0.52\textwidth]{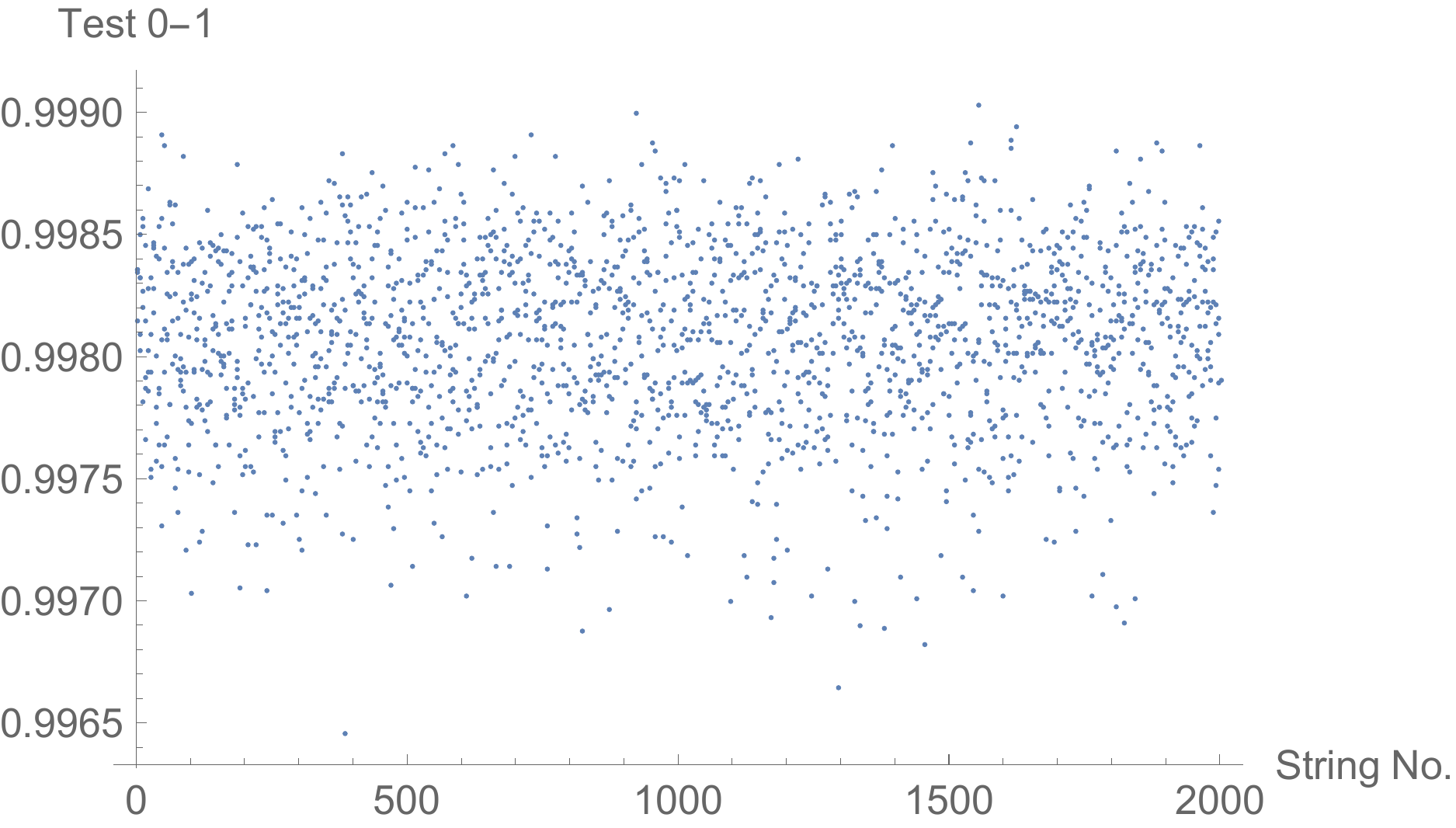}\label{fig0122}
}
\caption{Test chaos 0-1 of 
(a)~logistic equation,
(b)~Mersenne Twister generator,
(c)~standard PRNG, 
(d)~deterministic series (returned values are all around 0, i.e. evaluated process is deterministic),
(e)~Rule 22. All results are concentrated around 1. Chaos based on test chaos 0-1 was proved.
}
\label{chaostest}
\end{figure}
A new binary stream--cipher algorithm based on dual one-dimensional chaotic maps is proposed in 
\cite{Wang:Yang-2012}
with statistic properties showing that the sequence is of high randomness. Similar studies are also done in 
\cite{Caponetto:Fortuna:Fazzino:Xibilia-2003,Bucolo:Caponetto:Fortuna:Frasca:Rizzo-2002,Hu:Liu:Ding-2013,Pluchino:Rapisarda:Tsallis-2013}.
 For a long time various PRNGs were used inside evolutionary algorithms. During last few years deterministic chaos systems (DCHS) are used instead of PRNGs. As was demonstrated in 
 \cite{Pluhacek:Senkerik:Davendra:Oplatkova-2013,Pluhacek:Budikova:Senkerik:Oplatkova:Zelinka-2012},
very often the performance of EAs using DCHS is better or fully comparable with EAs using PRNGs. See for example 
 \cite{Pluhacek:Senkerik:Davendra:Oplatkova-2013}.

The chaos test 0--1 
\cite{Gottwald:Melbourne-2009} 
has been already and successfully used on various tasks as for example on experimental data from a bipolar motor 
\cite{Falconer:Gottwald:Melbourne:Wormnes-2007}, 
behavior of the cutting process
\cite{Litak:Syta:Wiercigroch-2009}, 
real experimental time series of laser droplet generation process 
\cite{Krese:Govekar-2012} 
and validated by applying to typical nonlinear dynamic systems, including fractional--order dynamic system 
\cite{Ke-Hui:Xuan:Cong-Xu-2010} amongst  others.

The same approach was used here. Test 0-1 was used on different series in order to verify and test the nature of Rule 22. Figure~\ref{chaostest} visualizes results of our experiments. For evaluation 2000 Rule 22 behavior strings of length 2000 have been used. The same was repeated for the Mersenne twister random number generator 
\cite{Matsumoto:Takuji-1998} 
(MTPRNG) (Fig.~\ref{PRNG}), chaos generated by logistic equation (LE) (Fig.~\ref{figHistEntr22}), and periodic series (PS): the sinus function generated from the randomly selected position (Fig.~\ref{sin}). As clearly visible, test 0-1 has clearly classified Rule 22, MTPRNG and LE as a non-deterministic series while series based on periodic pattern as deterministic. The random series were not distinguished from chaotic ones. This was probably caused by insufficient length (2000 is likely not enough) of series, however, this was not a matter of this experiment.
%
%
\subsection{Memory leads to complexity}
%
%
In this section, we show that ECA Rule 22 with memory is {\it strongly chaotic} 
\cite{Martinez:Adamatzky:Alonso-Sanz-2013}.

Conventional CA are ahistoric (memoryless): i.e., the new state of a cell depends on the neighborhood configuration solely at the preceding time step of $\varphi$. Thus, CA with {\it memory} (CAM) can be considered as an extension of the standard framework of CA where every cell $x_i$ is allowed to remember some period of its previous evolution. A memory is based on the state and history of the system, thus we design a memory function $\phi$, as follows: $\phi (x^{t-\tau}_{i}, \ldots, x^{t-1}_{i}, x^{t}_{i}) \rightarrow s_{i}$, such that $\tau < t$ determines the backwards degree of memory and each cell $s_{i} \in \Sigma$ is a function of the series of states in cell $x_i$ up to time-step $t-\tau$. To execute the evolution we apply the original rule as follows: $\varphi(\ldots, s^{t}_{i-1}, s^{t}_{i}, s^{t}_{i+1}, \ldots) \rightarrow x^{t+1}_i$.
\begin{figure}[!tbp]
\centering
\subfigure[]{\includegraphics[width=0.495\textwidth]{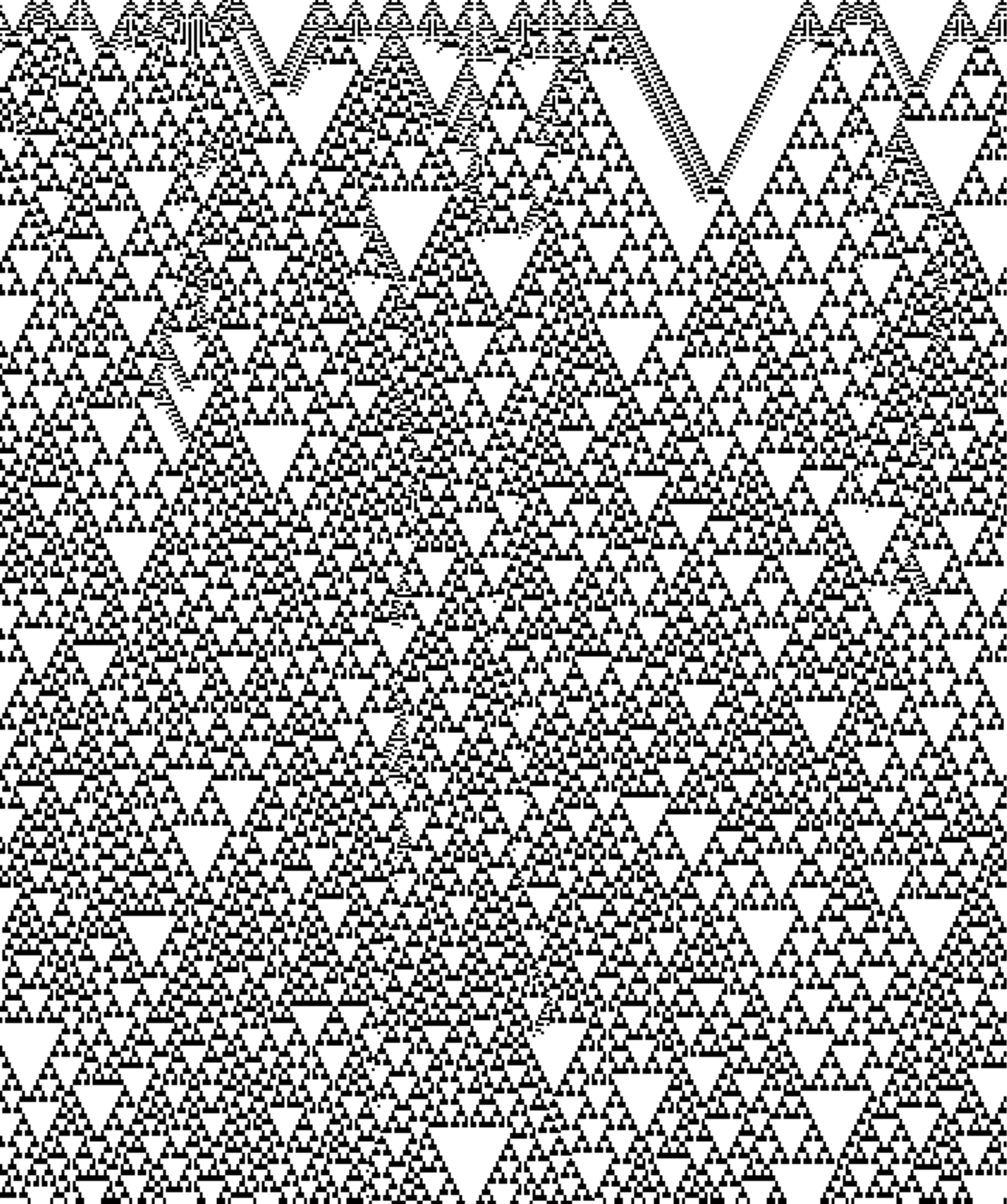}}
\subfigure[]{\includegraphics[width=0.495\textwidth]{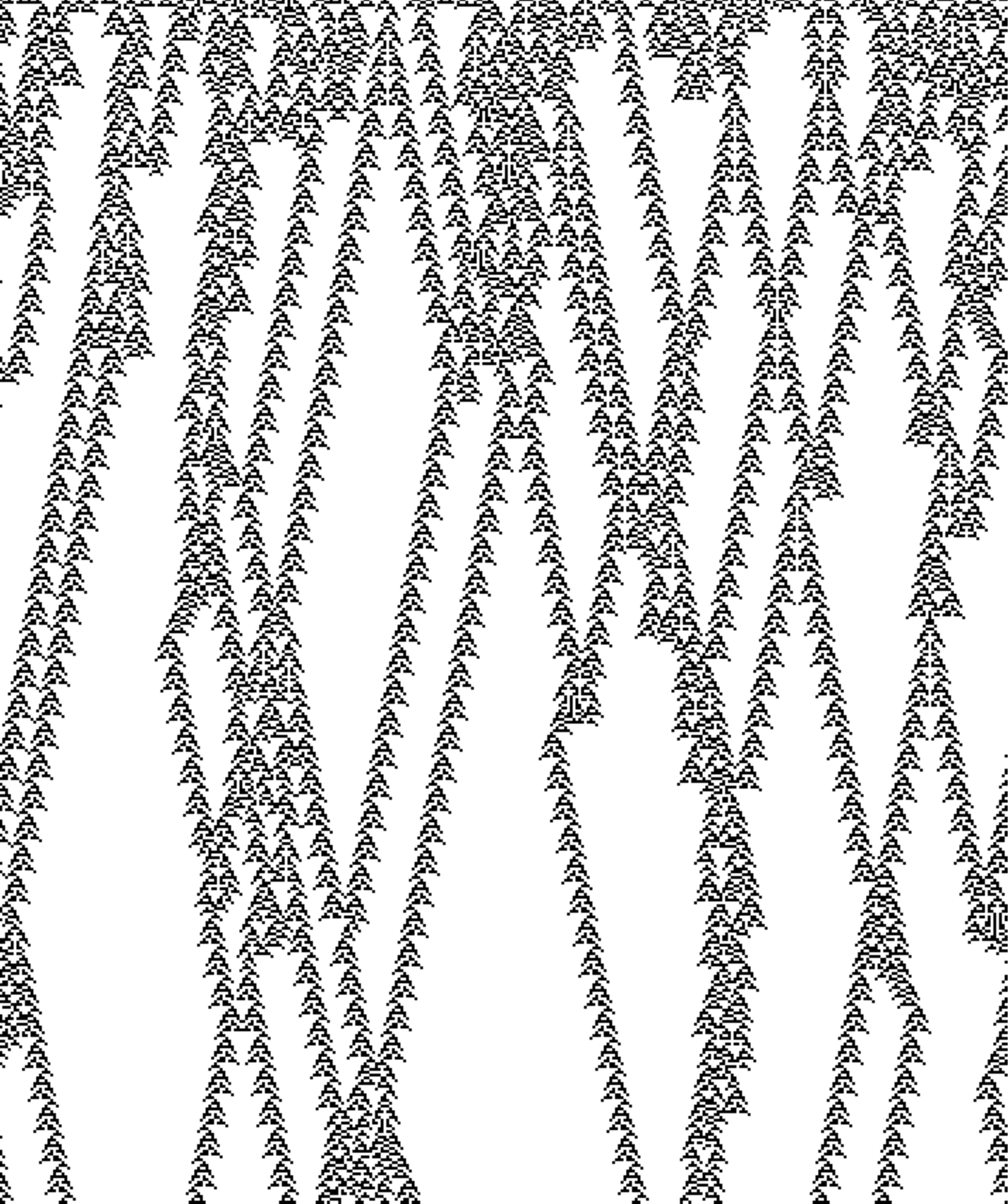}}
\subfigure[]{\includegraphics[width=0.495\textwidth]{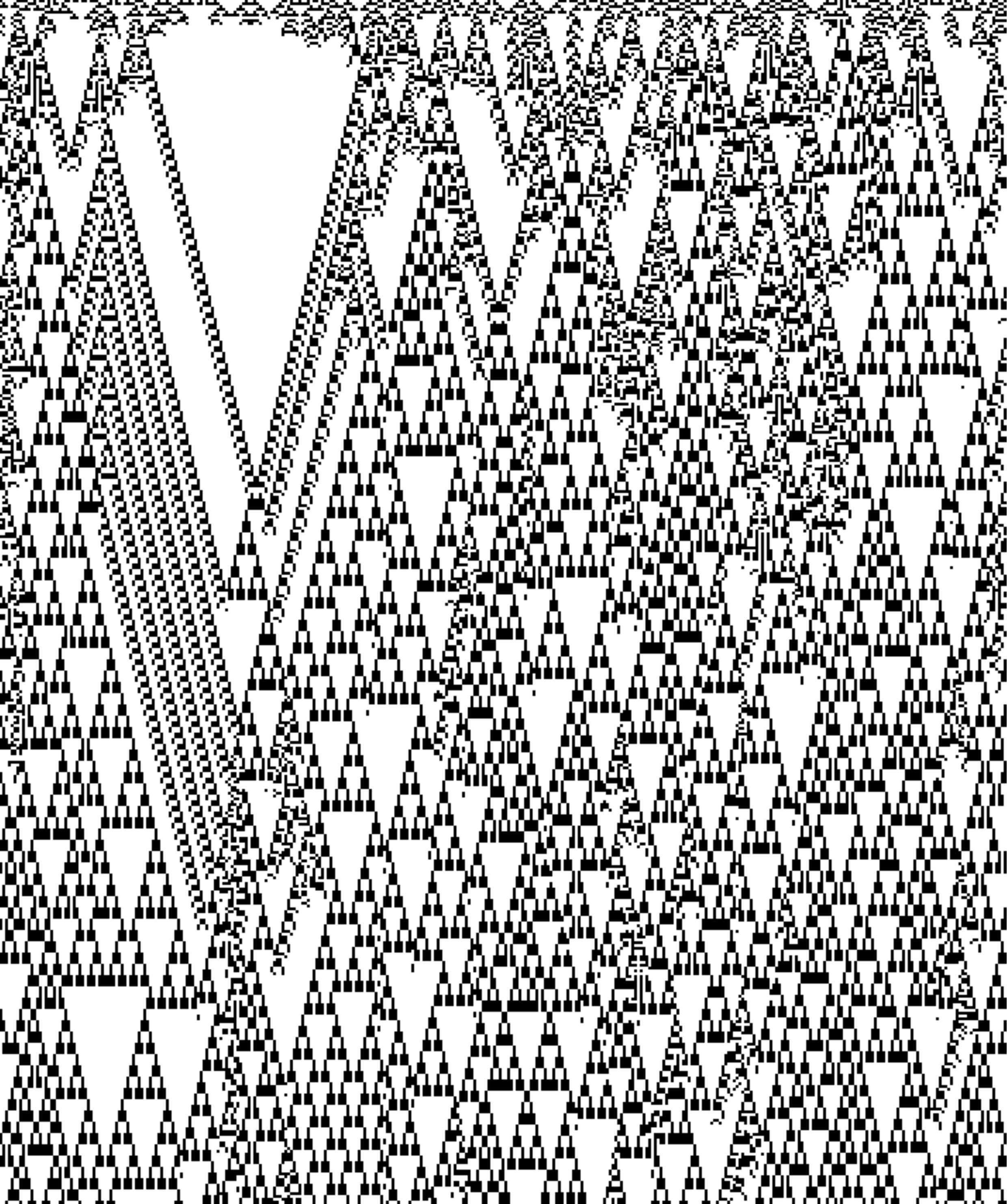}}
\subfigure[]{\includegraphics[width=0.495\textwidth]{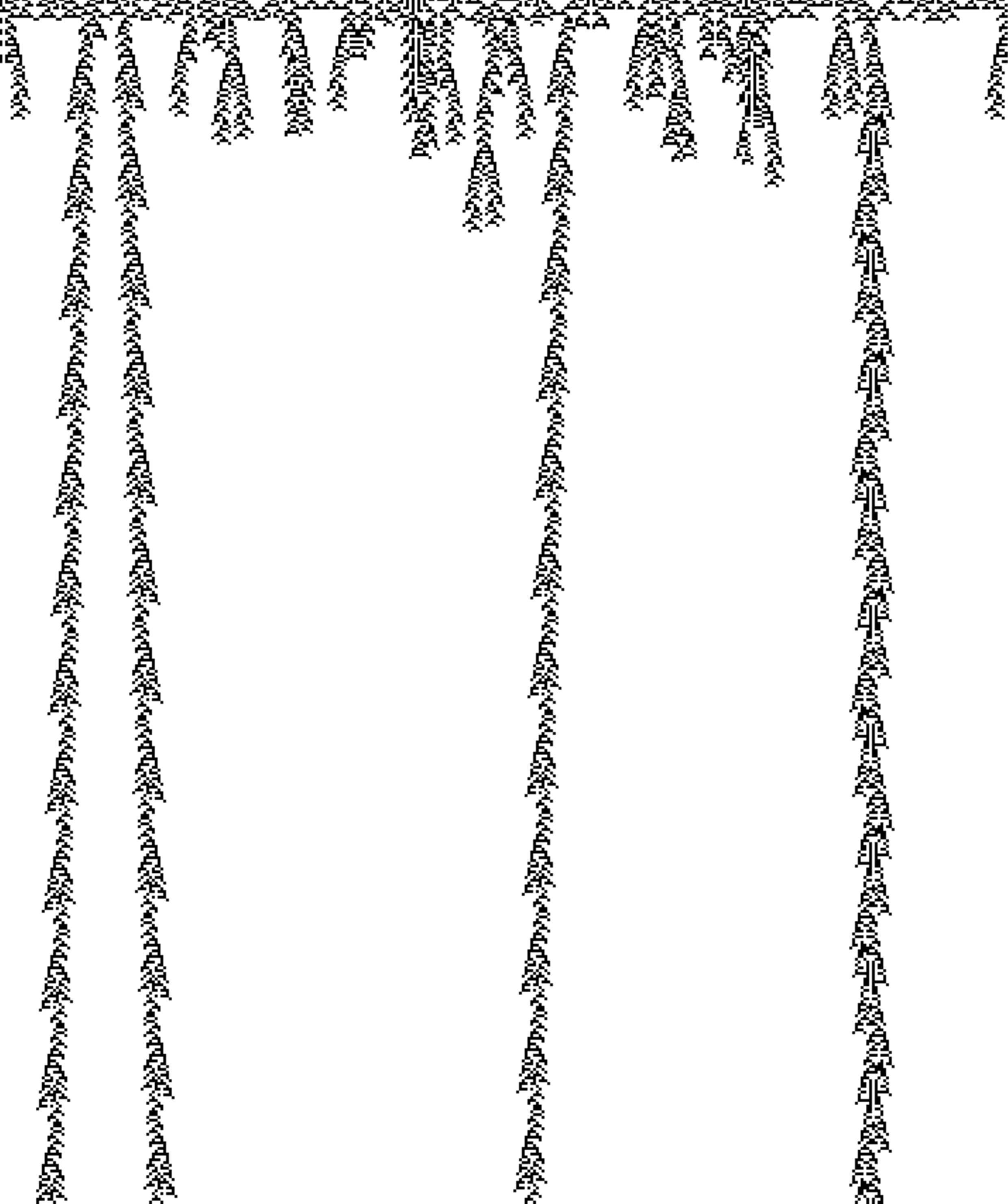}}
\caption{ECA Rule 22 with a memory function reveals complex behavior. 
(a) Evolution of the function $\phi_{R22maj:3}$. 
(b) The function $\phi_{R22maj:4}$ (recently proven to be logically universal by simulating the Fredkin gate in
 \cite{Martinez:Morita-2018,Martinez:Adamatzky:Morita-2018}). 
(c) The function $\phi_{R22maj:7}$ (a glider gun was discovered in this rule 
\cite{Martinez:Adamatzky:Alonso-Sanz-2013}). 
(d) The function $\phi_{R22maj:8}$ (particles with long period). 
}
\label{R22majMem}
\end{figure}

In CAM, while the mapping $\varphi$ remains unaltered, a historic memory of past iterations is retained by featuring each cell as a summary of its previous states; therefore cells {\it canalize} memory to the map $\varphi$. As an example, we can take the memory function $\phi$ as a {\it majority memory}: $\phi_{maj} \rightarrow s_{i}$, where in case of a tie given by $\Sigma_1 = \Sigma_0$ in $\phi$, we shall take the last value $x_i$. So $\phi_{maj}$ represents the classic majority function for three variables on cells $(x^{t-\tau}_{i}, \ldots, x^{t-1}_{i}, x^{t}_{i})$ and defines a temporal ring before calculating the next global configuration $c$. In case of a tie, it is allowed to break it in favor of zero if $x_{\tau-1}=0$, or to one whether $x_{\tau-1}=1$.

The representation of a ECAM is given as follows:
\begin{equation}
\phi_{CARm:\tau}
\end{equation}
where $CAR$ represents the decimal notation of a particular ECA rule and $m$ the kind of memory given with a specific value of $\tau$.

Note that memory is as simple as any CA, and that the global behavior produced by the local rule is rather unpredictable, it can lead to emergent properties and so can be classed as complex. Memory functions were developed and extensively studied by Alonso-Sanz in 
\cite{Alonso-Sanz-2009}. 
Memory in ECA have been studied, showing its potential to produce complex behavior from chaotic systems and beyond in 
\cite{Martinez:Adamatzky:Alonso-Sanz:Seck-Tuoh-Mora-2009,Martinez:Adamatzky:Alonso-Sanz-2013}, 
and recently in  
\cite{Chen:Chen:Martinez-2017} 
authors have included hybrid versions. Thus, we can conjecture that a memory function can produce complex behavior 
\cite{Martinez:Adamatzky:Alonso-Sanz-2013} 
as follows:
\begin{equation}
\phi(\varphi_{chaos}) \rightarrow complex.
\end{equation}
Eppstein 
\cite{Eppstein-WCA} 
demonstrates that a CA class IV is a system where mobile self--localizations emerge. 
We can relate type of classes with memory functions in ECA Rule 22 as:
\begin{align}
\phi_{R22maj:3} & \Rightarrow chaos \rightarrow chaos \\
\phi_{R22maj:4} & \Rightarrow chaos \rightarrow complexity \\
\phi_{R22maj:7} & \Rightarrow chaos \rightarrow complexity \\
\phi_{R22maj:8} & \Rightarrow chaos \rightarrow complexity
\end{align}
%
%
\subsection{Rareness and unpredictability}
%
%
The state transition function $\varphi(x_{i-r}^t, \ldots, x_{i}^t, \ldots, x_{i+r}^t) \rightarrow x_i^{t+1}$ can be re-written as Boolean formula with two {\sc xor} operations: $x_{i+r}^t=x_{i-r}^t \oplus x_{i}^t \oplus x_{i+r}^t$. The {\sc xor} gate is the most rare, most hard to find in natural non-linear systems, Boolean gate. 
Let gates $g_1$ and $g_2$ discovered with occurrence frequencies  $f(g_1)$ and $f(g_2)$, 
we say a gate $g_1$ is easier to develop or evolve than a gate $g_2$: $g_1 \rhd g_2$ if  $f(g_1) > f(g_2)$. 
The hierarchies of gates obtained using evolutionary techniques in liquid crystals 
\cite{Harding:Miller-2007}, 
light--sensitive modification of Belousov--Zhabotinsky system  
\cite{Toth:Stone:Adamatzky:Costello:Bull-2008}, 
slime mould {\em Physarum polycephalum}
\cite{Harding:Koutnik:Schmidhuber:Adamatzky-2018} 
and protein molecules
\cite{Adamatzky-2017a,Adamatzky-2017b}:
%
%
\begin{itemize}
\item Gates in liquid crystals: 
\hfill \{{\sc or}, {\sc nor}\} $\rhd$ {\sc and} $\rhd$ {\sc not} $\rhd$ {\sc nand} $\rhd$ {\sc xor}
\item Gates in Belousov--Zhabotinsky medium: 
\hfill {\sc and} $\rhd$ {\sc nand} $\rhd$ {\sc xor}
\item Gates in cellular automata
\cite{Adamatzky-2009}: 
\hfill {\sc or} $\rhd$ {\sc nor} $\rhd$ {\sc and} $\rhd$ {\sc nand} $\rhd$ {\sc xor}
\item Gates in Physarum: 
\hfill {\sc and} $\rhd$ {\sc or} $\rhd$ {\sc nand} $\rhd$ {\sc nor} $\rhd$ {\sc xor} $\rhd$ {\sc xnor}
\item Gates in protein molecules verotoxin
and actin:
 
\hfill  {\sc and} $\rhd$ {\sc or} $\rhd$ {\sc and-not} $\rhd$ {\sc xor}
\end{itemize}
%
%
The {\sc xor} gate is hard to find and space-time dynamics of Rule 22 automata is hard to predict. A strong link computational difficulty of a problem and its randomness was established by Yao
\cite{Yao-1982}. 
His famous lemma, rephrased by Impagliazzo and Wigderson
\cite{Impagliazzo:Wigderson-1997}, 
can be seen in the framework of predictability of Rule 22 ECA behavior:
\begin{quote}
Fix a non-uniform model of computation (with certain closure properties) 
and a Boolean function $f : \{0, 1\}^n \rightarrow \{0, 1\}$. Assume that any algorithm in the model of a certain complexity has a significant probability of failure when predicting $f$ on a randomly chosen instance $x$. Then any algorithm (of a slightly smaller complexity) that tries to guess the {\sc xor} $f(x_1) \oplus f(x_2) \oplus \ldots \oplus f(x_k)$ of $k$ random instances $x_1, \ldots, x_k$ won’t do significantly better than a random coin toss.
\end{quote}
Potential associations between dynamics in Rule 22 ECA and a role of {\sc xor} functions in communication complexity
\cite{Zhang:Shi-2009,Montanaro:Osborne-2010} 
could be explored in future. 
%
%


\begin{thebibliography}{99}
%
\bibitem{Wolfram-1983a} 
S. Wolfram, Statistical mechanics of cellular automata, 
Rev. Mod. Phys. 55(3) (1983) 601--644.
%
\bibitem{Wolfram-1994} 
S. Wolfram, {\em Cellular Automata and Complexity} (1994) 
Addison-Wesley Publishing Company.
%
\bibitem{Wolfram-1995} 
S. Wolfram, Twenty Problems in the Theory of Cellular Automata, 
Physica Scripta 9 (1985) 170--183.
%
\bibitem{Redeker:Adamatzky:Martinez-2013} 
M. Redeker, A. Adamatzky, G.J. Mart\'{\i}nez, Expressiveness of Elementary Cellular Automata, 
International Journal of Modern Physics C 24(3) (2013) 1350010--14.
%
\bibitem{Zabolitzky-1988} 
J.G. Zabolitzky, Critical Properties of Rule 22 Elementary Cellular Automata, 
Journal of Statistical Physics 50(5-6) (1988) 1255--1262.
%
%
\bibitem{McIntosh-2009} 
H.V. McIntosh, {\em One Dimensional Cellular Automata}, Luniver Press, UK (2009)
%
\bibitem{Jin:Chen:Yang-2009} 
W. Jin, F. Chen, C. Yang, Topological Chaos of Cellular Automata Rules, 
{\em Proceedings of International Workshop on Chaos--Fractals Theories and Applications} (2009) 216--220.
%
\bibitem{Gutowitz:Victor-1987} 
H.A. Gutowitz, J.D. Victor, Local structure theory in more that one dimension, 
Complex Systems 1 (1987) 57--68.
%
\bibitem{McIntosh-1990a} 
H.V. McIntosh, Wolfram's Class IV and a Good Life, 
Physica D 45 (1990) 105--121.
%
\bibitem{Grassberger-1986} 
P. Grassberger, Long-Range Effects in an Elementary Cellular Automaton, 
J. Stat. Phys. 45 (1-2) (1986) 27--39.
%
%
\bibitem{Sinha-2006} 
S. Sinha, Phase Transitions in the Computational Complexity of “Elementary” Cellular Automata, 
Minai, A.A., Bar-Yam,Y. (eds.) Unifying Themes in Complex Systems. Springer  (2011) 337--348.
%
\bibitem{Zenil-2013} 
H. Zenil, E. Villarreal-Zapata, Asymptotic Behaviour and Ratios of Complexity in Cellular Automata,
International Journal of Bifurcation and Chaos 23(9) (2013) 1350159.
%
\bibitem{Gravner:Griffeath-2011} 
J. Gravner, D. Griffeath, The One-Dimensional Exactly 1 Cellular Automaton: Replication, Periodicity, and Chaos from Finite Seeds, J. Stat. Phys. 142(1) (2011) 168--200.
%
\bibitem{Halmos-1956} 
P. Halmos, Lectures on ergodic theory, Amer. Math. Soc. 142, 1956.
%
\bibitem{Lesne-1998} 
A. Lesne, Renormalization Methods, John Wiley and Sons, 1998.
%
%
\bibitem {Deserable-2011}
D. D\'{e}s\'{e}rable, Embedding Kadanoff's scaling picture into the triangular lattice,
Acta Phys. Pol. B Proc. Suppl. 4(2) (2011) 249--265.
%
\bibitem{Wuensche:Lesser-1992} 
A. Wuensche, M. Lesser, 
{\em The Global Dynamics of Cellular Automata}, Addison-Wesley (1992)
%
\bibitem{Wuensche-2016} 
A. Wuensche, {\em Exploring Discrete Dynamics} 2nd. Edition. Luniver Press, Bristol (2016)
%
\bibitem{McIntosh-1993} 
H.V. McIntosh, Ancestors: Commentaries on The Global Dynamics of Cellular Automata,  (1993) 
 \url{http://delta.cs.cinvestav.mx/~mcintosh/cellularautomata/Working_Papers_files/global.pdf}
%
\bibitem{Martinez:Adamatzky:Chen:Chen:Mora-2017} 
G.J. Mart{\'i}nez, A. Adamatzky, B. Chen, F. Chen,  J.C. Seck-Tuoh-Mora,
Simple networks on complex cellular automata: From de Bruijn diagrams to jump-graphs, 
In: {\em Swarm Dynamics as a Complex Network}, 
I. Zelinka, G. Chen (Ed.), Springer  (2017) 241--264.
%
%
\bibitem{Adak:Mukherjee:Das-2018} 
S. Adak, S. Mukherjee, S. Das, 
{\em Do There Exist Non-linear Maximal Length Cellular Automata? A Study},
G. Mauri, S. El Yacoubi, A. Dennunzio, K. Nishinari, L. Manzoni eds) 
Cellular Automata LNCS 11115 (2018) 289--297.
%
\bibitem{deBruijn-1946} 
N.G. de Bruijn, A combinatorial problem, 
Proceedings of the Section of Sciences of the Koninklijke Nederlandse Akademie van Wetenschappen te Amsterdam 49(7) (1946) 758--764.
%
\bibitem{Golomb-1967} 
S.W. Golomb, 
{\em Shift Register Sequences}, Holden-Day, San Francisco (1967) 
%
\bibitem{Voorhees-1996} 
B.H. Voorhees, 
{\em Computational analysis of one-dimensional cellular automata}, World Scientific (1996) 
%
\bibitem{McIntosh-1991} 
H.V. McIntosh, Linear cellular automata via de Bruijn diagrams, 
\url{http://delta.cs.cinvestav.mx/~mcintosh/cellularautomata/Papers_files/debruijn.pdf} (1991)
%
%
\bibitem{Hopcroft:Ullman-1979} 
J.E. Hopcroft, J.D. Ullman,  
{\em Introduction to Automata Theory Languages, and Computation}, Reading, Addison-Wesley (1979)
%
\bibitem{von Neumann-1966} 
J. von Neumann, {\it Theory of Self-reproducing Automata} 
(edited and completed by A.W. Burks), University of Illinois Press, Urbana and London (1966)
%
\bibitem{Goucher-2010} 
A.P. Goucher, Universal Computation and Construction in GoL Cellular Automata, 
In: {\em Game of Life Cellular Automata} A. Adamatzky (Ed.) Springer (2010) 505--517.
%
\bibitem{Amoroso:Cooper-1970} 
S. Amoroso, G. Cooper, The Garden-of-Eden theorem for finite configurations, 
{\it Proceedings of the American Mathematical Society} {\bf 26(1)} (1970) 158--164.
%
\bibitem{Seck-Tuoh-Mora:Medina-Marin:Hernandez-Romero:Martinez:Barragan-Vite-2017} 
J.C. Seck-Tuoh-Mora, J. Medina-Mar{\'i}n, N. Hern{\'a}ndez-Romero, G.J. Mart{\'i}nez, I. Barrag{\'a}n-Vite, 
Welch sets for random generation and representation of reversible one-dimensional cellular automata, 
{\em Information Sciences} {\bf 382--383} (2017) 81--95.
%
%
\bibitem{Eppstein-Fractals} 
D. Eppstein, ``Fractals'', 
\url{http://www.ics.uci.edu/~eppstein/junkyard/fractal.html}.
%
\bibitem{Chen:Dong-1998} 
G. Chen, X. Dong, {\em From Chaos to Order}, 
World Scientific Series on Nonlinear Science, Series A, Vol. 24 (1998) 
%
\bibitem{Wolfram-1983b} 
S. Wolfram, Cellular Automata, 
Los Alamos Science 9 (1983) 2--21.
%
\bibitem{Wolfram-1984} 
S. Wolfram, Computation theory of cellular automata, 
Comm. Math. Phys. 96(1) (1984) 15--57.
%
\bibitem{McIntosh-1990b} 
H.V. McIntosh, Linear cellular automata, 
\url{http://matematicas.reduaz.mx/~cellularautomata/cellularautomata/Papers_files/lcau.pdf} (1990) 
%
%
\bibitem{Li:Packard-1990} 
W. Li, N.H. Packard, The Structure of the Elementary Cellular Automata Rule Space, 
Complex Systems 4(3) (1990) 281--297.
%
\bibitem{Martinez:Adamatzky:Alonso-Sanz:Seck-Tuoh-Mora-2009} 
G.J. Mart{\'{\i}}nez, A. Adamatzky, R. Alonso--Sanz, J.C. Seck Tuoh Mora, Complex Dynamics Emerging in Rule 30 with Majority Memory, 
Complex Systems 18(3) (2009) 345--365.
%
\bibitem{Martinez:Adamatzky:McIntosh-2014} 
G.J. Mart{\'{\i}}nez, A. Adamatzky, H.V. McIntosh, Complete Characterization of Structure of Rule 54, 
Complex Systems 23(3) (2014) 259--293.
%
\bibitem{Martin:Odlyzko:Wolfram-1984} 
O. Martin, A.M. Odlyzko, S. Wolfram, Algebraic Properties of Cellular Automata, 
Comm. Math. Phys. 93(2) (1984) 219--258.
%
\bibitem{Martinez:Adamatzky:Seck-Tuoh-Mora:Alonso-Sanz-2010} 
G.J. Mart{\'{\i}}nez, A. Adamatzky, J.C. Seck Tuoh Mora, R. Alonso--Sanz, How to make dull cellular automata complex by adding memory: Rule 126 case study, 
Complexity 15(6) (2010) 34--49.
%
%
\bibitem{Bandini:Bonomi:Vizzari-2012} 
S. Bandini, A. Bonomi, G. Vizzari, An analysis of different types and effects of asynchronicity in cellular automata update schemes, 
Nat. Comp. 11(2) (2012) 277--287.
%
\bibitem {Sierpinski-1916}
W. Sierpi\'{n}ski, O krzywej, kt\'{o}rej ka\.{z}dy punkt jest punktem rozga\l\c{e}zienia,
Prace Mat.--Fiz. 27(1) (1916) 77--86.
%
\bibitem{Barnsley-1988} 
M. Barnsley, Fractals everywhere, 
Acad. Press., San Diego, CA (1988) 
%
\bibitem {Deserable-1999}
D. D\'{e}s\'{e}rable, A family of Cayley graphs on the hexavalent grid,
Discrete Applied Math. 93 (1999) 169--189.
%
\bibitem {Deserable-2014}
D. D\'{e}s\'{e}rable, Systolic Dissemination in the Arrowhead Family,
Cellular Automata, W\c{a}s, J., Sirakoulis, G., Bandini, S. eds, 
LNCS 8751 (2014) 75--86.
%
%
\bibitem{Meinhardt-2009} 
H. Meinhardt, The Algorithmic Beauty of Seashells, 
The Virtual Laboratory, Springer (2009) 
%
\bibitem {Gierer:Meinhardt-1972}
A. Gierer, H. Meinhardt, A theory of biological pattern formation,
Kybernetik 12(1) (1972) 30--39
%
\bibitem {Turing-1952}
A. Turing, The Chemical Basis of Morphogenesis,
Phil. Trans. Royal Soc. London 237(641) (1952) 37--72
%
\bibitem{Gottwald:Melbourne-2009} 
G.A. Gottwald, I. Melbourne, On the implementation of the 0–1 test for chaos,
SIAM Journal on Applied Dynamical Systems 8(1) (2009) 129--145
%
\bibitem{IZelinka:Celikovsky:Richter:Chen-2010} 
I. Zelinka, S. Celikovsk\'{y}, H. Richter, G. Chen, 
Evolutionary Algorithms and Chaotic Systems, (Eds), Studies in Computational Intelligence, Springer (2010) 
%
%
\bibitem{Caponetto:Fortuna:Fazzino:Xibilia-2003}
R. Caponetto, L. Fortuna, S. Fazzino, M.G. Xibilia, 
Chaotic sequences to improve the performance of evolutionary algorithms, 
IEEE Trans. Evol. Comput. 7(3) (2003) 289--304.
%
\bibitem{Pluhacek:Senkerik:Davendra:Oplatkova-2013} 
M. Pluhacek, R. Senkerik, D. Davendra, Z. Oplatkova, 
On the Behavior and Performance of Chaos Driven PSO Algorithm with Inertia Weight, 
Computers \& Mathematics with Applications 66(2) (2013) 122--134
%
\bibitem{Pluhacek:Budikova:Senkerik:Oplatkova:Zelinka-2012} 
M. Pluhacek, V. Budikova, R. Senkerik, Z. Oplatkova, I. Zelinka, 
On the Performance of Enhanced PSO algorithm with Lozi Chaotic Map – An Initial Study, 
In: Proceedings of 18th International Conference on Soft Computing -- MENDEL 2012, (2012) 40--45
%
\bibitem{Lozi-2012} 
R. Lozi, Emergence of Randomness From Chaos, 
International Journal of Bifurcation and Chaos 22(2) World Scientific (2012) 1250021 
%
\bibitem{Wang:Yang-2012} 
X.-Y. Wang, L. Yang, Design of Pseudo-Random Bit Generator Based on Chaotic Maps, 
International Journal of Modern Physics 26(32) World Scientific (2012) 1250208 
%
%
\bibitem{Sun:Zhang:Gu-2012} 
Y. Sun, L. Zhang, X. Gu, A hybrid co-evolutionary cultural algorithm based on particle swarm optimization for solving global optimization problems, 
Neurocomputing 98 (2012) 76--89
%
\bibitem{Hong:Dong:Zhang:Chen:Panigrahi-2013} 
W-Ch. Hong, Y. Dong, W.Y. Zhang, L.Y. Chen; B.K. Panigrahi, 
Cyclic electric load forecasting by seasonal SVR with chaotic genetic algorithm, 
International Journal of Electrical Power and Energy Systems 44(1)   (2013) 604--614 
%
\bibitem{Senkerik:Davendra:Zelinka:Oplatkova:Pluhacek-2013} 
R. Senkerik, D. Davendra, I. Zelinka, Z. Oplatkova, M. Pluhacek, 
Optimization of the Batch Reactor by Means of Chaos Driven Differential Evolution, 
Soft Computing Models in Industrial and Environmental Applications, Springer, AISC 188 (2013) 93--102
%
\bibitem{Davendra:Zelinka:Senkerik-2010} 
D. Davendra, I. Zelinka, R. Senkerik, 
Chaos driven evolutionary algorithms for the task of PID control, 
Computers \& Mathematics with Applications 60(4) (2010) 1088--1104
%
\bibitem{Persohn:Povinelli-2012} 
K.J. Persohn, R.J. Povinelli, 
Analyzing logistic map pseudorandom number generators for periodicity induced by finite precision floating-point representation, 
Chaos, Solitons and Fractals 45(3) (2012) 238--245
%
%
\bibitem{Drutarovsky:Galajda-2007} 
M. Drutarovsky, P. Galajda, 
A robust chaos-based true random number generator embedded in reconfigurable switched-capacitor hardware, 
17th International Conference Radioelektronika (2007) 29--34
%
\bibitem{Wang:Qin-2012} 
X.Y. Wang, X. Qin, A new pseudo-random number generator based on CML and chaotic iteration, 
Nonlinear Dynamics 70(2) (2012)1589--1592 
%
\bibitem{Pareek:Patidar:Sud-2010} 
N.K. Pareek, V. Patidar, and K.K. Sud, 
A Random Bit Generator Using Chaotic Maps, 
International Journal of Network Security 10!1) !2010) 32--38
%
\bibitem{Bucolo:Caponetto:Fortuna:Frasca:Rizzo-2002} 
M. Bucolo, R. Caponetto, L. Fortuna, M. Frasca, A. Rizzo,
Does chaos work better than noise?
IEEE Circuits and Systems Magazine 2(3) (2002) 4--19 
%
\bibitem{Hu:Liu:Ding-2013} 
HP. Hu, LF. Liu, N. Ding,
Pseudorandom sequence generator based on the Chen chaotic system, 
Computer Physics Communications 184(3) (2013) 765--768
%
%
\bibitem{Pluchino:Rapisarda:Tsallis-2013} 
A. Pluchino, A. Rapisarda, C. Tsallis, 
Noise, synchrony, and correlations at the edge of chaos, 
Phys. Rev. E 87(2) (2013) 022910
%
\bibitem{Falconer:Gottwald:Melbourne:Wormnes-2007} 
I. Falconer, G.A. Gottwald, I.Melbourne, K. Wormnes, 
Application of the 0-1 test for chaos to experimental data,
SIAM Journal on Applied Dynamical Systems 6(2) (2007) 395--402
%
\bibitem{Litak:Syta:Wiercigroch-2009} 
G. Litak, A. Syta, M. Wiercigroch, 
Identification of chaos in a cutting process by the 0–1 test,
Chaos, Solitons \& Fractals 40(5) (2009) 2095-2101
%
\bibitem{Krese:Govekar-2012} 
B. Krese, E. Govekar, 
Nonlinear analysis of laser droplet generation by means of 0–1 test for chaos,
Nonlinear Dynamics, 67(3) (2012) 2101--2109
%
\bibitem{Ke-Hui:Xuan:Cong-Xu-2010} 
S. Ke-Hui, L. Xuan, Z. Cong-Xu, 
The 0-1 test algorithm for chaos and its applications,
Chinese Physics B, 19(11) (2010) 110510
%
%
\bibitem{Matsumoto:Takuji-1998} 
M. Matsumoto, N. Takuji,
Mersenne twister: a 623-dimensionally equidistributed uniform pseudo-random number generator,
ACM Transactions on Modeling and Computer Simulation (TOMACS) 8(1) (1998) 3--30
%
\bibitem{Martinez:Adamatzky:Alonso-Sanz-2013} 
G.J. Mart{\'i}nez, A. Adamatzky, A., R. Alonso-Sanz,  Designing Complex Dynamics with Memory, 
{\em Int. J. Bifurcation and Chaos} {\bf 23(10)} (2013) 1330035--131
%
\bibitem{Alonso-Sanz-2009} 
R. Alonso-Sanz, {\em Cellular Automata with Memory}, 
Old City Publishing, Inc. (2009) 
%
\bibitem{Chen:Chen:Martinez-2017} 
B. Chen, F. Chen, G.J. Mart{\'i}nez,  Glider Collisions in Hybrid Cellular Automaton with Memory Rule (43,74), 
{\em International Journal of Bifurcation and Chaos} {\bf 27(6)} (2017) 1750082
%
\bibitem{Eppstein-WCA} 
D. Eppstein, ``Wolfram's Classification of Cellular Automata'', 
\url{https://www.ics.uci.edu/~eppstein/ca/wolfram.html}.
%
%
\bibitem{Martinez:Morita-2018} 
G.J. Mart{\'i}nez, K. Morita,  Conservative Computing in a One-dimensional Cellular Automaton with Memory, 
{\em Journal of Cellular Automata} {\bf 13(4)} (2018) 325--346
%
\bibitem{Martinez:Adamatzky:Morita-2018} 
G.J. Mart\'{\i}nez, A. Adamatzky, K. Morita, Logical Gates via Gliders Collisions, 
In: {\em Reversibility and Universality}, A. Adamatzky (Ed.), Springer (2018) 199--220
%
\bibitem{Harding:Miller-2007} 
S. Harding, J.F. Miller, Evolution in materio: Evolving logic gates in liquid crystal,
International Journal of Unconventional Computing {\bf 3(4)} (2007) 243--257
%
\bibitem{Toth:Stone:Adamatzky:Costello:Bull-2008} 
R. Toth, C. Stone, A. Adamatzky, B.d.L. Costello, L. Bull, 
Dynamic control and information processing in the Belousov–Zhabotinsky reaction using a coevolutionary algorithm,
The Journal of Chemical Physics, {\bf 129(18)}  (2008) 184708.
%
\bibitem{Harding:Koutnik:Schmidhuber:Adamatzky-2018} 
S. Harding, J. Koutnik, J. Schmidhuber, A. Adamatzky, Discovering Boolean Gates in Slime Mould,
{\em In} Inspired by Nature -- Emergence, Complexity and Computation 28, S. Stepney, A. Adamatzky (Eds.) (2018) 323--337 
%
%
\bibitem{Adamatzky-2017a} 
A. Adamatzky, Computing in Verotoxin. 
ChemPhysChem 18(13),  (2017) 1822--1830.
%
\bibitem{Adamatzky-2017b} 
A. Adamatzky, Logical gates in actin monomer,
Scientific reports 7(1)  (2017) 11755
%
\bibitem{Adamatzky-2009} 
A. Adamatzky, L. Bull, Are complex systems hard to evolve?
Complexity 14(6) (2009) 15--20.
%
\bibitem{Yao-1982} 
A.C. Yao, Theory and application of trapdoor functions,
SFCS'1982, 23rd Annual Symposium on Foundations of Computer Science, IEEE (1982) 80--91
%
\bibitem{Impagliazzo:Wigderson-1997} 
R. Impagliazzo, A. Wigderson, P=BPP unless E has sub-exponential circuits: Derandomizing the XOR Lemma,
Proceedings of the 29th Symposium on Theory of Computing STOC (1997).220-229
%
%
\bibitem{Zhang:Shi-2009} 
Z. Zhang, Y. Shi, Communication complexities of symmetric XOR functions,
Quantum Information \& Computation 9(3) (2009) 255--263 
%
\bibitem{Montanaro:Osborne-2010} 
A. Montanaro, T. Osborne, On the communication complexity of XOR functions,
Computational Complexity, arXiv:0909.3392 (2010) 1--18
%
%
\end{thebibliography}
\end{document}